\RequirePackage{fix-cm}
\documentclass[smallextended]{svjour3}       
\smartqed  
\usepackage{graphicx}
\usepackage{cite}
\usepackage{amsmath,amssymb,amsfonts}
\usepackage{appendix}
\usepackage{textcomp}
\usepackage{algorithm}
\usepackage{algorithmic}
\usepackage{xcolor}
\begin{document}

\title{Parameterised-Response Zero-Intelligence Traders
}


\author{Dave Cliff}


\institute{Dave Cliff \at
              Department of Computer Science,
              University of Bristol, 
              Bristol BS8 1UB, UK.\\
              \email{csdtc@bristol.ac.uk}     \hspace*{15em}
              Copyright \copyright 2023 Dave Cliff.      
}

\date{Received: 28 June 2021 / Accepted: 14 April 2023}

\maketitle

\begin{abstract}
I introduce PRZI (Parameterised-Response Zero Intelligence), a new form of zero-intelligence (ZI) trader intended for use in simulation studies of the dynamics of continuous double auction markets. Like Gode \& Sunder's classic ZIC trader, PRZI generates quote-prices from a random distribution over some specified domain of discretely-valued allowable quote-prices. Unlike ZIC, which uses a uniform distribution to generate prices, the probability distribution in a PRZI trader is parameterised in such a way that its probability mass function (PMF) is determined by a real-valued control variable $s$ in the range $[-1.0, +1.0]$ that determines the {\em strategy} for that trader. When $s=0$, a PRZI trader behaves identically to the ZIC strategy, with a uniform PMF; but when $s \approx \pm 1$  the PRZI trader's PMF becomes maximally skewed to one extreme or the other of the price-range, thereby making it more or less ``urgent'' in the prices that it generates, biasing the quote-price distribution toward or away from the trader's limit-price.
To explore the co-evolutionary dynamics of populations of PRZI traders that dynamically adapt their strategies, I show initial results from long-term market experiments in which each trader uses a simple stochastic hill-climber algorithm to repeatedly evaluate alternative $s$-values and choose the most profitable at any given time. In these experiments the profitability of any particular $s$-value may be non-stationary because the profitability of one trader's strategy at any one time can depend on the mix of strategies being played by the other traders at that time, which are each themselves continuously adapting. Results from these market experiments demonstrate that the population of traders' strategies can exhibit rich dynamics, with periods of stability lasting over hundreds of thousands of trader interactions interspersed by occasional periods of change.   


Python source-code for PRZI traders, and for the stochastic hill-climber, have been made publicly available on GitHub.


\end{abstract}


\section*{Acronyms and Abbreviations}
\begin{center}
\begin{tabular}{c l l}
\hline
Characters & Expansion & First Use \\
\hline
AA & Adaptive Aggressive (adaptive trading strategy) & Section~\ref{sec:intro} \\
AF & Asymptotic Form ({\em PMF} envelope) & Section~\ref{sec:przi-motivation} \\
AI & Artificial Intelligence & Section~\ref{sec:intro} \\
AKA & Also Known As & Section~\ref{sec:intro} \\
ABM & Agent-Based Model & Section~\ref{sec:intro} \\
ACE & Agent-Based Computational Economics & Section~\ref{sec:intro} \\
BG & Balanced Group (experiment design) & Section~\ref{sec:coevolve} \\
BSE & Bristol Stock Exchange (simulation platform) & Section~\ref{sec:shvr} \\
CDA & Continuous Double Auction & Section~\ref{sec:background} \\
CDF & Cumulative Distribution Function & Section~\ref{sec:przi-details} \\
CIAO & Current Individual vs.\  Ancestral Opponent & Section~\ref{sec:prsh_coev} \\
GD & Gjerstad-Dickhaut (adaptive trading strategy) & Section~\ref{sec:intro} \\
GDX & {\em GD} Extended (adaptive trading strategy) & Section~\ref{sec:intro} \\
GVWY & Giveaway (nonadaptive trading strategy) & Section~\ref{sec:intro} \\
HBL & Heuristic-Based Learning (adaptive trading strategy) & Section~\ref{sec:intro} \\
IBM & International Business Machines Corp. & Section~\ref{sec:intro} \\
IID & Independent and Identically Distributed & Section~\ref{sec:prsh_solo} \\
IPRZI & Imbalance-sensitive {\em PRZI} (nonadaptive trading strategy) & Section~\ref{sec:impact} \\
ISHV & Imbalance-sensitive {\em SHVR} (nonadaptive trading strategy) & Section~\ref{sec:impact} \\
JPE & Journal of Political Economy & Section~\ref{sec:background} \\
LOB & Limit Order Book & Section~\ref{sec:background} \\
LOI & Line of Identity & Section~\ref{sec:coevolve} \\
LUT & Look-Up Table & Section~\ref{sec:przi-details} \\
MAB & Multi-Armed Bandit & Section~\ref{sec:prsh_defn} \\
MGD & Modified {\em GD} (adaptive trading strategy) & Section~\ref{sec:intro}  \\
MI & Minimal Intelligence & Section~\ref{sec:intro} \\
ML & Machine Learning & Section~\ref{sec:intro} \\
MLOFI & Multi-Level Order-Flow Imbalance & Section~\ref{sec:impact} \\
NZI & Near-Zero Intelligence (nonadaptive trading strategy) & Section~\ref{sec:opinions} \\ 
OD & Opinion Dynamics & Section~\ref{sec:opinions} \\
OIM & One-in-Many (experiment design) & Section~\ref{sec:coevolve} \\
ONZI & Opinionated {\em NZI} (nonadaptive trading strategy) & Section~\ref{sec:opinions} \\
OPRZI & Opinionated {\em PRZI} (nonadaptive trading strategy)  & Section~\ref{sec:opinions} \\
OZIC & Opinionated {\em ZIC} (nonadaptive trading strategy) & Section~\ref{sec:opinions} \\
PMF & Probability Mass Function & Section~\ref{sec:ZIC} \\
PRSH & {\em PRZI} Stochastic Hillclimber (adaptive trading strategy) & Section~\ref{sec:coevolve} \\
PRZI & Parameterized-Response {\em ZI} (nonadaptive trading strategy)& Section~\ref{sec:intro} \\
RDA & Replicator Dynamics Analysis & Section~\ref{sec:coevolve} \\
RP & Recurrence Plot & Section~\ref{sec:prsh_coev} \\
RMS & Root Mean Square & Section~\ref{sec:ZIC} \\
RQA & Recurrence Quantification Analysis & Section~\ref{sec:prsh_coev} \\ 
SHVR & Shaver (nonadaptive trading strategy) & Section~\ref{sec:intro} \\
SNPR & Sniper (nonadaptive trading strategy) & Section~\ref{sec:intro} \\
ZI & Zero Intelligence (class of trader-agents) & Section~\ref{sec:intro} \\
ZIC & {\em ZI} Constrained (nonadaptive trading strategy) & Section~\ref{sec:intro} \\
ZIP & {\em ZI} Plus (adaptive trading strategy) & Section~\ref{sec:intro} \\
ZIU & {\em ZI} Unconstrained (nonadaptive trading strategy) & Section~\ref{sec:ZIC} 
\end{tabular}
\end{center}

\section*{Notations}
\begin{center}
\begin{tabular}{c l l}
\hline
Symbol & Denotes & First Use \\
\hline
$\alpha$ & Smith's equilibration metric & Section~\ref{sec:ZIC} \\
$\delta_p$ & The market's tick-size & Section~\ref{sec:3strats} \\
$\Delta_m(t)$ & Top-of-LOB imbalance metric $\Delta_m(t)=p_\mu(t) - p_m(t)$ & Section~\ref{sec:impact} \\
$\Delta_P$ & Difference between minimum and maximum price on a PMF & Section~\ref{sec:przi-motivation} \\
$\Delta_s$ & Step-size in strategy space when mapping fitness landscape & Section~\ref{sec:coevolve} \\
$\Delta_t$ & Timestep duration in the BSE simulation: $\Delta_t=1/N_T$ seconds & Section~\ref{sec:3strats} \\
$\epsilon$ & Tiny threshold value to avoid divide-by-zero errors in $\theta(x)$ & Section~\ref{sec:przi-details} \\
$F_P(p)$ & Cumulative Distribution Function (CDF) for PRZI trader & Section~\ref{sec:przi-details} \\
$ \widehat{F_P^{-1}}(c)$ & Approximation to inverse CDF, via reverse table-lookup & Section~\ref{sec:przi-details} \\
${\cal F}(\cdot)$ & Fitness function used in stochastic hillclimber & Section~\ref{sec:coevolve} \\
${\cal G}(\cdot)$ & Genesis function used in stochastic hillclimber to create ${\cal S}_{i,0}$ & Section~\ref{sec:coevolve} \\
${\cal I}_i(\cdot)$ & Market-impact function for trader $i$ & Section~\ref{sec:impact} \\
$k$ & Number of different strategies held by a PRSH trader & Section~\ref{sec:coevolve} \\
$\lambda_b$ & Buyer's limit-price  & Section~\ref{sec:3strats} \\
$\lambda_s$ & Seller's limit-price  & Section~\ref{sec:3strats} \\
$\lambda_{s(i:\text{max})}(t_0)$ & Largest limit price assigned to seller $i$ at any time $t\leq t_0$  & Section~\ref{sec:3strats} \\
${\cal M}(\cdot)$ & Mutation function used in stochastic hillclimber & Section~\ref{sec:coevolve} \\
$N_\text{Buy}$ & Number of buyers in the market & Section~\ref{sec:prsh_solo} \\
$N_\text{Sell}$ & Number of sellers in the market & Section~\ref{sec:prsh_solo} \\
$N_R$ & Number of IID repetitions of an experiment & Section~\ref{sec:coevolve} \\
$N_S$ & Number of discrete strategies available to choose between  & Section~\ref{sec:coevolve} \\
$N_T$ & Number of traders in the market: $N_T = N_\text{Buy} + N_\text{Sell}$ & Section~\ref{sec:coevolve} \\
$\Omega$ & Opinionated limit price & Section~\ref{sec:opinions} \\
$\omega_i$ & Opinion value for trader $i$ & Section~\ref{sec:opinions} \\ 
$\pi_B$ & Total profit/surplus extracted by the set of buyers & Section~\ref{sec:coevolve} \\
$\pi_S$ & Total profit/surplus extracted by the set of sellers & Section~\ref{sec:coevolve} \\
$\pi_T$ & Total profit/surplus extracted by all traders: $\pi_T = \pi_B + \pi_S$ & Section~\ref{sec:coevolve} \\
$p_0$ & Equilibrium price & Section~\ref{sec:background} \\
$p^*_{\text{ask}}(t)$ & Price of the best ask on the LOB at time $t$ & Section~\ref{sec:background} \\\
$p^*_{\text{bid}}(t)$ & Price of the best bid on the LOB at time $t$ & Section~\ref{sec:background} \\
$p_{\text{max}}$ & Arbitrary maximum price allowed in the market & Section~\ref{sec:ZIC} \\
$p_m(t)$ & Mid-price on the LOB at time $t$ & Section~\ref{sec:impact} \\
$p_\mu(t)$ & Micro-price on the LOB at time $t$ & Section~\ref{sec:impact} \\
$P_{bq(\text{STRAT})}(t)$ & Price quoted at time $t$ by a buyer of strategy-type {\sc strat} & Section~\ref{sec:3strats} \\
$P_{sq(\text{STRAT})}(t)$ & Price quoted at time $t$ by a seller of strategy-type {\sc strat} & Section~\ref{sec:3strats} \\
${\mathbb P}(P=p)$ & Probability that random variable $P$ is equal to price $p$ & Section~\ref{sec:przi-motivation} \\
${\cal P}(\cdot)$ & PMF envelope profile function & Section~\ref{sec:przi-details} \\
${\cal PMF}_i(\cdot)$ & PMF for trader $i$ & Section~\ref{sec:przi-details} \\ 
$q_0$ & Equilibrium quantity & Section~\ref{sec:background} \\
$q^*_{\text{ask}}(t)$ & Quantity available at price $p^*_{\text{ask}}(t)$ on the LOB at time $t$ & Section~\ref{sec:impact} \\\
$q^*_{\text{bid}}(t)$ & Quantity available at price $p^*_{\text{bid}}(t)$ on the LOB at time $t$ & Section~\ref{sec:impact} \\\
$s_i$ & PRZI strategy-value for trader $i$: $s_i \in [-1,+1] \in {\mathbb R}$ & Section~\ref{sec:intro} \\
$\hat{s}$ & moving average of $s$ & Section~\ref{sec:prsh_solo} \\
$\widehat{S_{T}}$  & The set of terminal $\hat{s}$ values from a population of traders & Section~\ref{sec:prsh_solo} \\
$\vec{S}(t)$ & Strategy-vector for all-PRSH market: $|\vec{S}(t)|=N_T$; $[\vec{S}(t)]_i=s_i$ & Section~\ref{sec:coevolve} \\
${\cal S}_{i,t} $ & Set of $k$ different $s_i$-values at time $t$ for individual PRSH trader $i$ & Section~\ref{sec:coevolve} \\
$t$ & Time: $t \geq 0 \in {\mathbb R}$ & Section~\ref{sec:background} \\
$\theta(x)$ & Linear rectifier function & Section~\ref{sec:przi-details} \\
${\cal U}(a, b)$ & Draws from a uniform random distribution over the range $[a,b]$ & Section~\ref{sec:3strats} \\
\end{tabular}
\end{center}

\section{Introduction}
\label{sec:intro}

In attempting to understand and predict the fine-grained dynamics of financial markets, there is a long tradition of studying simulation models of such markets. Simulation studies nicely complement the two primary alternative lines of enquiry: analysis of real market data recorded at fine-grained temporal resolution, as is studied in the branch of finance known as  {\em market microstructure}; and running carefully planned experiments where human subjects interact in artificial markets under controlled laboratory conditions, i.e.\ {\em experimental economics}. Simulation modelling of financial markets very often involves creating agent-based models (ABMs) that populate a market mechanism with some number of {\em trader-agents}: autonomous entities that have ``agency'' in the sense that they are empowered to buy and/or sell items within the particular market mechanism that is being simulated. This approach, known as {\em agent-based computational economics} (ACE), has a history stretching back for more than 30 years. Over that multi-decade history, a small number of specific zero-intelligence (ZI) and/or minimal-intelligence (MI) trader-agent algorithms, i.e. precise mathematical and procedural specifications of particular trading strategies, have been frequently used for modelling various aspects of financial markets, and the convention that has emerged is to refer to each such strategy via an acronym or short sequence of letters, reminiscent of a stock-market ticker-symbol.\footnote{Notable trading strategies in this literature include (in chronological sequence): SNPR ({\sc aka} {\em Kaplan's Sniper}, as described in \cite{rust_etal_1992}), ZIC ({\em Zero Intelligence Constrained}: \cite{gode_sunder_1993});  ZIP ({\em Zero Intelligence Plus}: \cite{cliff_1997_zip}); RE ({\em Roth-Erev}: \cite{erev_roth_1998}); GD ({\em Gjerstad-Dickhaut}:  \cite{gjerstad_dickhaut_1998}); MGD ({\em Modified GD}: \cite{tesauro_das_2001}); GDX ({\em GD eXtended}: \cite{tesauro_bredin_2002}); HBL ({\em Heuristic Based Learning}: \cite{gjerstad_2003});  AA ({\em Adaptive Aggressive}: \cite{vytelingum_etal_2008}) and IEL ({\em Individual Evolutionary Learning}: \cite{arifovic_ledyard_2011}); several of which are explained in more detail later in this paper. } 
Of these, ZIC \cite{gode_sunder_1993} is notable for being both highly stochastic and extremely simple, and yet it gives surprisingly human-like market dynamics; GD \cite{gjerstad_dickhaut_1998} and ZIP \cite{cliff_1997_zip} were the first two strategies to be demonstrated as superior to human traders, a fact first established in a landmark paper by IBM researchers \cite{das_etal_2001} (see also: \cite{deluca_cliff_2011_icaart,deluca_cliff_2011_ijcai,deluca_etal_2011_foresight}), which is now commonly pointed to as initiating the rise of algorithmic trading in real financial markets; and until very recently AA \cite{vytelingum_etal_2008} was widely considered to be the best-performing strategy in the public domain. With the exception of SNPR \cite{rust_etal_1992} and ZIC, all later strategies in this sequence are adaptive, using some kind of machine learning (ML) or artificial intelligence (AI) method to modify their responses over time, better-fitting their trading behavior to the specific market circumstances that they find themselves in, and details of these algorithms were often published in major AI/ML conferences and journals. 

The supposed dominance of AA has recently been questioned in a series of publications \cite{vach_2015,cliff_2019,snashall_cliff_2019,rollins_cliff_2020,cliff_rollins_2020} which demonstrated AA to have been less robust than was previously thought. Notably, \cite{rollins_cliff_2020,cliff_rollins_2020} report on trials where AA is tested against two novel nonadaptive algorithms that each involve no AI or ML at all: these two newcomer strategies are known as GVWY and SHVR \cite{cliff_2012_bse,cliff_2018_bse}, and each share the pared-back minimalism of Gode \& Sunder's ZIC mechanism. In the studies that have been published thus far, depending on the circumstances, it seems (surprisingly) that GVWY and SHVR can each outperform not only AA but also many of the other AI/ML-based trader-agent strategies in the set listed above. Given this surprising recent result, there is an appetite for further zero-intelligence ACE-style market-simulation studies involving GVWY and SHVR. One compelling issue to explore is the co-adaptive dynamics of markets populated by traders that can choose to play one of the three strategies from GVWY, SHVR, and ZIC, in a manner similar to that studied by \cite{walsh_etal_2002} who employed `replicator dynamics' modelling techniques borrowed from theoretical evolutionary biology to explore the co-adaptive dynamics of markets populated by traders that could choose between SNPR, ZIP, and GD. 

One way of studying co-adaptive dynamics in markets where the traders can choose to either deploy GVWY, SHVR, or ZIC is to give each trader a discrete choice of one from that set of three strategies, such that at any one time any individual trader is either operating according to GVWY or SHVR or ZIC. However it is appealing to instead design experiments where the traders can {\em continuously vary} their trading strategy, exploring a potentially infinite range of differing strategies, where the space of possible strategies includes GVWY, SHVR, and ZIC; and that is the motivation for this paper. Here, I introduce a new trading strategy that has a {\em parameterised response}: that is, its trading behavior is determined by a strategy parameter $s \in [-1,+1] \in {\mathbb R} $: when $s=0$, the trader behaves identically to ZIC; and when $s = \pm 1$ it behaves the same as GVWY or SHVR; but $s$ can also take on any other value in its range, such as $-0.75$ or $+0.5$, which gives novel ``hybrid'' trading behavior part-way between ZIC and either GVWY or SHVR.  As is explained in more detail later in this paper, GVWY, ZIC, and SHVR are each members of the class of {\em zero intelligence} (ZI) trading strategies, and hence I've named the new strategy described here as the Parameterized-Response Zero-Intelligence (PRZI) trading strategy. The acronym PRZI is pronounced like ``pressie''.

To provide a zero-intelligence model trader for studying evolutionary or adaptive markets (as discussed in, for example:  \cite{friedman_1991_evolecon,blume_easley_1992,friedman_1998_evolecon,lo_2004,lo_2019,nelson_2020_evolecon}), each PRZI trader needs some adaptation mechanism, allowing it to adjust its individual $s$-value over time, to be better suited to the prevailing market conditions that the particular trader finds itself in. There are many potential adaptation mechanisms that could be used, but the results in this paper come -- in the minimalist style of ZI traders -- from a very basic adaptive algorithm (arguably, the simplest possible): a $k$-point stochastic hill-climber, the operation of which is described in detail below, in Section~\ref{sec:coevolve}. I refer to traders with this adaptation mechanism, {\bf PR}ZI with {\bf S}tochastic {\bf H}ill-climbing, as PRSH traders (pronounced ``pursh'').  PRSH is offered here as an absolutely minimal model of an adaptive trader -- it has only a single parameter (unlike, for example, ZIP which has between 8 and 60 parameters, depending on which version is used: see \cite{cliff_2009_zip60}), and it does usefully adapt the value of that parameter over time (although, as discussed further below, there are many better ways of doing adaptation). Section~\ref{sec:coevolve} presents results and analysis from multiple experiments with markets populated entirely by PRSH traders -- these are zero-intelligence {\em adaptive markets}, in the sense popularised by Lo \cite{lo_2004,lo_2019}.

While one motivation for devising PRZI was just described: to enable explorations of co-adaptive dynamics in continuous strategy-spaces, that is not the only motivation. Two other compelling reasons for wanting a ZI-style trader with a variable response are as follows:
\begin{itemize}
\item
Recent publications \cite{church_cliff_2019,zhang_cliff_2021} have described methods for making these simple trader-agents exhibit a form of sensitivity to temporary imbalances between supply and demand in the marketplace, and that imbalance-sensitivity then gives rise to so-called ``market impact'' effects, in which the prices quoted by traders shift in the {\em anticipated} direction of future transaction prices, where the shift is anticipated from the degree of supply/demand imbalance instantaneously evident in the market. Market impact is a significant issue for traders in real markets who are looking to buy or sell an unusually large amount of some asset, and major exchange operators such as the London Stock Exchange have introduced specialised mechanisms to try to reduce the ill-effect of market impact (see e.g. \cite{church_cliff_2019} for further discussion). For markets populated by ZI trader-agents to be able to exhibit impact effects, the traders need to be able to modulate their trading activity according to the direction and degree of imbalance in the market, becoming either more ``urgent'' or more ``relaxed'' in their trading: varying PRZI's strategy-parameter $s$ implements exactly this kind of tuneable response, as is illustrated in Section~\ref{sec:impact}.
\item
Shiller has recently proposed \cite{shiller_2017,shiller_2019} that certain economic phenomena which defy easy explanation via classical assumptions of individual economic rationality can be better understood by reference to the {\em narratives} (i.e., the stories) that economic agents tell themselves and each other about current and future economic factors. Shiller refers to this new approach as {\em narrative economics}. Noting that stories are merely externalisations of an agent's internally-held {\em opinions}, a recent publication \cite{lomas_cliff_2021} described an agent-based modelling platform for studying issues in narrative economics, in which two types of ZI traders were extended to also each include a real-valued {\em opinion} variable (the value of which could be altered by interactions with other agents, thereby modelling the way in which an agent's opinions are shifted by the narratives it is exposed to) and adapted so that their trading strategies alter as a function of their individual opinion: this also gives rise to a need for ZI traders that can smoothly vary the nature of their trading behavior, and PRZI was designed with the intention of being used in such opinion dynamics modelling work: exploring the use of PRZI in ACE-style agent-based models is a topic of current research, discussed further in Section~\ref{sec:opinions}.
\end{itemize} 

This paper is intended merely as an introduction to PRZI; it is beyond the scope of this document to provide a comprehensive and detailed literature review. Readers seeking further details of market microstructure are referred to \cite{ohara_1998,harris_2002,lehalle_laruelle_2018}, and for overviews of experimental economics see for example \cite{kagel_roth_1997,smith_2000,plott_smith_2008}. For reviews of ACE research, see \cite{tesfatsion_judd_2006,chen_2011_jedc,chen_2018,hommes_lebaron_2018}; and for discussions of ZI traders in finance research, see e.g.\ \cite{farmer_etal_2005,ladley_2012,cartlidge_etal_2012}.


\section{Background: Experimental Economics}
\label{sec:background}

In a landmark 1962 paper \cite{smith_1962} published in the {\em Journal of Political Economy} (JPE), Vernon Smith described a seminal set of experiments in which human traders interacted within a {\em continuous double auction} (CDA), the mechanism embodied in most major real-world financial markets, under laboratory conditions. The introduction to Smith's 1962 JPE paper rightly cited the earlier work of Chamberlin who had described results from an experimental market in a JPE paper published in 1948 \cite{chamberlin_1948}. Smith's 1962 paper, and Chamberlin's before that, are widely regarded as marking the birth of experimental economics, and Smith's work in this field led to him being awarded the 2002 Nobel Prize in Economics. 

In the simplest case, such experimental economics work involves a market with only one type of a tradeable asset (think of it as a stock market on which only one stock is listed), and each human trader can buy or sell a specific quantity of the asset by issuing one or more {\em quotes} to the market's central exchange. A quote would specify: the trader's desired {\em direction}\/ (i.e., buy or sell) for the transaction; the quantity (number of units) that the trader is seeking to transact; and the price-per-unit that they want to pay or be paid. Each trader would be given instructions, referred to here as {\em assignments}, that are private and specific to that individual trader, and each trader is told to keep these instructions secret. Some traders will be instructed to buy some quantity of the asset, paying no more than a trader-specific maximum price per unit (i.e., these traders are {\em buyers} each with a specific {\em limit price}); other traders will have been instructed to sell some quantity of the asset, accepting no less than some trader-specific minimum price per unit (so they are {\em sellers}, again each with their own limit-price). In this way, instructing some traders to be buyers and other traders to be sellers, and by varying the limit-prices in each trader's instructions, the market's underlying supply and demand curves could be controlled, along with that market's competitive equilibrium price (denoted by $p_0$) and equilibrium quantity ($q_0$). 
Smith's 1962 paper (which reported on a sequence of experiments that he had commenced several years earlier) was notable for establishing a set of experiment methods that have since been reproduced and replicated by researchers around the world, and also for being the first empirical demonstration that CDA markets could show reliable equilibration even when populated with only very small numbers of buyers and sellers. 

In many experimental economics studies, equilibration is not the only factor of interest. Another significant question is how much {\em surplus} is extracted from the market by the specific trading behaviors of the traders. In everyday language, surplus can be thought of a a seller's profit or a buyer's saving: if a seller has been given a limit-price of \$10 per unit, but manages to agree a transaction at \$15, then the \$5 difference between the seller's limit-price and the sale-price is that seller's surplus, her profit; similarly if a buyer is given a limit-price of \$10 but manages to instead buy for \$8, then her saving, her surplus, is \$2. 

The initial experiments reported by Smith in 1962 were conducted with entirely manual issuing of assignments to traders, and with the traders simply shouting out their quote-prices in a laboratory version of an open-outcry trading-pit, a common sight on the trading floors of major exchanges for many decades prior to the arrival of automated market-trading technology. More recently, as in real financial exchanges, so in experimental economics: most experimental economics studies for the past 30 years or more have involved the human traders interacting with one another by each being sat at an individual trader-terminal (e.g. a PC running specialised trader-interface software, networked to a central exchange-server PC). The display-screen on a trader-terminal (whether in a real market or in a laboratory experiment) will often show a summary of all the currently outstanding bid-quotes received from buyers, and all the currently outstanding ask-quotes received from sellers, in a tabular form known as a {\em Limit Order Book} (LOB). Whole books have been written on the LOB (see e.g.\ \cite{osterrieder_2009,nolte_etal_2014,abergel_etal_2016}) but for the purposes of this paper, all we need to know is that the LOB allows all traders in the market to see the best (lowest) ask-price from a seller and the best (highest) bid-price from a buyer. In the rest of this paper I'll use $p^*_{\text{ask}}(t)$ to denote the price of the best ask at time $t$, and $p^*_{\text{bid}}(t)$ to denote the price of the best bid.

Any study in experimental economics where human traders interact with one another, via some market mechanism, subject to the constraints imposed by the design of the experiment, gives rise to the question of just how big a role the intelligence of the human traders plays in determining the market's equilibration behavior. A genuinely shocking answer to that question was provided in 1993 by Gode \& Sunder, also publishing in the JPE \cite{gode_sunder_1993}, who presented results which appeared to show that the answer was simple: zero. That is, Gode \& Sunder showed that markets populated entirely by so-called {\em zero-intelligence} (ZI) traders could give rise to market dynamics, to equilibration behavior, which was statistically indistinguishable from that of human traders, when measured by the then-standard metric known as {\em allocative efficiency}, i.e. how much of the available surplus in the market was extracted by the traders. Gode \& Sunder's ZI traders manifestly have no intelligence at all, and are discussed in more detail in the next section.

\section{Three ZI Trading Strategies}
\label{sec:3strats}

This section describes the three trading strategies {\em Zero-Intelligence-Constrained} (ZIC), {\em Shaver} (SHVR), and {\em Giveaway} (GVWY). As will be seen, all three of these can very reasonably be described as ZI trading strategies. 

In the following text, ${\cal U}(a,b)$ is used to denote draws from a uniform random distribution over the range $[a, b]$;  the integer $\delta_p$ denotes the market's {\em tick-size}, the minimum price change allowed in the market (very often --but not always -- one cent of the national currency in real financial exchanges, see e.g.\  \cite{darley_outkin_2007,chung_lee_roesch_2020,chung_chuwonganant_2022,cartea_chang_penalva_2022}); all prices are integer multiples of $\delta_p$, and hence members of ${\mathbb Z}$; and $P_{bq(\text{\sc strat})}(t)$ and $P_{sq(\text{\sc strat})}(t)$ denotes the prices quoted by a buyer and a seller, respectively, at time $t$, by a trader of strategy-type {\sc strat} (where {\sc strat} is one of a set of known strategy-types, for example {\sc strat} $\in$ \{GD, ZIC, ZIP\}). Note that a capital $P$ is used to denote a price that is (or could be, in principle) randomly generated, i.e.\ a random variable; and the various lower-case $p$ values are nonrandom. Time proceeds in discrete steps of $\Delta_t \in {\mathbb R}$, and each strategy is summarised as an equation that specifies the price that will be quoted by a trader at time $t+\Delta_t$ on the basis of information available, the state of the market, at time $t$. The limit-price assigned to a buyer is denoted by $\lambda_b$, and the seller's limit-price is $\lambda_s$. The subscripted index $i$ is used where necessary to distinguish values that are specific to trader $i$.

\subsection{ZIC}
\label{sec:ZIC}

In their seminal 1993 JPE paper \cite{gode_sunder_1993}, Gode \& Sunder presented results from three sets of experimental economics studies: in the first, groups of human traders interacted in an electronic CDA, the mechanism found in real-world financial markets, under laboratory conditions, as described above. As is commonplace in much experimental economics work, Gode \& Sunder fixed the limit-quantity to one for all traders, so that transactions always involved a single unit of the asset changing hands, and hence the primary variable of interest was the transaction prices in the market. This first set of experiments established baseline data from the human traders, and Gode \& Sunder recorded a key metric, $\alpha$, first introduced in Smith's 1962 paper, which measures the variation of transaction prices around the market's $p_0$ value, as the RMS difference between $p_0$ and the transaction prices over some period, i.e. $\alpha$ is a measure of equilibration; and they also recorded the {\em allocative efficiency} for each market, which is a measure of how much of the total theoretically available surplus is actually extracted by the traders in that market. 

In their second set of experiments, they replaced all the human traders with simple autonomous software agents that could electronically issue quotes to the market's central exchange mechanism: as with the human traders in the first experiments, the software agents were each assigned a direction (buy or sell) and given a limit price for their transactions. Gode \& Sunder performed two sets of experiments with these software agents: one set with a type of trader-agent that they named {\em ZI-Unconstrained} (ZIU); and another set with a modified version of the ZIU strategy, one that involved imposition of an additional constraint, and so those traders were named {\em ZI-Constrained} (ZIC). If ever a seller ZI trader issued a quote-price below the current best bid-price (i.e., in the terminology of financial markets, if the quote is for a price that {\em crosses the spread}), that seller sold its unit the buyer who had issued that best bid; and similarly if ever a ZI buyer issued a quote-price that was higher than the current best ask-price issued by a seller, the buyer would buy from that seller, because the buyer's quote crossed the spread. (The {\em spread} is the difference between the current best ask price and the current best bid price).

ZIU traders were very basic: they generated a randomly-selected quote-price drawn from ${\cal U}(\delta_p, p_{\text{max}})$, where $p_{\text{max}}$ is an arbitrarily-chosen maximum-allowable price in the market. That is, ZIU traders were designed to ignore their limit prices. Unsurprisingly, the time-series of transaction prices in markets populated by ZIU traders looked much like random noise, and ZIU traders would often enter into loss-making deals, because they were buying at prices above their limit price or selling at prices below their limit price. 

Gode \& Sunder then modified the ZIU strategy, giving the ZIC strategy, by introducing a just one constraint: to not quote prices that were potentially loss-making. ZIC traders still quote random prices drawn from a uniform distribution over some range, but now there is a difference depending on whether the ZIC trader is a buyer or a seller:
\begin{equation}
P_{bq(\text{ZIC})}(t+\Delta_t) = {\cal U}(\delta_p, \lambda_b);
\end{equation}
\begin{equation}
P_{sq(\text{ZIC})}(t+\Delta_t) = {\cal U}(\lambda_s, p_{\text{max}}).
\end{equation}

That is, a ZIC buyer randomly generates its quote-prices equiprobably from $[\delta_p, \lambda_b]$, where $\lambda_b$ is that buyer's limit price; and a ZIC seller with limit-price $\lambda_s$ generates its quote-prices equiprobably over the range $[\lambda_s, p_{\text{max}}]$. Surprisingly, markets populated by ZIC traders showed equilibration behaviors (as measured by Smith's $\alpha$ metric) and allocative effriciency scores that were virtually indistinguishable from the comparable human-populated markets. This notable result quickly became highly-cited, as it seemed to demonstrate that if there was any `intelligence' in the system at all, it was in the CDA market mechanism rather than residing in the traders. Gode \& Sunder were also careful to note that a different measure of surplus-extraction, called {\em profit dispersion}, was worse for ZIC traders than for human traders. . 

\begin{figure}
\begin{center}
\includegraphics[width=0.5\linewidth]{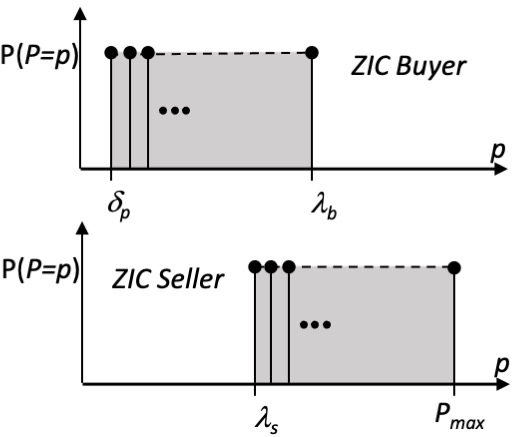}
\end{center}
\caption{Illustrative Probability Mass Function (PMF) for the uniformly-distributed discrete quote-prices generated by a ZIC buyer (upper figure) and seller (lower figure). In this and all other PMF graphs in this paper, the horizontal axis is price, and the vertical axis is probability.}
\label{fig:ZIC_uniform}
\end{figure}

Because quote-prices in any market are quantized by that market's tick-size $\delta_p$, the prices quoted by a ZIC trader are samples of a discrete random variable, and the probability mass function (PMF) for that variable has a rectangular profile, because the distribution is uniform, as illustrated in Figure~\ref{fig:ZIC_uniform}.

One problematic aspect of ZIC traders that becomes clear to anyone who actually implements them to run live in an operational market is that while ZIC buyers have a PMF domain that is bounded from below by the smallest nonzero price $\delta_p$ and bounded from above by the trader's limit price $\lambda_b$, the PMF domain for a ZIC seller is bounded from below by its limit price $\lambda_s$ but the upper bound is an arbitrary exogenous system limit $p_\text{max}$, the highest price allowed in the market. In theory, $p_\text{max}$ might appear to be unimportant, but if its value is set to a large multiple of the largest buyer's limit price $\lambda_{b:\text{max}}$ then the ZIC sellers will spend an awful lot of their time quoting at prices $P_{sq}(t)>\lambda_{b:\text{max}} $, i.e. at prices that can never lead to a transaction, and so the market is flooded with unfulfillable ask-quotes, and you have to wait a long time before a ZIC seller just happens to randomly generate a plausible ask, i.e. one for which $P_{sq}(t)\leq \lambda_{b:\text{max}}$. If the only issue of interest in the market is the temporally-ordered sequence of transaction prices, this is not a problem; but if you care about the actual time intervals between successive transactions, that can be greatly affected by setting $p_\text{max}$ too high. Experimenters also need to be careful to ensure that $p_\text{max}$ is not set below the highest limit-price assignable to a seller in the market, which can be a problem in practice if the seller limit-prices are generated by an unbounded random-walk process such as geometric Brownian motion with positive drift. We'll return to these issues, the ``$p_\text{max}$ problem'', in Section~\ref{sec:PRZI}.

\subsection{SHVR}
\label{sec:shvr}
Source-code for the {\em Shaver} strategy, abbreviated to SHVR, was made public in 2012 \cite{cliff_2012_bse} when it was released as one of the several trader-agent strategies available in the {\em Bristol Stock Exchange} (BSE) which is a freely-available open-source simulation of a LOB-based financial exchange, written in Python. BSE was initially developed as a teaching resource, used by masters-level students studying on a Financial Technology module taught at the University of Bristol. In the years since it was first released, it has been used by hundreds of students and has increasingly also found use as a trusted and stable test-bed for exploring research questions in ACE.  

Like ZIC, SHVR is minimally simple, involving no intelligence at all. Unlike ZIC, SHVR is entirely deterministic. SHVR can be explained very simply: a SHVR buyer sets its quote-price at time $t+\Delta_t$, denoted by $P_{bq(\text{SHVR})}(t)$, to be one tick (i.e., $\delta_p$) more than the current best bid $p^*_{\text{bid}}(t)$, so long as that does not exceed its limit price $\lambda_b$, i.e.:
\begin{equation}
P_{bq(\text{SHVR})}(t+\Delta_t) = \min(p^*_{\text{bid}}(t)+\delta_p, \lambda_b);
\label{eq:SHVRbid} 
\end{equation}
and similarly a SHVR seller sets its quote-price to be: 
\begin{equation}
P_{sq(\text{SHVR})}(t+\Delta_t) =\max(p^*_{\text{ask}}(t)-\delta_p, \lambda_s). 
\label{eq:SHVRask} 
\end{equation}

If the best bid or ask is undefined at time $t$, e.g. because no quotes have yet been issued, a SHVR buyer will start with a very low $P_{bq}(t)$, and a SHVR seller with a very high $P_{sq}(t)$. If there is no prior trading data available in the market, the low value could be $\delta_p$ and the high value $p_{\text{max}}$, i.e. the same lower and upper bounds as for ZIC traders; if there is prior trading data, the low and high initial values could instead be the lowest and highest prices seen in prior trading over some recent time-window.  

SHVR was introduced to BSE as something of a joke, as a tongue-in-cheek approximation to a high-frequency trading algorithm. It does serve the purpose of being a minimal illustrative implementation of a trader that actually uses information available from the LOB. The surprising result (discussed in more detail in \cite{rollins_cliff_2020,cliff_rollins_2020}) is that if the circumstances are in its favour then SHVR can out-perform well-known strategies like ZIP or AA, which had previously been hailed as examples of AI-powered super-human robot-trader systems. And, in one further surprise, the same study that revealed SHVR can outperform ZIP and AA also revealed that an even simpler strategy, called GVWY, can do just as well as SHVR and sometimes better. 

\subsection {GVWY}

Like SHVR, source-code for the {\em Giveaway} strategy (GVWY) was made public when the first version of BSE was released as open-source in 2012 \cite{cliff_2012_bse}.
The correlates of Equations~\ref{eq:SHVRbid} and~\ref{eq:SHVRask} for GVWY are as follows:
\begin{equation}
P_{bq(\text{GVWY})}(t+\Delta_t) = \lambda_b;
\label{eq:GVWYbid} 
\end{equation}
\begin{equation}
P_{sq(\text{GVWY})}(t+\Delta_t) = \lambda_s. 
\label{eq:GVWYask} 
\end{equation}
As can be seen, a GVWY trader simply sets its quote price to whatever its currently-assigned limit price is, regardless of the time. As the name implies, {\em prima facie} this trading strategy gives away any chance of surplus, because there is no difference between its quote price and its limit price: if its quote results in a transaction at that price, it yields zero surplus. 

However, the spread-crossing rule (which is standard in most LOB-based markets) means that it is possible for a GVWY trader to enter into surplus-generating trades. For example, consider a situation in which a GVWY buyer has a limit price $\lambda_b= \$10$, and the current best ask is $p^*_\text{ask}=\$7$: when the GVWY buyer quotes its limit price (i.e., $P_{bq}(t)=\lambda_b$), the \$10 price on the quote crosses the spread and so the GVWY buyer is matched with whichever seller issued that best ask, and the transaction goes through at a price of \$7, yielding a \$3 surplus for the GVWY buyer (and yielding whatever surplus is arising for the seller, dependent on that seller's value of $\lambda_s$). 

In the next section, I explain how PRZI traders have a continuously-variable strategy space that includes GVWY, ZIC, and SHVR: by setting a strategy-parameter $s$ to an appropriate value, PRZI traders will act either as one of those three ZI strategies, or as some kind of novel hybrid, intermediate between two of them; that is, a PRZI trader's response is a parameterised version of one or more of these three ZI trading strategies, hence the name.

\subsection{Summary}
\label{sec:3strats_summary}

Table~\ref{tbl:stratcompare} summarises the three ZI strategies (GVWY, SHVR, and ZIC) that are integrated within PRZI, and also includes Gode \& Sunder's ZIU, for completeness. 

\begin{table}[h]
\begin{center}
\begin{tabular}{|l|l|l|l|l|}
 & \multicolumn{2}{l}{\bf Buyer}  &  \multicolumn{2}{l}{\bf Seller}   \\
 \hline
{\bf Strategy} & $p_{\text{lo}}$    & $p_{\text{hi}}$    & $p_{\text{lo}}$     & $p_{\text{hi}}$     \\
\hline
GVWY & $\lambda_{b}$    &  $p_\text{lo} $  &  $\lambda_{s}$  &    $p_\text{lo} $      \\
SHVR  &     $\min(p^*_{\text{bid}}(t)+\delta_p,\lambda_b)$ &  $p_\text{lo} $    & $\max(p^*_{\text{ask}}(t)-\delta_p,\lambda_s)$    &  $p_\text{lo}$            \\
ZIC 	& $\delta_p$	& $\lambda_{b}$	& $\lambda_{s}$	 & $p_{\text{max}}$     \\        

ZIU 	& $\delta_p$	& $p_\text{max}$	& $\delta_p$	 & $p_{\text{max}}$     \\        
\hline
\end{tabular}
\end{center}
\caption{Summary of the three ZI strategies (GVWY, SHVR, and ZIC) that are integrated within PRZI, and also of Gode \& Sunder's ZIU. Each ZI strategy generates quote-prices at random from a uniform distribution ${\cal U}(p_\text{lo}, p_\text{hi})$, although for both GVWY and SHVR $p_\text{lo}=p_\text{hi}$. The bounds on the quote-price generator distribution are different for buyers and sellers: $\lambda_b$ is the buyer's limit price; $\lambda_s$ is the seller's; $\delta_p$ is the market's tick-size; $p^*_\text{bid}(t)$ and $p^*_\text{ask}(t)$ are the best bid-price and best ask-price on the LOB at time $t$, respectively; $p_\text{max}$ is an arbitrary system constant, the largest price quotable in the market.}
\label{tbl:stratcompare}
\end{table}

\section{Parameterised-Response ZI Traders}
\label{sec:PRZI}

\subsection{Motivation}
\label{sec:przi-motivation}

The initial motivation for developing PRZI came from a desire to give ZI traders a sense of {\em urgency}, of how keen they are to find a counterparty for a transaction. Intuitively, there is a tradeoff between time-to-transact and the expected surplus (profit or saving) on that transaction. In human-populated markets, over time, while working a trade, an individual buyer will typically announce increasing bid prices, in the hope that the better prices increase the chance of finding a counterparty seller to transact with, but each of those increases in the bid-price reduces the surplus that the transaction will generate for that buyer; the situation is the same for sellers, gradually reducing their ask prices, again in the hope that each price-cut makes it more likely that a buyer will come forward, but each cut slices away at the seller's final profit for this transaction. In both cases, if the trader is urgent to get a deal done they can increase their chances of finding a counterparty by cutting their potential surplus, i.e. by making bigger step-changes in the prices that they're quoting.

Lacking the intelligence of human traders, ZIC traders just issue their next quote price by drawing from a uniform distribution over a specified range of prices; for an individual ZIC trader, the change in price from one quote to the next may be positive or may be negative, and there is no control over the step-size. ZIC traders have no sense of urgency. 

In contrast, a SHVR agent {\em can} be given some approximation to urgency: in \cite{church_cliff_2019}, SHVR was extended by applying a multiplying coefficient $k \in {\mathbb Z}^+ $ to the $\delta_p$ term in Equations~\ref{eq:SHVRbid} and~\ref{eq:SHVRask}, such that when the market circumstances dictate it, a value of $k>1$ allows SHVR to make larger step-changes in its quote prices. Work on PRZI started as a response to asking: how can we do something similar for ZIC?

\begin{figure}
\begin{center}
\includegraphics[width=0.5\linewidth]{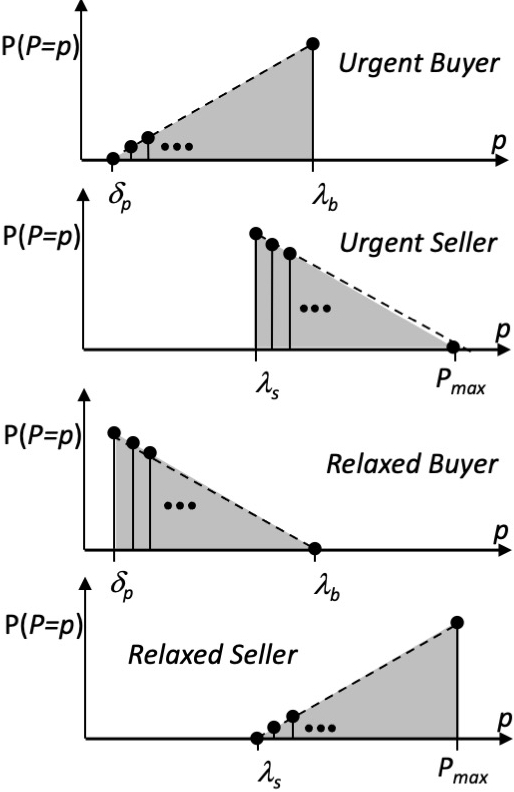}
\end{center}
\caption{Illustrative PMF for the variants of ZIC traders that are either {\em urgent}  (upper two figures) or {\em relaxed} (lower two figures).}
\label{fig:ZIC_triangles}
\end{figure}

Figure~\ref{fig:ZIC_triangles} illustrates the reasoning underlying PRZI, when viewed as a way of making ZIC traders quote in a way that is more or less likely to lead to a transaction within any particular time-window: the rectangular PMF from the uniform distribution of the original ZIC can be replaced by a PMF that has either a right- or left-handed right-angled-triangle; depending on which way the triangle is oriented, the ZIC trader is either more or less likely to generate a quote-price that leads to a transaction. Let's refer to these two right-triangle PMFs as the {\em urgent} and {\em relaxed} variants of ZIC. 

\begin{figure}
\begin{center}
\includegraphics[width=0.5\linewidth]{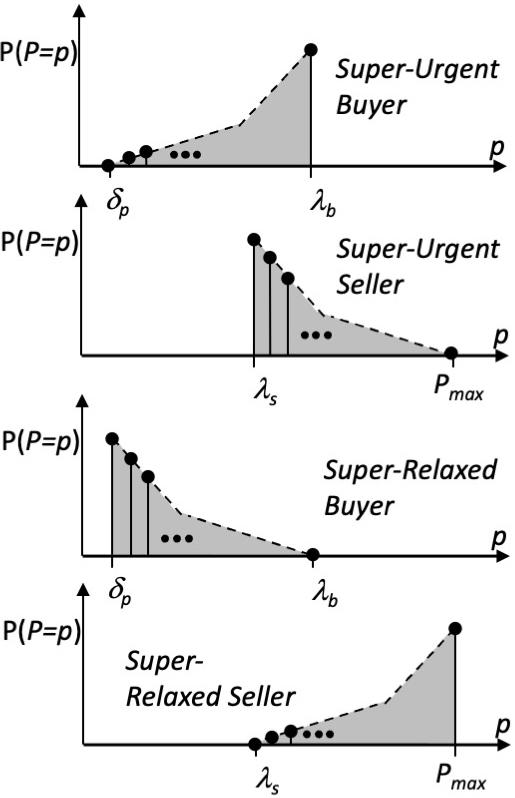}
\end{center}
\caption{Illustrative PMF for the variants of ZIC traders that are either {\em super-urgent}  (upper two figures) or {\em super-relaxed} (lower two figures).}
\label{fig:ZICcurves}
\end{figure}

But why stop at these variants? Figure~\ref{fig:ZICcurves} illustrates two more extreme ZIC variants, where the PMF profile is nonlinear: let's refer to these as the  {\em super-urgent} and {\em super-relaxed} variants of ZIC.

\begin{figure}
\begin{center}
\includegraphics[width=0.5\linewidth]{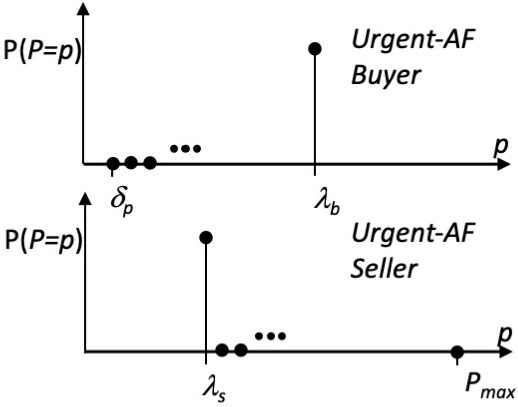}
\end{center}
\caption{Illustrative PMF for the {\em urgent-AF} variant of ZIC: a single price (the trader's limit-price) is quoted with probability one; the probability for all other prices is zero. This is identical to GVWY, as discussed in the text.} 
\label{fig:ZIC-AF-GVWY}
\end{figure}

\begin{figure}
\begin{center}
\includegraphics[width=0.5\linewidth]{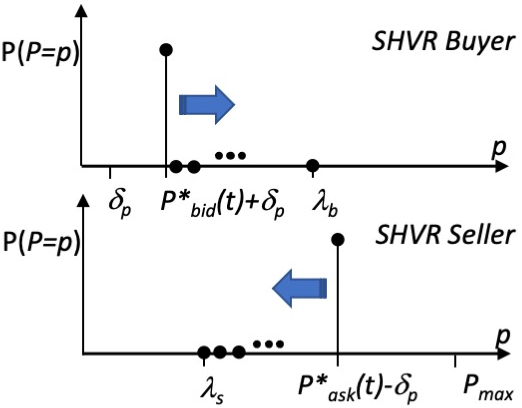}
\end{center}
\caption{Illustrative PMF for SHVR which can be considered as the equilbrating {\em relaxed-AF} variant of ZIC. The only price with a nonzero probability is a one-tick improvement on the best bid or ask at time $t$, and that price has a probability of one. The arrow pointing right (left)  indicates the direction of travel of the lower (upper)-bound on the PMF domain; the upper (lower) bound is given by the trader's limit-price $\lambda$.}
\label{fig:ZIC-AF-SHVR}
\end{figure}

Clearly, the degree of nonlinearity in the variant-ZIC PMFs shown in Figure~\ref{fig:ZICcurves} could be made even more extreme, and eventually when the degree of curvature is at its most extreme the PMF would have only one price at nonzero probability: the price that had the highest probability in the triangular PMF.  And, as there is only one price with nonzero probability, that probability must be one (i.e., 100\%, total certainty).  Let's call these absolute extremes the {\em urgent-AF} and {\em relaxed-AF} (`AF' might stand for {\em asymptotic form}); they're illustrated in Figures~\ref{fig:ZIC-AF-GVWY} and~\ref{fig:ZIC-AF-SHVR}.

But the shapes of the urgent-AF PMFs in Figures~\ref{fig:ZIC-AF-GVWY} and~\ref{fig:ZIC-AF-SHVR} are familiar: the urgent-AF PMFs are simply a probabilistic representation of GVWY, because equations \ref{eq:GVWYbid} and \ref{eq:GVWYask} can be expressed in (redundantly) probabilistic language as: {\em with probability one, the price quoted by a GVWY trader is its current limit-price}.  

This then prompts the question: can a useful link be made from the relaxed-AF form of ZIC to SHVR? The PMFs of SHVR and relaxed-AF ZIC have essentially the same shape, the difference is in where they occur on the number-line of prices: SHVR as specified in Equations~\ref{eq:SHVRbid} and~\ref{eq:SHVRask} quotes a price that is a one-tick (i.e. $1 \times \delta_p$) improvement on the current best price on the LOB; whereas a relaxed-AF ZIC would quote the most extreme price available, either the lowest nonzero price (i.e., $\delta_p$) for a buyer, or the arbitrary system maximum $P_{max}$ for a seller, neither of which is going to do anything at all to help equilibrate the market's transaction prices. Because the relaxed-AF versions of ZIC would not equilibrate, it makes most sense to move the relaxed-AF price away from the most extreme value, and toward the best price on the LOB, as the degree of nonlinearity in the variant-ZIC PMF increases. This would mean that by the time the nonlinearity is maximally extreme, i.e. when the PMF has the -AF shape, the variant ZIC is doing the same thing as a SHVR. 

\begin{figure}[ph]
\begin{center}
\includegraphics[width=0.75\linewidth]{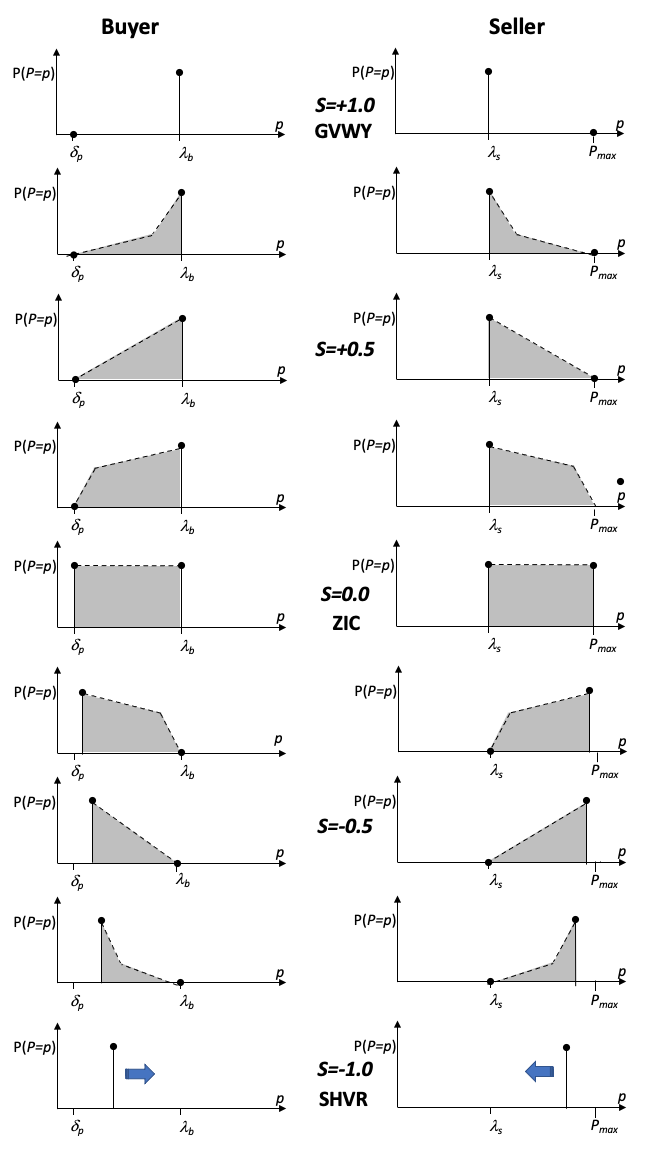}
\end{center}
\caption{The full spectrum of quote-price PMFs for PRZI: left-hand column is Buyer PMFs; right-hand column is Seller PMFs. Top row (where $s=+1.0$) is the {\em urgent-AF} PMF, equivalent to GVWY; then on each successive row the PMF envelope warps to become closer to the shape of the original ZIC PMF, which is in the middle row ($s=0.0$); after that, each lower row warps closer to the {\em relaxed-AF} PMF that implements SHVR (at $s=-1.0$). Note that the scale of the vertical axis is the same for the two graphs in each row, but varies between rows. At the upper and lower extremes where $s= \pm1$, the PMF is a single point at ${\mathbb P}(P=p)=1$; in the middle row ($d.s=0$), the PMF is a rectangle of height ${\mathbb P}(P=p)=1/\Delta_P$ where $\Delta_P$ is the difference between the minimum and maximum prices that form the bounds of the PMF on the horizontal axis. }
\label{fig:PRZI-PMFs}
\end{figure}

And that is the motivation for creating PRZI: now all that is needed is a function that smoothly varies from the urgent-AF variant that is GVWY, on through the nonlinear super-urgent variant, on through the triangular-PMF of the urgent variant, then on to the original ZIC, then onwards to the triangular-PMF of the relaxed variant, and then on through the super-relaxed nonlinear PMF to the relaxed-AF PMF that implements SHVR: this progression is controlled by PRZI's strategy parameter, denoted by $s \in [-1,+1] \in {\mathbb R}$, as illustrated in Figure~\ref{fig:PRZI-PMFs}.

\subsection{Details}
\label{sec:przi-details}

The various seller PMFs shown in Figure~\ref{fig:PRZI-PMFs} have a domain that is bounded from below (the left-hand limit of the PMF) by the individual seller's limit-price $\lambda_s$, but the upper bound on that domain, the right-hand limit of the PMF, brings us back to ``the $p_{\text{max}}$ problem'' discussed previously in Section~\ref{sec:ZIC}. Each PRZI  seller $i$ addresses this problem by determining its own private estimate of the highest plausible price, denoted by $p_{i:\text{max}}$, according to the following set of heuristic criteria:
\begin{itemize}
\item
If the PRZI seller has no other information available to it (e.g., it is the start of the first session in a market experiment, and the LOB is empty) then the only available information it has is the highest limit price it has been assigned thus far, denoted by $\lambda_{s(i:\text{max})}(t)$ (and if it has not been assigned a limit price, it is unable to participate in the market). In this situation, the PRZI seller sets $p_{i:\text{max}}=c_i \lambda_{s(i:\text{max})}(t)$ for some coefficient $c_i>1$. Informally, this models a naive and uninformed seller making a wild guess at the highest price that the market might tolerate. Different sellers can have different values of $c_i$ -- that is, some might guess cautiously while others might be much more optimistic.
\item
If ever another seller issues an ask-quote at a price $P_{sq} > p_{i:\text{max}}$ then seller $i$ sets $p_{i:\text{max}}=P_{sq}$. Informally, this models a seller realising that her current best guess of the highest price the market will tolerate was too low, because there is some other seller in the market who has quoted a higher price. 
\end{itemize}

In principle, if reliable price information is available from an earlier market session, then that information could be used instead. However, in any one previous market session there are likely to be multiple potential candidate values (e.g: the highest ask-price quoted in that session, or the highest transaction-price in that session, or the final transaction price, or the linear-regression prediction of the next transaction price in the market, or a nonlinear prediction, and so on), and that would soon take us away from the appealing minimalism of ZI traders.  

This approach, of letting each seller form its own private estimate of the highest tolerable ask-price could be criticised as merely swapping one arbitrary system parameter (the global constant $p_{\text{max}}$) for a bunch of new arbitrary exogenously-imposed parameters (the set of individual $c_i$ values, one per trader). If all we care about is minimising parameter-count, that would be a valid criticism. However, this approach has the advantage that only uses information that is locally available to an individual trader (i.e., it does not require knowledge of a system-wide constant) and it never requires manual re-calibration to the highest price in the market's supply curve. In practice setting the $c_i$ values at random over a moderate range is sufficient to generate useful results: the results presented here used  $c_i={\cal U}(1,10)^{0.5}$; this makes it unlikely that all traders will have the same $c_i$ value, and the fractional exponent gives a nonlinear bias toward smaller $c_i$ values.

Finally note that, if introduced alone, this solution to the $p_\text{max}$ problem can give rise to behavioral asymmetries between PRZI buyers and sellers when $s_i \approx -1$, (i.e., SHVR-style strategies) because PRZI buyers still have the lower-bound of their PMF set to the system minimum price $\delta_p$. That is, unlike the sellers, the buyers are not using some multiple of their lowest limit-price or any prices observed in the market to form their initial lowest bid-price. To illustrate the problem, consider a market populated entirely by PRZI traders all with $s_i=-1$ and in which the supply and demand schedules are set such that the equilibrium price is a large multiple of $\delta_p$ relative to the number of buyers: the sellers, no longer limited by having to start their quotes at some arbitrary system maximum price, will drop their quote prices to near-equilibrium values relatively quickly, but in contrast the SHVR-style buyers will start by quoting $\delta_p$, then $2\delta_p$, then $3\delta_p$ and so on, potentially making much slower progress toward the equilibrium if that lies at, say, $10,000\delta_p$: there is, in this sense, also a $p_\text{min}$ problem. To counter this, PRZI buyers can set their $p_\text{min}$ in the same style that PRZI sellers set their $p_\text{max}$: initially using $p_{i:\text{min}} = \text{max}(\frac{1}{c_i}\lambda_{b(i:\text{min})},\delta_p)$ in the absence of any other information, and using the lowest price quoted by another buyer if that quote is less than $i$'s initial $p_{i:\text{min}}$ value.\footnote{To illustrate the behavioral asymmetry that arises when the $p_{\text{min}}$ problem is not remedied in this way, in all the experiments described in Section~\ref{sec:coevolve}, the PRZI buyers simply use the original ZIC-style $p_{\text{min}}=\delta_p$: the strategies of the buyer population then consistently diverge from those of the sellers. This nevertheless does yield rich co-evolutionary dynamics, serving to illustrate issues in the visualization and analysis of such high-dimensional co-evolutionary systems. }  

As described thus far, we have a method for setting the upper limit of the PRZI seller PMF when $s \geq 0$, 
i.e., for the PRZI range of strategies from ZIC to GVWY. However, for the $s=-1$ seller-case that implements 
SHVR, we need the upper limit of the PMF to be set such that $p_{i:\text{max}}=p^*_{\text{ask}}-\delta_p$, and 
we need to get there smoothly from the $s=0$ case where PRZI is implementing ZIC and $p_{i:\text{max}} $ and  $p_{i:\text{min}} $ are 
set by the method just described. The simplest way of doing that is to have $p_{i:\text{max}}$ be a linear 
combination of the two. For a PRZI seller, let $p_{i:\text{max:ZIC}}$ denote the $p_{i:\text{max}} $ value at $s=0$, then for $s \in [-1,0]$ 
we use:
\begin{equation}
p_{i:\text{max}} = (1+s)p_{i:\text{max:ZIC}} - s.\max(p^*_{\text{ask}}-\delta_p,\lambda_{s:i})
\end{equation}
\noindent
and the same form of linear combination for PRZI buyers, to pull the lower limit on the buyer PMFs progressively away from $P_{i:\text{min:ZIC}} \rightarrow \min(p^*_\text{bid}+\delta_p,\lambda_{b:i})$. In extremis, this approach will narrow the PMF interval to just a single discrete price, i.e.\ $p_\text{min}=p_\text{max} = \lambda_{b|s}$,  generated with probability one. 

Next, let $s_i$ denote the strategy-value for trader $i$; and let $p_{i:\text{min}} \in {\mathbb Z}^+$ and $p_{i:\text{max}} \in {\mathbb Z}^+$ be the bounds on trader $i$'s discrete-valued price-range $[p_{i:\text{min}}, p_{i:\text{max}}]$, with $p_{i:\text{min}} < p_{i:\text{max}}$ and let the extent of that range be $r_i=p_{i:\text{max}}-p_{i:\text{min}}$. Also define a price-range normalization function $N(p)$:
\[
N: [p_{i:\text{min}}, p_{i:\text{max}}] \in {\mathbb Z}^+ \mapsto [0,1] \in {\mathbb R} ; N(p)=(p-p_{i:\text{min}})/r_i
\]

Then note that over the domain $x \in [0,1] \in {\mathbb R}$, the function ${\cal P}$ in Equation~\ref{eq:P-curves} has the right profile, makes the right shapes, for the buyer PMFs that we want (as were illustrated in Figure~\ref{fig:PRZI-PMFs}):
\begin{equation}
{\cal P}(x,s_i) = 
\begin{cases}
	\frac{e^{cx}-1}{e^c-1}  & \text{if } s_i>0 \\
	\frac{1}{r_i}  & \text{if } s_i=0 \\
 	1-\frac{e^{cx}-1}{e^c-1} & \text{if }s_i<0
\end{cases}
\label{eq:P-curves}
\end{equation}
\noindent
where
\begin{equation} 
c = \theta(m\tan(\pi(s_i+\frac{1}{2})))
\label{eq:c_fn}
\end{equation}
and $\theta(x)$ is the linear-rectifier threshold function symmetrically bounded by a cutoff constant $\theta_0$, which also (to avoid divide-by-zero errors) clips near-zero values at $\pm \epsilon$ for some sufficiently small value of $\epsilon$ (e.g $\epsilon=10^{-6}$):

\begin{equation}
\theta(x) = 
\begin{cases}
	\max(-\theta_0, \min(\theta_0, x))  & \text{if } |x|>\epsilon \\
	\epsilon & \text{if } 0<x<\epsilon \\
 	-\epsilon & \text{if } -\epsilon<x<0
\end{cases}
\label{eq:threshold_fn}
\end{equation} 

\noindent
and then finally use: 
\begin{equation}
{\cal PMF}_i(p,s_i) = 
\begin{cases}
	{\cal P}(N(p),s_i) & \text{when $i$ is a buyer}  \\
	{\cal P}(1-N(p),s_i) & \text{when $i$ is a seller}
\end{cases}
\label{eq:PMFs}
\end{equation}

\begin{figure}
\includegraphics[width=0.99\linewidth]{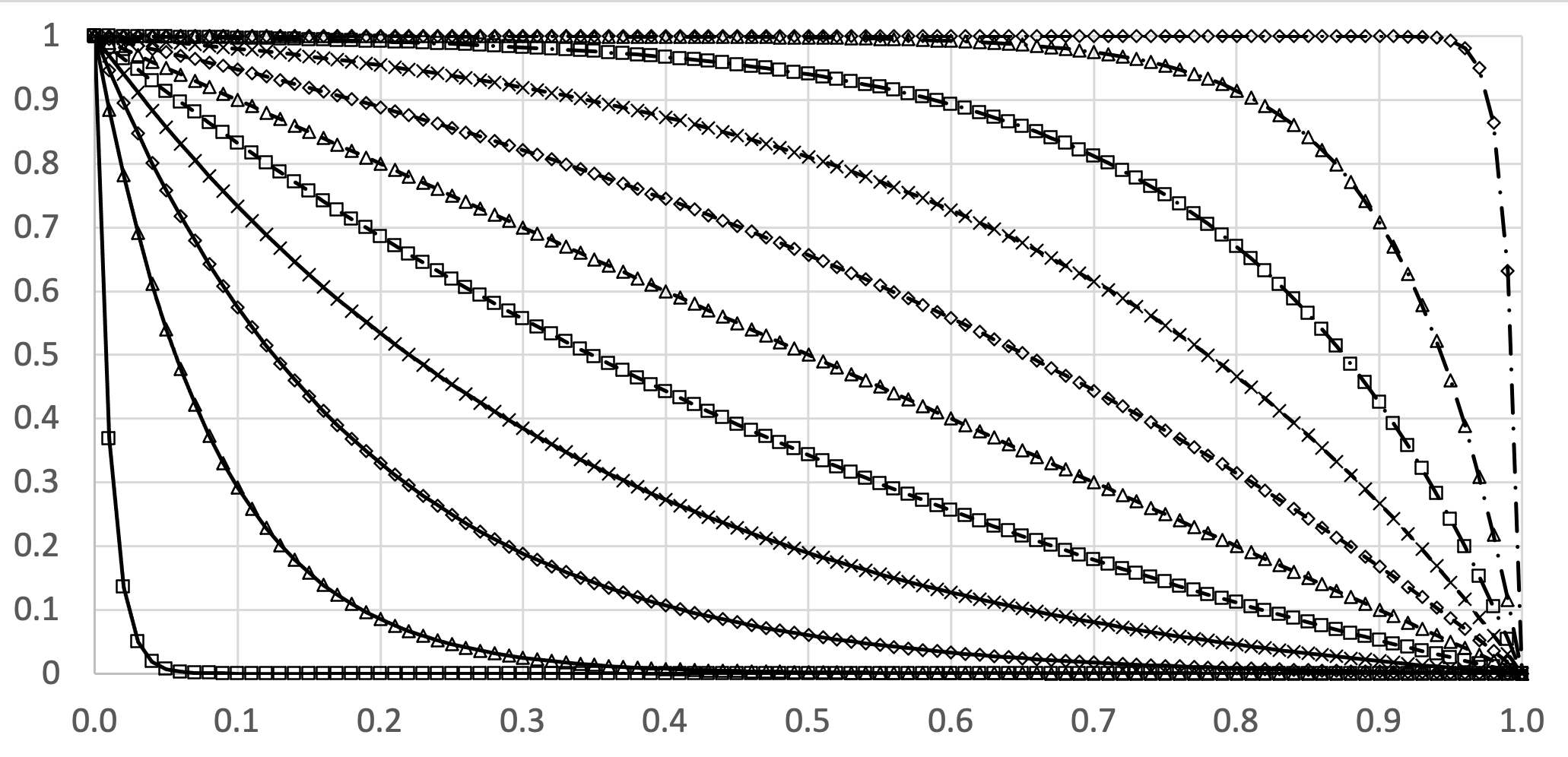}
\caption{Plots for ${\cal P}(p, s_i)$ (Equation~\ref{eq:P-curves}) for values of $s_i$ over the range $[0,1]$ in steps of $0.1$: horizontal axis is $p$ in steps of 0.01; vertical axis is ${\cal P}$. The ${\cal P}$ curves have the desired shape, but require normalisation before use as PMF envelopes.}
\label{fig:P_raw}
\end{figure}

After a little trial-end-error exploration, it was found that values for the constants $m=4$ and $\theta_0=100$ give the desired shapes, i.e.\ similar to the qualitative PMF envelopes  illustrated in Figure~\ref{fig:PRZI-PMFs}: these are illustrated in Figure~\ref{fig:P_raw}.  Also, turning the ${\cal PMF}_i$ envelope into a usable PMF requires scaling and normalisation such that:
\begin{equation}
{\mathbb P}(P=p)=\frac{{\cal PMF}_i(p, s_i)}{\sum\limits_{j=0}^{r_i}{\cal P}(j/r_i,s_i)}: p \in [p_{i:\text{min}}, p_{i:\text{max}}]
\label{eq:P-PMF}
\end{equation}

\begin{figure}
\includegraphics[width=0.99\linewidth]{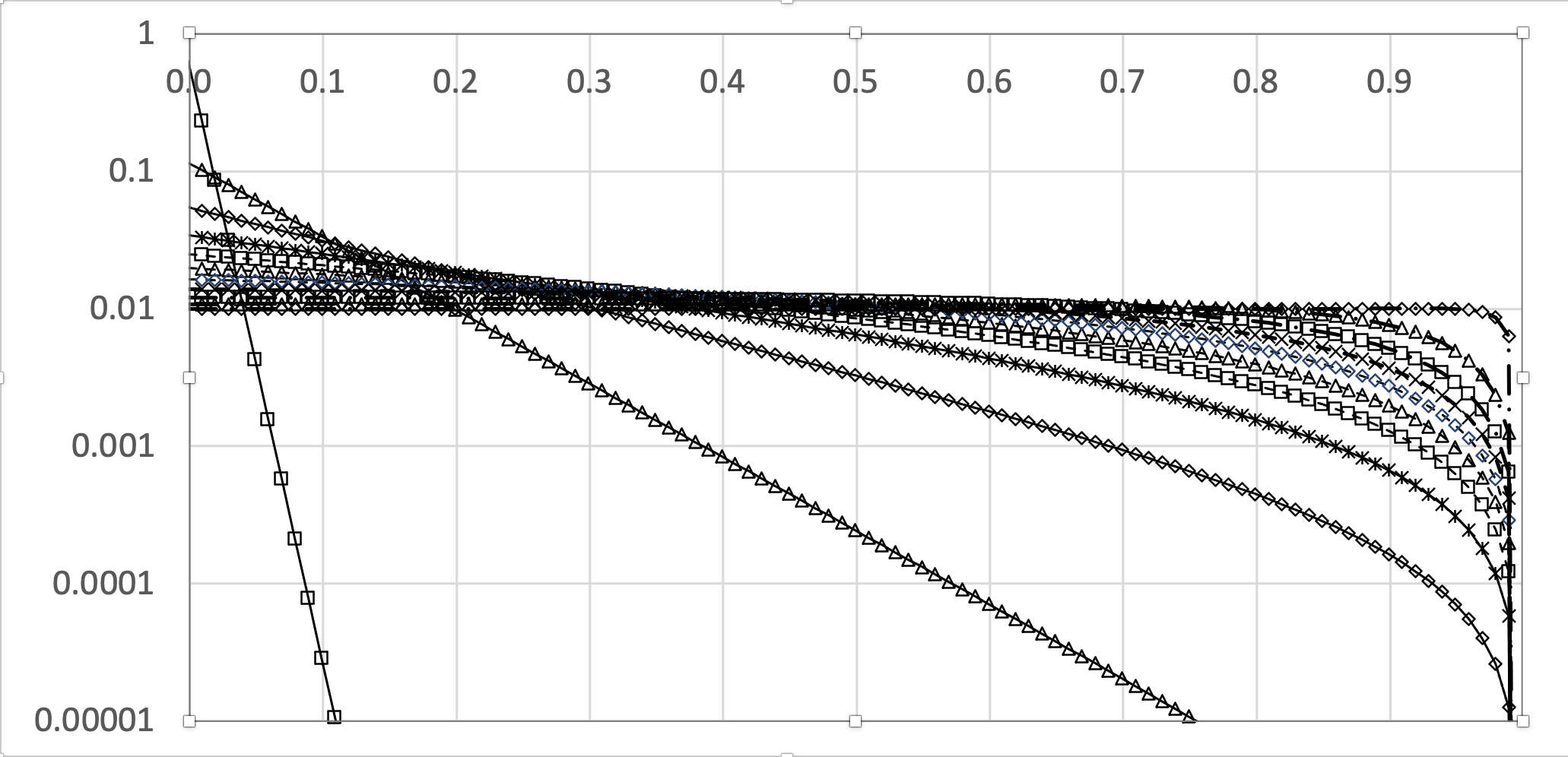}
\caption{Log-linear plots for ${\mathbb P}(P=p)$ (Equation~\ref{eq:P-PMF}) for values of $s_i$ varying from 0 to 1 in steps of $0.1$: horizontal axis is the interval $p \in [0,1]$ in steps of 0.01; vertical axis is ${\mathbb P}$. The ${\mathbb P}$ curves are normalised to have a discrete definite integral over $p=[0,1]$ equal to 1.0, and hence are valid as PMF envelopes. }
\label{fig:P_PMF}
\end{figure}

From this we can then compute the cumulative distribution function (CDF) for trader $i$ as $F_P(p)=P(P\leq p)$ which is defined over the domain $[p_{i:\text{min}}, p_{i:\text{max}} ]$ as:
\begin{equation}
F_P(p)=\sum\limits_{k=0}^{p-p_{i:\text{min}}}{\mathbb P}(P=p_{i:\text{min}}+k)
\label{eq:PRZI_CDF}
\end{equation}

The $F_P(p)$ CDF in Equation~\ref{eq:PRZI_CDF} maps from its domain $[p_{i:\text{min}}, p_{i:\text{max}} ]$ to cumulative probabilities in the interval $[0,1]\in{\mathbb R}$: for each of the $r_i$ discrete points in the domain, exact values of $F_P(p)$ can be computed and stored in a look-up table (LUT). It is then simple to use reverse-lookup on the LUT to give an inverse CDF function $\widehat{F_P^{-1}}(c)$ such that:
\begin{equation}
\widehat{F_P^{-1}}(c):[0,1]\in {\mathbb R}  \mapsto [p_{i:\text{min}}, p_{i:\text{max}} ]\in {\mathbb Z}
\end{equation} 
\noindent
Separate versions of $\widehat{F_P^{-1}}(c)$ need to be generated for buyers and sellers, which will be denoted by subscripted parenthetic $s$ and $b$ characters, and  can then be fed samples from a uniformly distributed pseudo-random number generator to produce quote-prices for PRZI traders:
\begin{equation}
P_{bq({\text {PRZI}})}(t+\Delta_t)=\widehat{F_{P(b)}^{-1}}({\cal U}(0,1))
\label{eq:inv_cdf_buy}
\end{equation}
\begin{equation}
P_{sq({\text {PRZI}})}(t+\Delta_t)=\widehat{F_{P(s)}^{-1}}({\cal U}(0,1))
\label{eq:inv_cdf_sell}
\end{equation}

And note that in the special case when $s=0$,  $\widehat{F_{P}^{-1}}$ reduces to the identity function. 
That completes the definition of PRZI traders. Section~\ref{sec:implement} discusses implementation issues, and Section~\ref{sec:results} presents illustrative results.

\section{PRZI Implementation}
\label{sec:implement}

A reference implementation of PRZI written in {\em Python 3.8} has been added to the BSE sourcecode repository on GitHub \cite{cliff_2012_bse}. For ease of intelligibility, the implementation follows the mathematics laid out in the previous section of this paper, computing an individual LUT for each trader:
this approach is conceptually simple, but is manifestly inefficient in space and in time. To illustrate this, consider the case where all traders in the market have the same value $s$ for their PRZI strategy parameter, and all sellers are assigned the same limit price $\lambda_s$ and all buyers the same $\lambda_b$: in such a situation, only two LUTs are needed: one to be shared among all buyers, and another to be shared among all sellers; but the current implementation wastes time and space by blindly computing an entire LUT for each trader. A more efficient implementation could be built around compiling any one LUT for a particular instance of an $(s, p_{\text{min}}, p_{\text{max}})$ triple only once, when it is first required, and storing it in some central shared key-value store or document database where the triple is the key and the LUT is the associated value or document. This would be considerably less inefficient, but would add considerably to the complexity of the code. As the Python code in BSE is intended to be simple, as an illustrative aid for non-expert programmers, the current BSE implementation does not use this approach. 

\section{Example Use-Cases}
\label{sec:results}

Section~\ref{sec:intro} listed three motivations for developing PRZI: to provide a mechanism for ZI traders to give a market-impact response; to enable ZI traders to be {\em opinionated}, thereby enabling the creation of ACE models exploring matters arising from Shiller's notion of {\em narrative economics}; and to facilitate the study of coevolutionary dynamics in markets populated by adaptive agents that can smoothly vary their trading strategies through a continuous space. Here I briefly summarise current work-in-progress on all three of those fronts, in sequence.

\subsection{PRZI as a Generator of Market-Impact Responses}
\label{sec:impact}

In \cite{church_cliff_2019} I introduced an altered version of the SHVR zero-intelligence trader strategy that is extended to be imbalance-sensitive, altering its behavior in response to instantaneous imbalances in market supply and demand, thereby giving ZI-populated ACE models in which the population of traders exhibit a market-impact effect. Market impact is here defined as the situation where the prices quoted by traders in a market shift in the direction of anticipated change in the equilibrium price, before any transactions have occurred, where the coming change in equilibrium price is anticipated because of an imbalance in the orders in the market, a sudden shift to excess demand or supply; and where it's safe to assume that actual market transaction prices are typically close to the equilibrium price. For an extensive and insightful analysis of market impact in financial markets, see \cite{farmer_etal_2013}. 
As the basic SHVR was made sensitive to {\em imbalance} for the purpose of market {\em impact}, the extended SHVR was named ISHV (pronounced ``eye-shave''). Here I briefly show how exactly the same mechanism developed for ISHV can be used to create an impact-sensitive version of PRZI, which I'll refer to as IPRZI. 

ISHV's impact-sensitivity is based on the difference between the current market {\em mid-price}, denoted here by $p_m(t)=(p^*_{\text{bid}}(t)+p^*_{\text{ask}}(t))/2$, and the current market {\em micro-price}, denoted here by $p_\mu(t)$, where
\begin{equation}
p_\mu(t)=\frac{ p^*_{\text{ask}}(t) q^*_{\text{bid}}(t) + p^*_{\text{bid}}(t) q^*_{\text{ask}}(t) } {q^*_{\text{bid}}(t)+q^*_{\text{ask}}(t)}
\label{eq:p_mu}
\end{equation}
\noindent
in which $p^*_{\text{ask}}(t)$ is the price of the best ask at time $t$ -- i.e., it is the price at the top of the bid-side of the CDA market's limit order book (LOB); $p^*_{\text{bid}}(t)$ is the price of the best bid at time $t$ -- i.e. the price at the top of the ask side of the LOB; $q^*_{\text{ask}}(t)$ is the total quantity available at $p^*_{\text{ask}}(t)$; and $q^*_{\text{bid}}(t)$ is the total quantity available at $p^*_{\text{bid}}(t)$. Equation~\ref{eq:p_mu} is how the micro-price is defined by \cite{cartea_etal_2015}.

When there is zero supply/demand imbalance at the top of the LOB  (i.e., $q^*_{\text{bid}}(t) = q^*_{\text{ask}}(t)$), Equation~\ref{eq:p_mu} reduces to the equation for the market mid-price, and hence the difference between the two prices, denoted by $\Delta_m(t)=P_\mu(t) - P_m(t)$, is zero.  However if $\Delta_m(t)>>0$ then the imbalance indicates that subsequent transaction prices are likely to increase (which should increase urgency in IPRZI buyers, and reduce it in IPRZI sellers); and if $\Delta_m(t)<<0$ then the indication is that subsequent transaction prices are likely to fall (so IRPZI sellers should increase urgency, while buyers relax). The mapping from $\Delta_m(t)$ to IRPZI $s$-value is achieved by giving each IPRZI trader $i$ an {\em impact function}, denoted here by ${\cal I}_i$, such that $s_i(t)={\cal I}_i(\Delta_m(t))$ and ${\cal I}_i : {\mathbb Z} \mapsto [-1.0,+1.0] \in {\mathbb R}$. This form of IPRZI was recently implemented by my student Owen Coyne in his Masters thesis \cite{coyne_2021}.

\begin{figure}
\begin{center}
\includegraphics[width=0.6\linewidth]{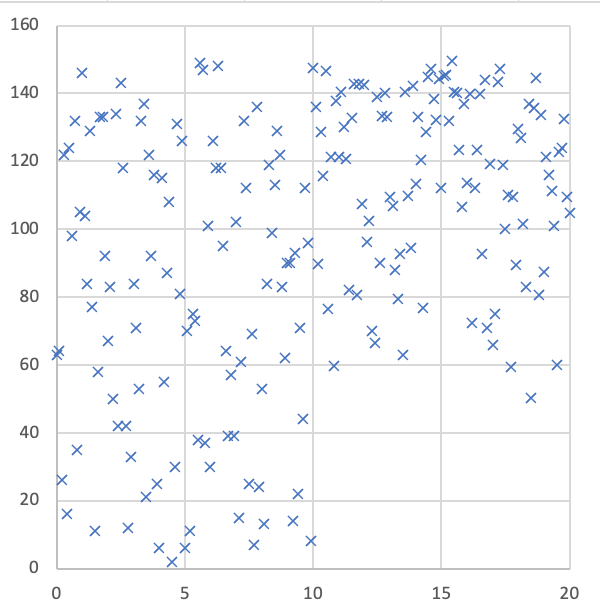}
\end{center}
\caption{Illustration of shift in quote-prices from an IPRZI buyer reacting as the supply/demand imbalance changes. The horizontal axis is time $t$ in seconds, and the vertical axis is price. Each marker shows a price quoted by a single IPRZI buyer with a current limit price $\lambda=\$150$. Initially there is no imbalance, so the IPRZI buyer has $s=0$ and is generating quote-prices from a uniform distribution, i.e.\ it is playing the ZIC strategy. At $t=10$ an imbalance is deliberately introduced into the market, a sudden injection of excess demand, and the IPRZI buyer reacts by shifting its $s$ value to increase {\em urgency}, with the PMF becoming heavily skewed toward $\lambda$: the resultant change in the distribution of  actual quote prices is manifest. 
}
\label{fig:przi_mktimpact}
\end{figure}

As illustration, Figure~\ref{fig:przi_mktimpact} shows the change in strategy of an IPRZI buyer when it reacts to a sudden change in imbalance, a sudden injection of excess demand at the top of the LOB. 

This version of IPRZI is the simplest to articulate, but it suffers from the vulnerability that Equation~\ref{eq:p_mu} is sensitive only to imbalances at the very top of the LOB -- any imbalance at deeper levels of the LOB is simply ignored, and hence this is quite a fragile measure of imbalance. This is discussed at more length by \cite{zhang_cliff_2021} who instead use {\em multi-level order-flow imbalance}\/ (MLOFI) as a more robust imbalance metric. Briefly, MLOFI is a specific implementation of the novel findings of 
\cite{cont_cucuringu_zhang_2021} who used a principal component analysis of {\em order-flow imbalance} (i.e., whether the number/amount of buy orders submitted to the exchange/LOB is in balance with the number/amount of sell orders, or not) to show that taking into account multiple levels of the LOB when defining order-book supply/demand imbalance leads to higher explanatory power for the short-term predictions of market-impact price movements. 
It is straightforward to extend IPRZI by replacing the ISHV-like  $s_i(t)={\cal I}_i(\Delta_m(t))$ method described here with the MLOFI method developed by Zhang \& Cliff, thereby making IPRZI more robustly sensitive to order imbalances: early results from exploring this approach are presented in \cite{cliff_zhang_taylor_2022}.

\subsection{PRZI as Opinionated Traders}
\label{sec:opinions}

Lomas \& Cliff \cite{lomas_cliff_2021} describe results from extending two well-known types of ZI trader, Gode \& Sunder's ZIC \cite{gode_sunder_1993} and Duffy \& {\"U}nver's NZI \cite{duffy_unver_2006}, where in both cases the extension adds an {\em opinion}\/ variable to each trader. This forms a novel intersection between work on ZI traders and work in {\em opinion dynamics} (see e.g. \cite{krause_2000,meadows_cliff_2012,meadows_cliff_2013}. Lomas \& Cliff prefix the ZIC and NZI acronyms with an O (for `opinionated') to give OZIC and ONZI. As was discussed in Section~1 of this paper, Lomas \& Cliff's work was motivated by the desire to develop ZI-trader agent-based models (ABMs) in the ACE tradition that would facilitate study of Shiller's recent notion of {\em narrative economics} \cite{shiller_2017,shiller_2019}; and one of the motivations for developing PRZI was to address some deficiencies in the original OZIC model. 

To recap, in brief: Shiller argues that traditional economic analyses too often under-emphasise (or wholly ignore) the extent to which the buying and/or selling behavior of agents within a particular market-based system is influenced by the {\em narratives} (i.e., the stories) that the agents tell each other -- and themselves -- about the past, present, and future states of that market system: if the narratives that the agents are telling each other are consistent with conventional economic theory, then there is nothing much to report; but if the stories in circulation run counter to the predictions of theory, then sometimes the actual market outcomes are difficult or impossible to explain by reference only to those factors favoured in orthodox economic analyses. Shiller's 2019 book discusses at length the extent to which the phenomenal rise in prices of cryptocurrencies such as Bitcoin is much better explained by reference to the narratives circulated and believed by the active participants in the markets for such crypto ``assets'' than it is by any conventional analysis or estimates of ultimate value.\footnote{Shiller's 2019 book pre-dates the explosive rise of trading in blockchain-backed non-fungible tokens  and the roller-coaster price gyrations in ``meme stocks'' such as GameStop (see e.g.\ \cite{jakab_2022}) and hence now seems ripe for a second edition, or at least for an epilogue to the first edition.} Schiller argues for an empirical approach to studying narrative economics: gathering as much data (e.g., records of the texts of news-articles and social-media discussions) as is practicable about the narratives circulating among agents in real economic systems, and then tying analysis of these narratives to analyses of the actual market dynamics and eventual outcomes. As Kenny Lomas and I argued in \cite{lomas_cliff_2021}, what Shiller proposes is solely an {\em a posteriori} analysis of the roles of narratives in economics systems, but there is an alternative approach, which is to build {\em constructive} models of economic systems in which narratives are an important factor, using ZI/MI ABM/ACE methods. This can be done by recognising that what Shiller describes as {\em narratives} are nothing more than the external expressions, the verbalizations, of agents' internally-held {\em opinions}, and hence that issues in narrative economics can be studied by developing ABMs in which the trader-agents each hold an opinion that can to some extent influence the opinion of other traders in the system, and that can in turn to some extent be influenced by the opinions of other agents that the agent interacts with; so long as each agent's opinion then also to some extent influences its economic behavior, the overall ABM can act as a test-bed for exploring aspects of narrative economics, in much the same way as laboratory experiments involving a few tens of human subjects acting as traders in a CDA can provide genuine insights on the dynamics of real financial markets. The \cite{lomas_cliff_2021} paper was our first report on extending ZI/MI traders so that they also held opinions that not only influenced their own trading behaviors but also could influence the opinions of other traders too; but, as is the case with very many exploratory first-attempts, analysis of our initial results revealed problems that needed to be addressed in further work. 

\begin{figure}
\begin{center}
\includegraphics[width=0.5\linewidth]{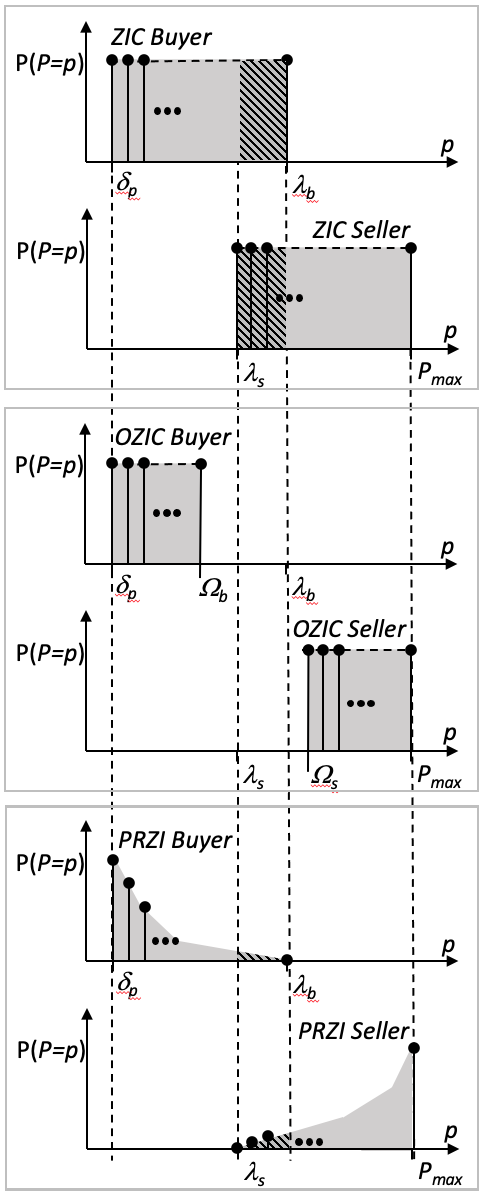}
\end{center}
\caption{Comparison of PMFs for ZIC, OZIC, and PRZI buyers and sellers. Three pairs of illustrative PMFs are shown here, stacked vertically for ease of comparison: $\lambda_s$ is the seller's limit price and $\lambda_b$ is the buyer's limit price; $\delta_p$ is the minimum price quotable in the market (here, one tick above zero) and $P_{\text max}$ is the market's maximum allowable price. The upper pair of PMFs show  ZIC: transactions can only occur between the buyer and the seller if $\lambda_s < \lambda_b$: this overlap zone in the two PMFs is illustrated by the area of diagonal hatching. The middle pair of PMFs show OZIC: the two PMFs are each truncated by the trader's {\em opinionated limit price}, denoted here by $\Omega_b$ and $\Omega_s$ for the buyer and the seller respectively: this truncation can readily eliminate the PMF overlap zone, thereby setting the probability of any future transactions to zero. The lower pair of PMFs show how PRZI PMFs can be set to attenuate toward the overlap zone, diminishing its relative proportion of the overall PMF, while retaining a nonzero probability of transactions occurring.
}
\label{fig:zic_ozic_przi}
\end{figure}

Figure~\ref{fig:zic_ozic_przi} illustrates the problem with OZIC traders that PRZI is intended to remedy: OZIC uses the value of the trader's opinion variable to set an {\em opinionated limit price} (here denoted by $\Omega$) which becomes a bound on the trader's PMF, introducing a region on the PMF between $\Omega$ and the trader's original limit price $\lambda$ in which the PMF is at the zero probability level, rather than at the positive uniform-probability level that it would be in a ZIC trader. While this does give a desirable link between the trader's opinion and the prices that it can quote, it is too easy for the opinion-dependent ranges of zero-probability in the buyer and seller PMFs to eliminate the overlap that is required for there to be any likelihood of the randomly-generated ZIC quote-prices crossing and leading to transactions. When this happens, each side issues quotes that are unacceptable to the counterparty side, and the market simply grinds to a halt, with traders on both the buyer-side and the seller-side quoting prices that their respective counterparty side can no longer transact at.

This problem is at its most acute if the supply and demand schedules are each perfectly elastic as used e.g. by \cite{smith_1965_swastika} (in this case both the supply and demand curves are horizontal and flat over the range of available quantities, giving a graph of the supply and demand curves that various authors have referred to as {\em swastika-} or {\em box-} shaped: see e.g. \cite{smith_1994}). As illustration, consider such a perfectly elastic pair of schedules and assume the best-case overlap in buyer and seller PMFs such that all buyers have the same limit price $\lambda_b$ that is set to the system maximum price ($\bar{M}$ in Lomas \& Cliff's terminology, or $P_{\text max}$ here) and all sellers have the same limit price $\lambda_s$ that is set to the system minimum price ($\underbar{\em M}$ in Lomas \& Cliff's terminology, or $\delta_p$ here). It is easy to prove that in these circumstances, if the market is populated entirely by OZIC traders and each trader holds a neutral opinion (i.e. have zero for their opinion value) then the buyer and seller PMFs cease to overlap for all traders; at which point -- given that we're talking about a situation in which all buyers have the same PMF and all sellers have the same PMF -- transactions cease to occur.  

This issue in OZIC is a consequence of what could be characterised as the binary thresholded nature of OZIC's implementation of opinion-influenced quote-price generation: the PMF for an OZIC is starkly divided into two zones by the trader's current value of $\Omega$: in one zone the PMF is a simple uniform distribution (as in ZIC) and in the other the PMF is a constant zero. PRZI's smoothly-varying PMFs obviate this problem because they allow a graded response, with probabilities being reduced to much lower values in the OZIC ``zero-zone" than in the OZIC ``uniform zone" while maintaining a nonzero possibility of a transaction actually occurring.  

As in Lomas \& Cliff's work, here we give PRZI traders a real-valued opinion variable, denoted here as $\omega_i$ for trader $i$, s.t. $\omega \in [-1,+1] \in {\mathbb R}$ where negative values of $\omega$ represent the opinion that prices are set to fall, and positive values represent the opinion that prices will rise. The linkage between $\omega$ and the PRZI $s$-value is straightforward: if trader $i$'s opinion is that prices will rise, then if $i$ is a buyer it needs to bias its PMF toward {\em urgency}\/ but if it is a seller then the rational thing to do when prices look set to rise is to bias the PMF toward {\em relaxed}; and if the trader's opinion is that prices will fall, similar reasoning applies {\em mutatis mutandis}. 

As should by now be obvious, we need some function $F_i$ that maps from trader $i$'s opinion $\omega_i$ to its PRZI strategy $s_i$, i.e.\ $s_i = F_i(\omega_i)$. In the very simplest case, given that both $s_i$ and $\omega_i \in [-1,+1] \in {\mathbb R}$, the mapping can be the identity function, or the negative of the identity function, depending on whether $i$ is  buyer or a seller. For a seller, the simplest $F_i$ is identity: $F_i(+1)=+1$; $F_i(-1)=-1$. For a buyer, the simplest $F_i$ is negative identity: $F_i(+1)=-1$; $F_i(-1)=+1$. However this fairly rapidly moves the trader's strategy to extremes (either SHVR or GVWY) as $|\omega| \rightarrow 1$, which may not always be desirable: instead, some nonlinear mapping resembling a logit/probit function is generally a better choice for $F_i$.

The link $F_i$ provides between $\omega_i$ and $s_i$ is, as stated thus far, necessary but not entirely sufficient: it provides a linkage between a trader's opinion and its PRZI strategy, and if this work was conducted in the manner familiar from much of the opinion dynamics (OD) literature, that {\em would}\/ be sufficient, because a lot of research in OD has been directed at the study of shifts in opinion among a population of agents where the {\em only}\/ factor that might change an individual's opinion is some set of one or more interactions with one or more other agents in the population. That approach makes sense if the opinions being modelled are {\em solely}\/ matters of individual subjective choice, such as personal religious or political views, where there is no external referent, no absolute {\em ground truth} that could prove the individual's opinion to in fact be wrong. But in financial markets there {\em is}\/ a ground truth: the actual dynamics of the actual market; actual prices, actual volumes. If an entire population of agents believes that the price of some asset will rise tomorrow, that can be a self-fulfilling prophecy because all traders will act in a way consistent with the collective belief (this takes us onto the well-trodden path toward sunspot equilibria \cite{cass_shell_1983} and all the way back to Merton's classic work in the 1940's e.g.\ \cite{merton_1948}). 

However if half the population believes that the price will rise while the other half believes it will fall and the next day the price does actually rise, then half the population got it wrong and may need to revise their faith in their own opinions to reduce the mismatch between their opinions and the ground truth. Given that another motivation for designing PRZI, another use-case (as discussed in Section~\ref{sec:impact}), was to explore market-impact effects by giving PRZI traders a sensitivity to supply/demand imbalances, currently I'm working with students to explore ABMs in which the PRZI trader's opinion-value is influenced by an appropriate mix of its OD-style interactions with other agents in the population, and its analysis of the currently observable market situation (e.g. its calculation of simple imbalance metrics such as the ISHV-style $\Delta_p$ value, or MLOFI). In the spirit of ZI and minimal-intelligence modelling, a simple weighted linear combination of the two is as good a place to start as any, but this is a rich seam for further research with many alternative approaches to explore.

\subsection{Co-evolutionary Dynamics in Markets of Adaptive PRZI Traders}
\label{sec:coevolve}

The continuously-variable strategy parameter $s$ in PRZI allows for studies of co-evolutionary dynamics in markets populated entirely by adaptive ZI agents. To do this, we need to set up markets populated entirely by PRZI traders where each individual trader can alter/adapt its $s$ value in response to market conditions, always trying to fine-tune it to generate higher trading profits. 

Doing this will then provides a ZI-style ABM/ACE test-bed for exploring issues in evolutionary economics -- the adaptive PRZI traders are manifestly engaged in a form of evolutionary game, each adapting their strategies over time to try to maximise their local measure of fitness, a topic explored extensively (although not always using exactly that terminology) since the dawn of economics, predating even \cite{vonneumann_morgenstern_1944}: see for example the reviews by \cite{friedman_1991_evolecon,blume_easley_1992,friedman_1998_evolecon,lo_2004,lo_2019,nelson_2020_evolecon}.

If we were to  allow {\em only}\/ a single PRZI trader $i$ to adaptively vary its $s_i$ value, trying to find the best setting of $s_i$ relative to whatever distribution of $s_{j \neq i}$ values is present in the market (i.e., relative to the current mix of other strategies in the market) then we could say that $i$ is {\em evolving}\/ its value of $s_i$ to try to find an optimum, the most profitable setting for its strategy parameter, given the unchanging set of fixed strategies that it is pitted against in the market. But when {\em every}\/ PRZI trader in the market is simultaneously adapting its $s$-value, the system is {\em co-evolutionary}\/ because what is an optimal setting of the $s$ parameter for any one trader will likely depend on the $s$-values currently chosen by many or perhaps all of the other traders in the market. That is, the profitability of $i$ is dependent not only on its own strategy value $s_i$ but also on many or perhaps all other $s_{j \neq i}$ values in play at any particular time, and in principle all the strategy values will be altering all the time. 

A primary motivation for studying such co-evolutionary markets with adaptive PRZI traders is the desire to move beyond prior studies of markets populated by adaptive automated traders in which the ``adaptation'' merely involves selecting between one of typically only two or three fixed strategies (as in, e.g., \cite{walsh_etal_2002,vytelingum_etal_2008,vach_2015}. The aim here is to create minimal model markets in which the space of possible ZI strategies is infinite, as a better approximation to the situation in real financial markets with high degrees of automated trading. 

Prior researchers' concentration on markets in which the traders can choose one of only two or three fixed strategies can be traced back to the sequence of publications that launched the trading strategies MGD, GDX, and AA (i.e., \cite{tesauro_das_2001,tesauro_bredin_2002,vytelingum_etal_2008}), and the papers in which these strategies were shown to outperform human traders (i.e., \cite{das_etal_2001,deluca_cliff_2011_icaart,deluca_cliff_2011_ijcai}). All of these works relied on comparing the strategy of interest with a small number of other strategies in a series of carefully devised experiments. For example, GDX was introduced in \cite{tesauro_bredin_2002}, and was compared only to ZIP and GD. 

In aiming for a fair and informative comparison, experimenters were immediately faced with issues in {\em design of experiments}\/ (see e.g.\ \cite{montgomery_2019}): how best to compare strategy $S_1$ with strategies $S_2$ and $S_3$ (and $S_4$ and $S_5$ and so on), given the finite time and compute-power available for simulation studies, and the need to control for the inherent noise in the simulated market systems. 

Early comparative studies such as \cite{das_etal_2001} limited themselves to running experiments that studied the performance of a selection of trading strategies in three fixed experiment designs: {\em homogeneous}\/ (in which the market is populated entirely by traders of a single strategy-type); {\em one-in-many}\/ (OIM: in which a homogeneous market was altered so that all the traders were of strategy type $S_1$ except one, which was of type $S_2$); and {\em balanced-group} (BG: in which there was a 50:50 split of $S_1$ and $S_2$, balanced across buyers and sellers, with allocation of limit-prices set in such a way that for each trader of type $S_1$ with a limit price of $\lambda_1$ there would be a corresponding trader of type $S_2$ also assigned a limit price of $\lambda_1$).  There were good reasons for this experiment design, and the results were informative, but they rested on only ever comparing two strategies $S_1$ and $S_2$ in markets with a total number of traders $N_T$  where the ratio of $S_1$:$S_2$ was one of either $N_T$:$0$ (i.e., homogeneous); or $(N_T-1)$:$1$ (i.e., OIM); or $\frac{N_T}{2}$:$\frac{N_T}{2}$ (i.e., BG). This approach left open the question of whether the performance witnessed in one of these three special cases generalised to other possible ratios, other relative proportions of the two strategies in the market. 

A method by which that open question could be resolved was developed by \cite{walsh_etal_2002} who borrowed the technique of {\em replicator dynamics analysis} (RDA) from evolutionary game theory (see e.g.\ \cite{maynardsmith_1982}). In a typical RDA, the population of traders is initiated with some particular ratio of the $N_S$ strategies being compared, and the traders are allowed to interact in the market as per usual, but every now and again an individual trader will be selected via some stochastic process and will be allowed to {\em mutate} its current strategy $S_i$ to one of the other available strategies $S_{j\neq i}$ if that new strategy appears to be more profitable than $S_i$. In this way, given enough time, the market system can be started with any possible ratio of the $N_S$ strategies, and in principle it can evolve from that starting point through other system state-vectors (i.e., other ratios of the $N_S$ strategies) to any other possible ratio of those strategies. However in practice the nature of the evolutionary trajectories of the system, i.e. the paths traced by the time-series of state-vectors of the system, will be determined by the profitability of the various strategies that are in play: some points in the state-space (i.e., some particular ratios of $N_S$ strategies) will be unprofitable {\em repellors}, with the evolutionary system evolving away from them; others will be profitable {\em attractors}, with the system converging towards them; and if the system converges to a stable attractor then it's at an {\em equilibrium point}, or potentially on a repeating sequence of equilibrium points, i.e.\ a {\em limit cycle}. Walsh et al's 2002 paper showed the results of RDA for market systems in which $N_S=3$, comparing the trading strategies GD, SNPR, and ZIP, and visualised the evolutionary dynamics as plots of the two-dimensional {\em unit simplex}, an equilateral triangular plane with a three-variable barycentric coordinate frame. 

Similar plots of the evolutionary dynamics on the 2D unit simplex were subsequently used by other authors when comparing trading strategies: see e.g.\ \cite{vytelingum_etal_2008,vach_2015}, and those authors also limited themselves to studies in which the traders in the market could switch between one of only $N_S=3$ different discrete strategies. And, in this strand of research, three-way comparisons seem to then have become the method of choice primarily because evolutionary trajectories through state-space, and the location and nature of any attractors and repellors on the space, is readily renderable as a 2D simplex when dealing with a $N_S=3$ system, but rapidly gets very difficult, to the point of impracticability, as soon as $N_S>3$. Higher-dimensional simplices are mathematically well-defined, but very difficult to visualise: the four-variable simplex is a 3D volume, a tetrahedron; and more generally the $N_S$-variable simplex is an $(N_S-1)$-dimensional volume -- so if we wanted to study the evolutionary dynamics of a six-strategy system, we would need to find a way of usefully rendering projections of the 5-D simplex, or we need to find alternative methods of visualisation and analysis.

However, as first shown by \cite{vach_2015} and later confirmed in more detailed studies by \cite{snashall_cliff_2019,rollins_cliff_2020} and \cite{cliff_rollins_2020}, when the complete state-space of all possible ratios of discrete strategies is exhaustively explored, the dominance hierarchies indicated by the simple OIM/BG analyses are sometimes overturned. That is, if strategy $S_1$ outperformed strategy $S_2$ in both the $OIM$ and the $BG$ tests, that would usually be taken as evidence that $S_1$ generally outperformed $S_2$, that $S_1$ was ``dominant'' in that sense; but actually if markets were set up with some ratio of $S_1$:$S_2$ {\em other}\/ than the OIM or BG ratios, then in those markets $S_2$ would dominate $S_1$ -- that is, the direction of the dominance relationship between $S_1$ and $S_2$ can often depend on the ratio of $S_1$:$S_2$, their relative proportions of the overall population. Furthermore, while $S_1$ might dominate $S_2$ in two-strategy experiments (i.e., where $N_S=2$), plausibly $S_2$ would dominate $S_1$ in experiments where values of $N_S>2$: the indications are that as yet there is no single master-strategy that dominates all others in all situations; what strategy is best will depend on the specific circumstances. 

By populating a model market entirely with adaptive PRZI traders we create a minimal test-bed for exploring issues of market efficiency and stability in situations where all traders are simultaneously co-evolving in an infinite continuous space of strategies. The state at time $t$ of such a market with $N_T$ traders in it can be characterised as an $N_T$-dimensional vector of $s$-values, denoted by $\vec{S}(t)$, identifying a single point in the $N_T$-dimensional hypercube that is the space of all possible system states, and that point will move over time as the traders each adapt their $s$ values. We can attempt to identify attractors and repellors in this hypercube, but we will need new visualisation techniques: we'll need to leave simplices behind.

There are many ways in which a PRZI trader could be made to dynamically adapt its $s$-value in response to market conditions. Here, in the spirit of minimalism associated with studies of ZI traders, I use a crude and simple stochastic hill-climbing algorithm, of the sort that might be found as an introductory illustrative straw-man sketch in the opening chapter of a book on machine learning. To keep with the tradition of naming ZI/MI trading algorithms with short acronyms, I've named this {\em {\bf PR}ZI {\bf S}tochastic {\bf H}ill-Climber} as PRSH (pronounced ``pursh''). PRSH is defined in Section~\ref{sec:prsh_defn}, and then some illustrative baseline results from experiments in which a {\em single} PRSH trader adapting in markets where all other traders are playing fixed strategies are presented in Section~\ref{sec:prsh_solo}. After that, Section~\ref{sec:prsh_coev} shows results from experiments in which {\em all}\/ traders are PRSH, and hence in which the market is maximally co-evolutionary. The Python source-code for PRSH has been released as free-to-use open-source, in BSE (see \cite{cliff_2012_bse}) to enable other researchers to replicate and extend the preliminary results shown here.

\subsubsection{PRSH: a minimal PRZI Stochastic Hill-Climber}
\label{sec:prsh_defn}

At any time $t$, a PRSH trader $i$ has a set of strategies ${\cal S}_{i, t_m}$ that was created at time $t_m \leq t$ and that consists of $k\in {\mathbb Z}^{+}$ different PRZI strategy values $s_{0,t_m}$ to $s_{{k-1},t_m}$ (i.e., $|{\cal S}|=k>1$). Although $t$ is continuous in this model, alterations to $S_{i,t}$ happen only occasionally. After an initialisation step in which the $k$ strategies are each assigned a value $s_{i,t_0} \in [-1,+1] \in {\mathbb R}$ via a {\em genesis} function ${\cal G}(.)$, PRSH enters into an infinite loop: let $t_m$ denote the time at which a new iteration of the loop is initiated; in each cycle of the loop a PRSH trader first {\em evaluates} each of its $k$ strategies in turn, trading with each of them as the sole exclusive strategy for at least a minimum period of time $\Delta_t$, such that all $k$ have been evaluated by time $t_n \geq t_m + k\Delta_t$; after that, it {\em ranks} the strategies by some performance or {\em fitness} metric $\cal F$, and copies the top-ranked strategy (the {\em elite}) at time $t_n$ into $s_{0,t_n}$; it then creates $k-1$ new `mutants' of $s_{0,t_n}$, via a stochastic {\em mutation} function ${\cal M}(s_{0,t_n})$, and this set of new strategies $s_{j,t_n:1\leq j \leq k-1}$ then replaces the old ${\cal S}_{i,t_m}$, becoming  ${\cal S}_{i, t_n}$, at which point it loops back for the next iteration (and hence in that next iteration the value $t_m$ is what was $t_n$ in the prior iteration). 

This definition leaves the experimenter free to decide certain key details when implementing PRSH:
\begin{itemize} 

\item The choice of $k$ and of $\Delta_t$ together determine the speed of adaptation: PRSH will generate a new ${\cal S}_{t_i}$ at most once every $k\Delta_t$ seconds: i.e., $k\Delta_t$ is the minimum time-period between successive mutations, where each mutation is an {\em adaptive step} on the underlying {\em fitness landscape}. If you want a PRSH to make $N_{\text{steps}}$ adaptive steps on the fitness landscape in the course of an experiment, that experiment needs to run for $>k \Delta_t N_{\text{steps}}$ seconds. 

\item Exactly how the set ${\cal S}_0$ is created at initialisation is left open. Naturally $s_{i,0} = \mathcal{U}( -1, +1) \in {\mathbb R}; i \in \{0,\ldots,k-1\}$ is the least constrained, but there may be circumstances where it is informative to use some other method, e.g.  $s_{i,0} = c; \forall i$ for some constant $c$ such as zero or $\pm 1$. 

\item The stochastic function ${\cal M}: [-1, +1] \in {\mathbb R} \mapsto [-1, +1] \in {\mathbb R} $ that creates new mutants of the elite $s_{0,t_k}$ is similarly unspecified. Treating each mutation as the addition of a random draw from a distribution with zero mean and nonzero variance makes intuitive sense, and then either truncating or using ring-arithmetic to ensure that the function maps to $[-1,+1]$. In the experiments shown below, ${\cal M}(s_{0,t_k}) = s_{0,t_k} + \mathcal{N}(0,\sigma)$ with $\sigma=0.01$. Plausibly a simulated-annealing approach could be introduced, steadily reducing $\sigma$ as time progresses, but that is not explored here.  

\item For $k>2$, questions immediately arise over what is the best way of generating the $k$ mutants. For instance if $k=3$ we could arrange a set of two different ${\cal M}$ functions, one per mutant, such that $s_{1,t_k} < s_{0,t_k}$ and $s_{2,t_k} > s_{0,t_k}$  and hence PRSH is always sampling $s$-values at random magnitudes either side of the current elite strategy; and for $k=5$ we could similarly arrange the mutants such that two are generated either side of the elite, one a small random distance away, and the other a much larger random distance away; such decisions are left as an implementation issue. In the work reported here we simply generate $k-1$ mutants via $\cal M$ with no additional constraints.

\item  Finally, each iteration of the loop requires deciding which of the $k$ strategies is the current elite, via the fitness function $\cal F$, and there are many possible ways to do that. The method used here was to rank the $k$ strategies at time $t_k$ by the amount of profit generated per unit of time, denoted by {\sc pps} (profit per second), such that the elite $s_{0,t_k}$ strategy has the highest {\sc pps}. To help avoid the  hill-climber from becoming trapped on local maxima, if the difference between the {\sc pps} scores of the two highest-ranked $s$-values in $S$ is less than some threshold $\epsilon_s$, then one of the two is chosen at random to be the elite for that iteration of the loop. 

\end{itemize}

In essence, PRSH with $k$ strategies is a very primitive $k-$armed bandit, and all of the extensive multi-armed bandit (MAB) literature (such as \cite{gittins_etal_2011,myleswhite_2012,lattimore_szepesvari_2020}) is potentially of relevance here, but ignored: again, the intention here is not to create the best adaptive-PRZI trader, instead it is merely to have a simple minimal adaptive-PRZI algorithm to act as a proof of concept and to enable an initial set of exploratory and illustrative experiments involving populations of adaptive-PRZI traders: PRSH does that job.

\subsubsection{Adaptive Evolution of Strategy in a Single PRSH}
\label{sec:prsh_solo}

Before studying co-evolving populations of PRSH traders, it is informative to explore situations in which there is only a single PRSH trader in the market, and all other traders are one or more of the three ZI strategies that are spanned by PRSH/PRZI, i.e. GVWY, SHVR, and ZIC. In such situations we can talk of how the PRSH trader's strategy evolves over time, but not of co-evolution because the rest of the traders in the market are non-adaptive. A single-PRSH-trader market is sufficiently simple that it eases the introduction of concepts that become significantly more complex in fully co-evolutionary markets.

First, we can visualise the fitness landscape for a single PRSH trader by setting up a market in which, purely for the sake of generating appropriate visualization data, we give the PRSH a large $k$, and initialize $S_0$ to a set of regularly-spaced $s_{i,0}$ values across the range $[-1,+1]$, and then plot the {\sc pps} fitness of each strategy in the first evaluation. Specifically, set:
$ S_0 = \{ s_{i,0}:  s_{i,0}=\frac{2i}{k-1}-1 ; i \in \{0, \ldots, k-1\}\} $
And let $\Delta_S=2/(k-1)$, the step-size in our mapping of the fitness landscape.
So for example with $k=21$ we have $\Delta_S=0.1$ and $S_0=\{-1,-0.9,-0.8,\ldots,+0.9,+1.0\}$.

For brevity, and without loss of generality, the discussion that follows in the rest of this section concentrates only on the case of a single PRSH seller in a market that is otherwise entirely populated by traders running nonadaptive strategies. The arguments that are made here for a single PRSH seller could just as easily be made for a single PRSH buyer, but to do both here would be overkill. 
 
\begin{figure}
\begin{center}
\includegraphics[width=0.7\linewidth]{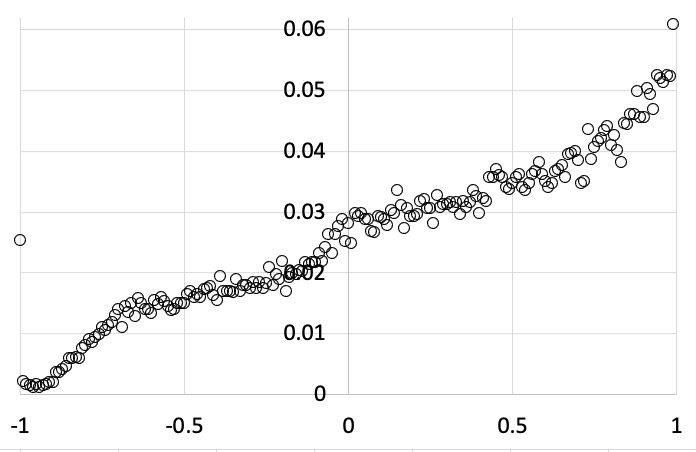}

\includegraphics[width=0.7\linewidth]{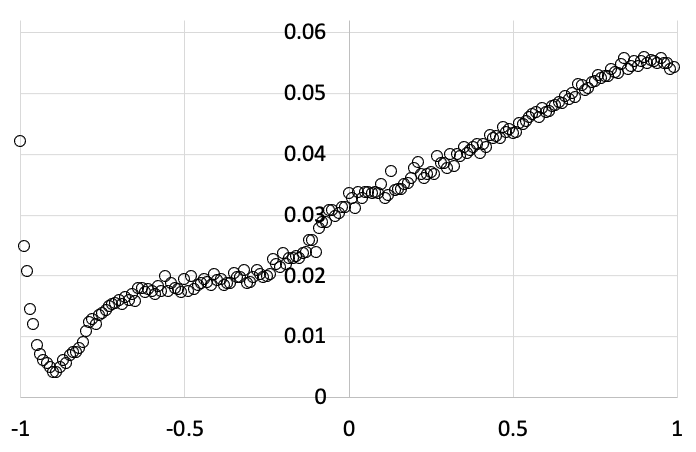}

\includegraphics[width=0.7\linewidth]{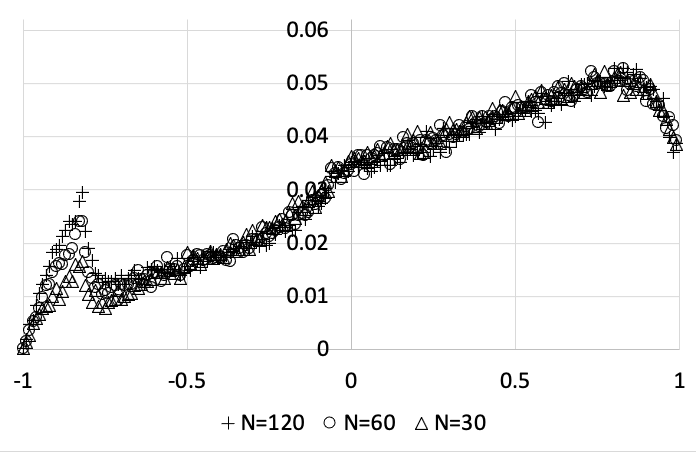}

\end{center}
\caption{Fitness landscapes for a single PRSH seller in a market where all other traders are homogeneously playing the same fixed strategy: horizontal axis is PRSH strategy value $s$; vertical axis is profit per second ({\sc pps}) recorded by the single PRSH trader using $s$ as its strategy. Strategy evaluation time $\Delta_t$ is 7200s. Data points are plotted at strategy-steps of $\Delta_S=0.01$.  Upper graph is when all other traders are playing the fixed SHVR strategy; middle graph is when all other traders are ZIC; lower graph is when all other traders are GVWY. In the lower graph only, data is shown for {\sc iid} repetitions of the experiment with the number of traders in the market (denoted by $N_T$) being set to 30 (data-points marked by open triangles), 60 (marked by open circles), and 120 (marked by plus-symbols).}
\label{fig:SHVRlandscape}
\end{figure}

Figure~\ref{fig:SHVRlandscape} shows fitness landscapes plotted  at $\Delta_S=0.05$ for a single PRSH seller when all other traders in the market are either (from top to bottom) SHVR, ZIC, or GVWY: i.e., a progression from all other traders in the market being maximally relaxed (SHVR) through to maximally urgent (GVWY). In all experiments reported in this paper, all buyers had the same limit price $\lambda_b$ and all sellers had the same limit price $\lambda_s < \lambda_b$, i.e. the supply and demand schedules were `box'  style, with perfect elasticity of supply and of demand.\footnote{Specifically, in all the experiments reported here, $\lambda_s=60$ for all sellers and  $\lambda_b=100$ 
for all buyers; and the number of buyers and sellers are the same (i.e., $N_{\text{Buy}}=N_{\text{Sell}}$), so there is no clearly defined equilibrium price in these market sessions: transactions can be expected to take place at any price in the range $[\lambda_s, \lambda_b]\in{\mathbb Z}$, and in principle all traders can expect to find a counterparty to transact with -- i.e., there are no extramarginal traders.} 
When generating the landscapes for SHVR and ZIC  the number of buyers ($N_{\text Buy}$) and the number of sellers ($N_{\text Sell}$) were each 30, i.e. $N_T=60$, but in the landscape for GVWY results from $N_T=60$ are overlayed with additional results from {\sc iid} repetitions of the same experiment where $N_T=30$ and where $N_T=120$ (in each case $N_{\text Buy} = N_{\text Sell} = N_T/2$), to demonstrate that the overall shape of the fitness landscape varies very little with respect to the $N_T=60$ case when the number of traders is halved or doubled.  
As can be seen from Figure~\ref{fig:SHVRlandscape}, in the single-PRSH case the fittest (most profitable) strategies --  i.e., the global maxima --  are all at the high end of the range, at or close to $s=+1$, but in each landscape there is also a local maxima at/near $s=-1$.

\begin{figure}
\begin{center}
\includegraphics[width=0.8\linewidth]{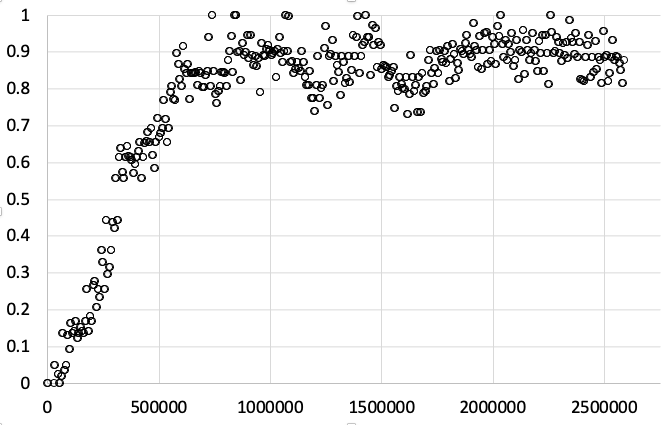}
\end{center}
\caption{Hourly strategy value over 30 days of round-the-clock trading for a single PRSH seller in a market populated with 29 GVWY sellers and 30 GVWY buyers: horizontal axis is time in seconds; vertical axis is the PRSH trader's strategy value $s$, which is initialized at the start of the experiment to $s=0$, i.e.\ to the ZIC strategy. The $s$-value evolves steadily toward a range of values close to the global optimum strategy identified in the bottom fitness-landscape plot of Figure~\ref{fig:SHVRlandscape}, and then stabilises to that range of values for the remainder of the experiment.
}
\label{fig:gvwy_evolve_raw}
\end{figure}

\begin{figure}
\begin{center}
\includegraphics[width=0.9\linewidth]{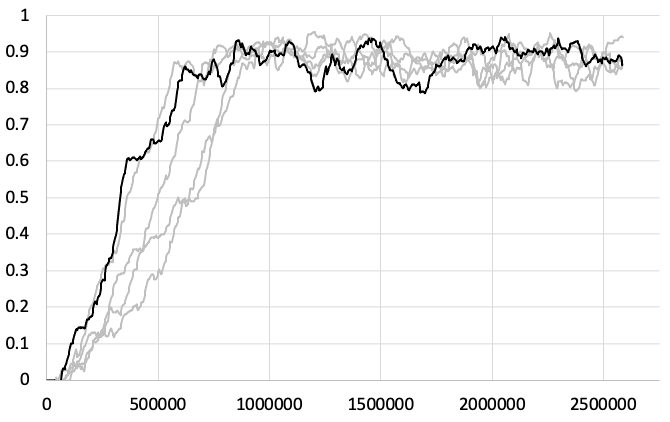}
\end{center}
\caption{Smoothed PRSH strategy values from multiple 30-day experiments, each with a single PRSH seller in a market populated by 29 GVWY sellers and 30 GVWY buyers: horizontal axis is time in seconds; vertical axis is 12-hour moving-average strategy value (denoted by $\hat{s}$). Black line is the $\hat{s}$ trace for the raw hourly $s$-data shown in Figure~\ref{fig:gvwy_evolve_raw}; the four grey lines are each the $\hat{s}$ traces from four {\sc iid} repetitions of the same experiment. After 100,000 seconds (roughly 11 days) of trading, all five $\hat{s}$ traces have evolved to a steady state close to the global optimum strategy identified in the bottom fitness-landscape plot of Figure~\ref{fig:SHVRlandscape}, and remain clustered around that value for the remainder of the experiment. The set of final $\hat{s}$ values recorded at the end of each experiment is referred to as the {\em terminal strategy set}, denoted by $\hat{S}_T$. Here, $\widehat{S}_T=\{ 0.86, 0.87, 0.88, 0.88,  0.93 \}$: see text for further discussion. 
}
\label{fig:gvwy_evolve_smooth}
\end{figure}

The GVWY fitness landscape for a single PRSH seller shown at the bottom of  Figure~\ref{fig:SHVRlandscape} clearly has a global maximum at $s \approx 0.8$. If the PRSH adaptation mechanism is operating as intended, when the single PRSH seller is initialised with $s=0$ and allowed to adapt for sufficiently long then its $s$ value should converge to roughly $0.8$, and then hold at that value. To demonstrate this, Figure~\ref{fig:gvwy_evolve_raw} shows the PRSH trader's $s$ value, plotted once per hour, in a simulation of 30 continuous days of 24-hour trading: as can be seen, from its initial value of zero there is a steady rise in $s$ over the first $\approx$750,000sec of trading (i.e., roughly the first 8.5 days), after which the system stabilises to $s$-values that noisily fluctuate around the $0.85$ level. To smooth out some of the noise, define $\hat{s}$ as the 12-hour simple moving average of the raw hourly $s$ data: Figure~\ref{fig:gvwy_evolve_smooth} shows the $\hat{s}$ line for the raw hourly data shown in Figure~\ref{fig:gvwy_evolve_raw}, along with $\hat{s}$ lines from a further four {\sc iid} repetitions of the same experiment. For the discussion that follows, let's call trader $i$'s $\hat{s_i}$ value at the end of an experiment the {\em terminal strategy} for $i$ in that experiment, and define the set $\widehat{S_{T}}$ as the set of terminal strategies from a population of PRSH traders that have co-evolved in a particular market environment. For the current discussion of the merely evolutionary (i.e., not co-evolutionary) adaptation of single PRSH traders, we can fill $\widehat{S_{T}}$ with the set of terminal strategy values arising from $N_R$ {\sc iid} repetitions of a particular experiment: in Figure~\ref{fig:gvwy_evolve_smooth}, we have $N_R=5$ and $\widehat{S}_T=\{ 0.86, 0.87, 0.88, 0.88,  0.93 \}$. As $N_R$ takes on larger values, it is natural to summarise values in the terminal strategy set $\widehat{S}_T$ as a frequency histogram or kernel density estimate, and from there to note whether the distribution of values in the terminal strategy set is unimodal or multimodal, either by eyeballing the distribution or density estimate, or by applying a test of modality such as those proposed by \cite{hartigan_hartigan_1985} or \cite{chasani_likas_2022}. 

\subsubsection{Co-Evolution of Strategies in All-PRSH Markets}
\label{sec:prsh_coev}

\begin{figure}
\begin{center}
\includegraphics[width=0.8\linewidth]{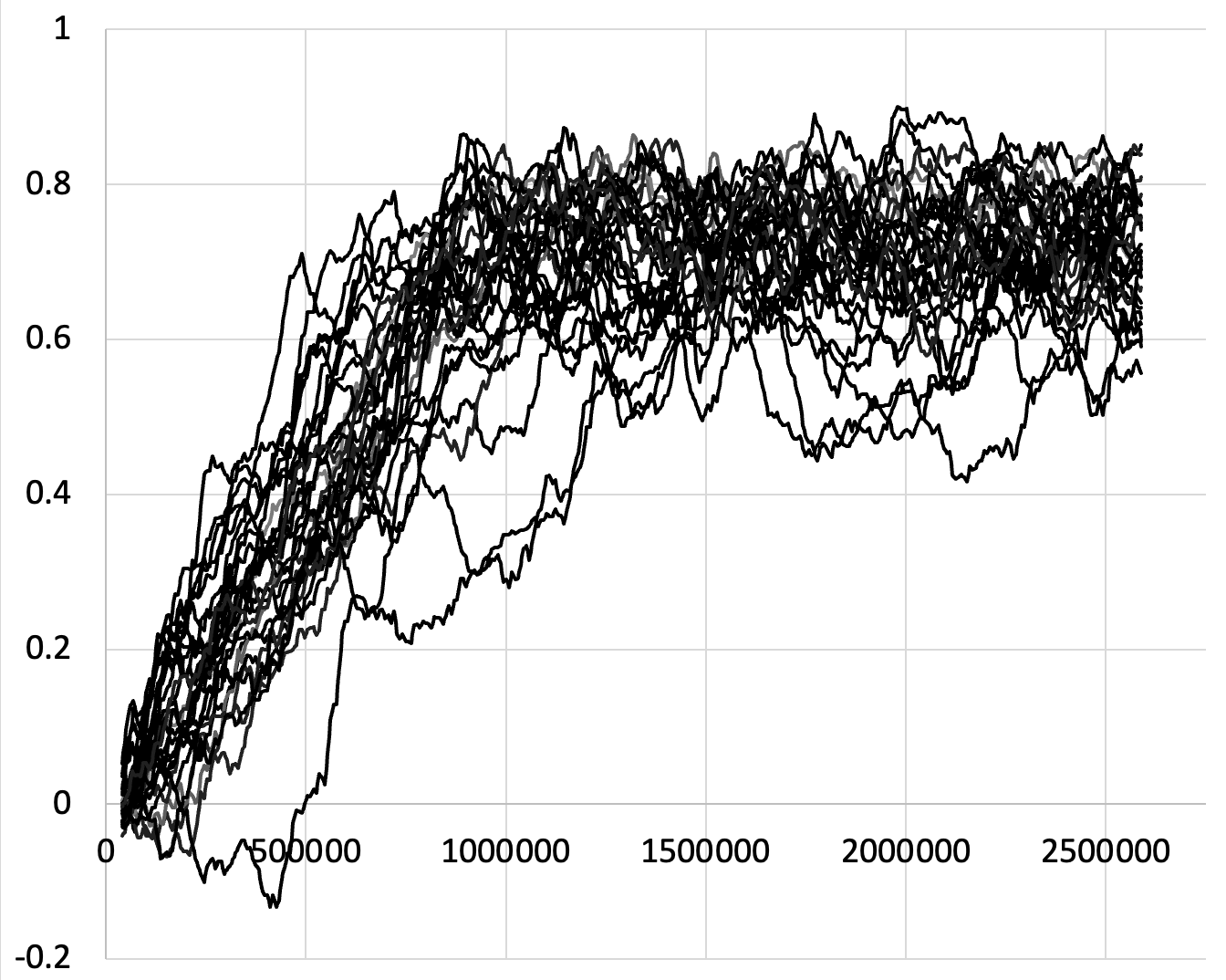}
\end{center}
\caption{Smoothed ($\widehat{s_{i,t}}$) strategy values for each of 30 PRSH sellers in a market experiment lasting for 30 days of continuous trading, where all traders are initialized to have $s_{i,0}=0$. Horizontal axis is time in seconds; vertical axis is the 12-hour moving average strategy $\widehat{s_{i,t}}$ of individual traders. The co-evolutionary dynamic is biphasic: in the initial ``adaptive transient'' phase over the $\approx12$ days (i.e., $\approx$1,000,000 seconds) the system settles to a unimodal steady-state centered on $s_i\approx0.7$; in the steady-state phase the strategy values of individual traders rise and fall but the overall distribution does not vary significantly.
}
\label{fig:prsh_coev_0init_s-hat}
\end{figure}

As a first illustration of the dynamics of a fully co-evolutionary ZI market system, Figure~\ref{fig:prsh_coev_0init_s-hat} shows the $\widehat{s_i}$ values over time for a 30-day experiment in which the market is populated by 30 PRSH sellers and 30 PRSH buyers, all of which are initialized to have $s_{i,0}=0$: i.e.\ an experiment directly comparable to the results from the zero-initialized single-PRSH system explored in the previous section, except that here the fitness landscape for any one trader will depend on the distribution of strategy-values for all the other traders in the market, and in which the fitness landscape will be varying over time, in principle altering each time any one PRSH trader changes its strategy to a new value. Again, a $\widehat{S_T}$ terminal strategy set can be assembled from the final $\widehat{s_i}$ values of the individual traders that co-evolved against each other in the single market experiment: the corresponding terminal strategy set distribution is again unimodal: in this experiment, all sellers converge on strategy-values in $[\approx+0.55, \approx+0.85]$; multiple {\sc iid} repetitions of this market experiment generate much the same results. 


Further investigation reveals that the unimodal distribution of terminal strategies in experiments like the one illustrated in Figure~\ref{fig:prsh_coev_0init_s-hat} is an artefact of the decision to initialize all traders with $s_{i,0}=0$: if instead we set $s_{i,0}={\cal U}(-1.0,+1.0)$ so that the initial set of strategy values in the population of traders is uniformly distributed over the entire range of possible strategies, we see qualitatively different results:
for both the buyers and the sellers the distribution of terminal strategy values is then multimodal. 



\begin{figure}
\begin{center}
\includegraphics[width=0.99\linewidth]{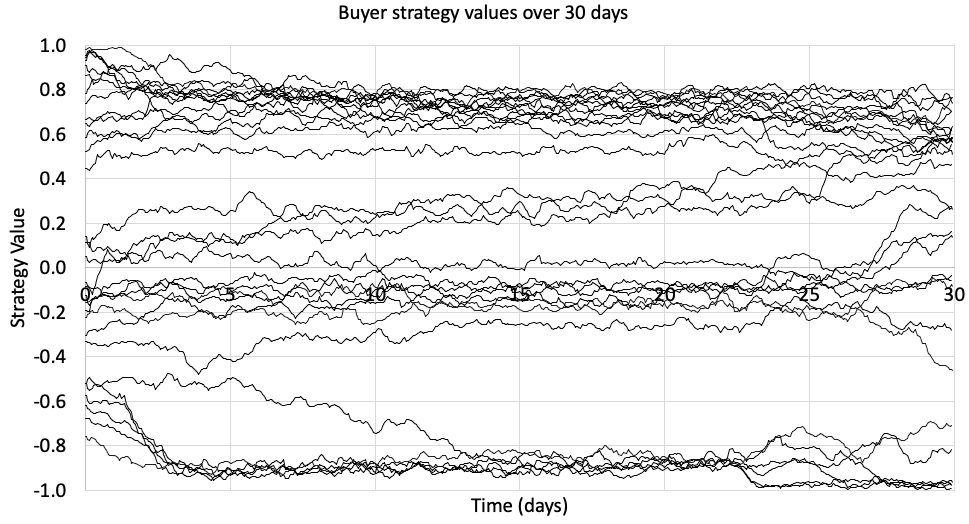}
\end{center}
\caption{Smoothed ($\widehat{s_{i,t}}$) strategy values for each of 30 PRSH buyers in a market experiment lasting for 30 days of continuous trading, simulated at 60Hz time-resolution, where all traders are initialized to have $s_{i,0}={\cal  U}(-1.0,+1.0)$. Horizontal axis is time $t$, with a vertical gridline every 5 days; vertical axis is the 12-hour moving average strategy $\widehat{s_{i,t}}$ of individual traders, with horizontal gridlines at $s$ intervals of 0.2: for $t\geq 0.5$~days (i.e., 12 hours) the trader's average strategy value over the preceding 12 hours is plotted; for $t<0.5$~days the trader's average strategy since the start of the experiment is plotted. By roughly Day~13 the system has settled into a state that then persists as a temporary equilibrium or stasis until roughly Day~22: during the equilibrium phase the modes are at roughly $s=-0.9$ $(n=6)$, $s=-0.1$ $(n=8)$, $s=+0.3$ $(n=3)$, and $s=+0.7$ $(n=13)$. After that, the equilibrium ``punctuates'', entering a new phase where first the mode at $-0.9$ loses its stability, then the mode at $+0.3$ seems to merge up into the mode that was at $+0.7$ but which now seems to be generally heading lower, and then the mode at $-0.1$ seems to dissipate in various directions. In the nine-days stasis/equilibrium, each trader would execute approximately 150,000 transactions. Clearly the dynamics have not reached a stable state after 30 days of trading, and longer simulations should be explored.  
}
\label{fig:prsh_coev_M1P1init_s-hat_30}
\end{figure}

The development of multimodal terminal strategy distributions is not the only change resulting from switching the initial state from $s_{\forall i}=0.0$ to $s_{\forall i}={\cal U}(-1.0,+1.0)$. In Figure~\ref{fig:prsh_coev_0init_s-hat}, over the 30 simulated days, the dynamics of the system's co-evolution through strategy space are biphasic: an initial {\em adaptive transient}\/ phase of roughly 12 days in which all traders increased their $s$ values from zero to $\approx0.7$; followed by a steady-state phase lasting for the remainder of the experiment where the population of $s$ values wandered randomly around the $0.7$ level In contrast, when $s_{\forall i}={\cal U}(-1.0,+1.0)$ the system shows no such long-term stability over the same time-period, as is illustrated in Figure~\ref{fig:prsh_coev_M1P1init_s-hat_30} and explained in the caption to that figure: even after the system's distribution of strategies has been relatively stable for a period of nine days, an equilibrium or stasis in which the traders have each executed roughly 150,000 transactions, chance co-evolutionary interactions can result in the stasis ending and the system entering a fresh period in which the strategies are in flux.

\begin{figure}
\begin{center}
\includegraphics[width=0.49\linewidth]{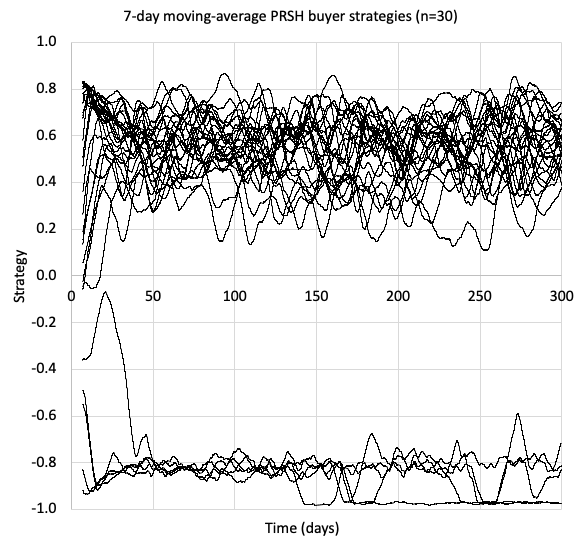}
\includegraphics[width=0.49\linewidth]{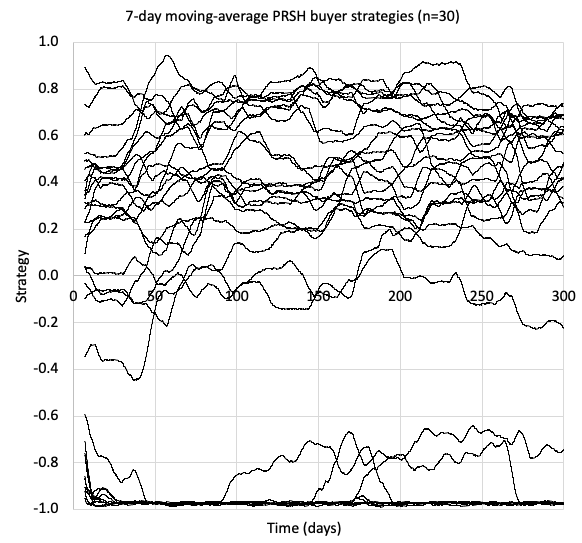}
\includegraphics[width=0.49\linewidth]{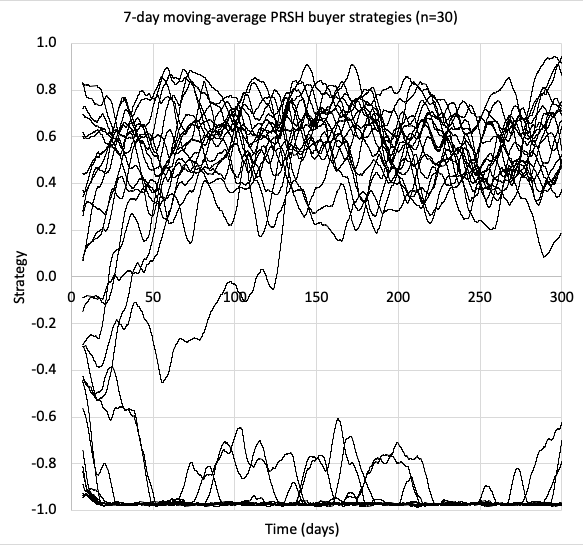}
\includegraphics[width=0.49\linewidth]{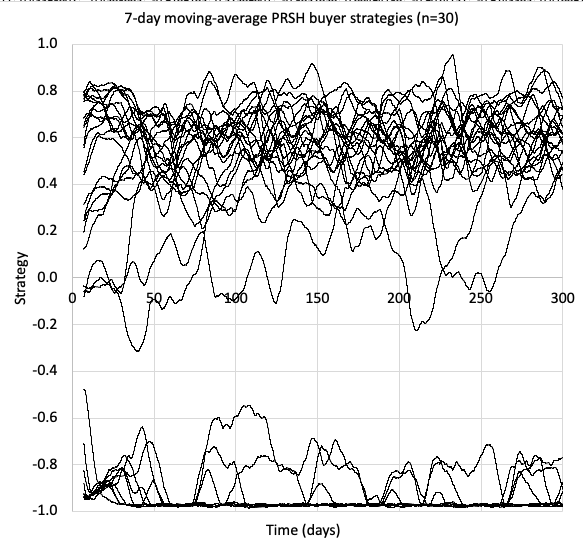}
\includegraphics[width=0.49\linewidth]{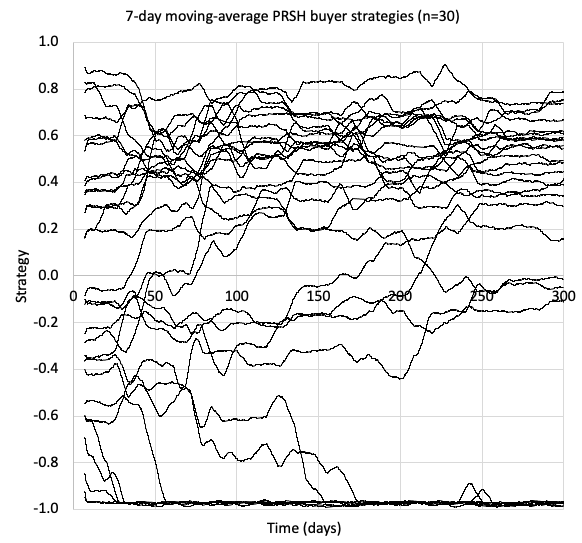}
\includegraphics[width=0.49\linewidth]{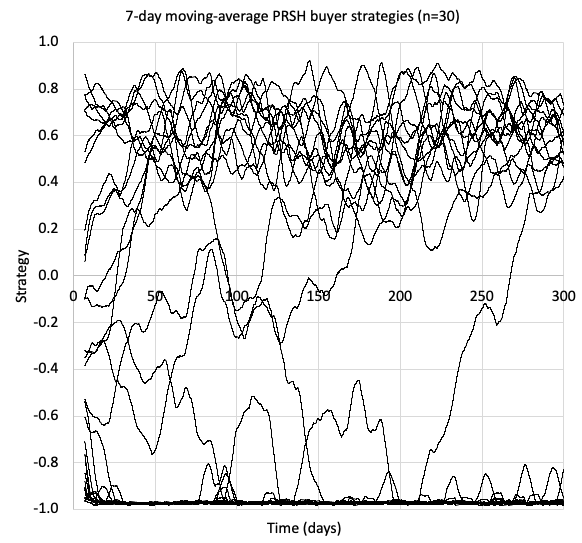}
\includegraphics[width=0.49\linewidth]{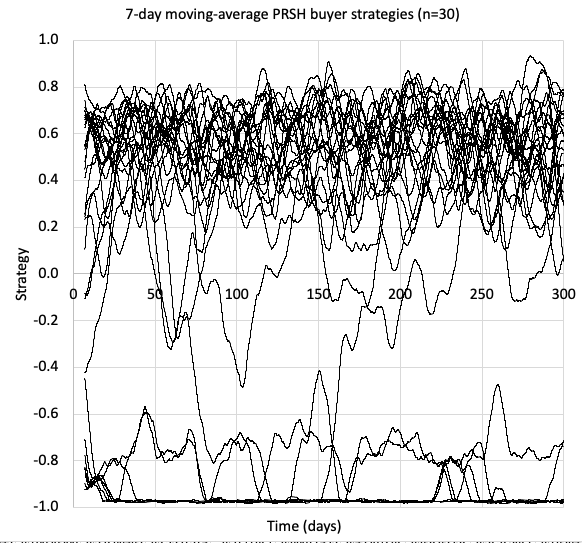}
\includegraphics[width=0.49\linewidth]{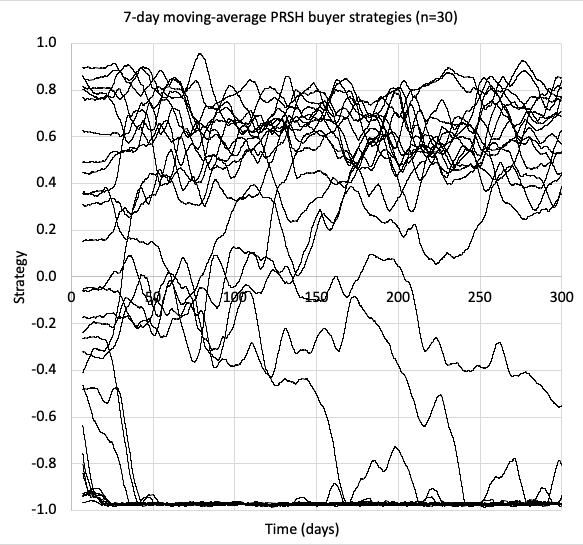}
\end{center}
\caption{Results from eight {\sc iid} experiments each otherwise the same as that illustrated in Figure~\ref{fig:prsh_coev_M1P1init_s-hat_30} but instead continued for 300 days. Data lines show smoothed ($\widehat{s_{i,t}}$) strategy values for each of 30 PRSH buyers in a market experiment over 300 days of continuous trading, simulated at 60Hz time-resolution, where all traders are initialized to have $s_{i,0}={\cal  U}(-1.0,+1.0)$. Horizontal axis is time $t$, with a vertical gridline every 50 days; vertical axis is the 7-day moving average strategy $\widehat{s_{i,t}}$ of individual traders, with horizontal gridlines at $s$ intervals of 0.2. See text for further discussion. 
}
\label{fig:prsh_coev_M1P1init_s-hat_300}
\end{figure}

To illustrate the longer-term dynamics of this system, Figure~\ref{fig:prsh_coev_M1P1init_s-hat_300} shows buyer-strategy co-evolutionary time series similar to that illustrated in Figure~\ref{fig:prsh_coev_M1P1init_s-hat_30} from eight {\sc iid} repetitions of an experiment that lasted 10 times longer, i.e. 300 simulated days. As is clear from the figure, although stable modes do occur in each experiment, individual trader's strategy-values will sometimes transition from one mode to another, with no clear pattern or predictability to the timing and/or direction of these transitions.  In particular,  The upper four graphs in Figure~\ref{fig:prsh_coev_M1P1init_s-hat_300} appear to show that, after an initial adaptive transient phase, the population of traders settles into a steady-state bimodal distribution; but the lower four graphs show that the system does not always quickly converge to such a steady-state distribution and that co-evolutionary interactions can result in major changes in the strategy distributions (e.g., a trader switching from one mode to another) even after 200 or more days of continuous trading, a period over which each trader would execute roughly 3,500,000 transactions.

\begin{figure}
\begin{center}
\includegraphics[width=0.75\linewidth]{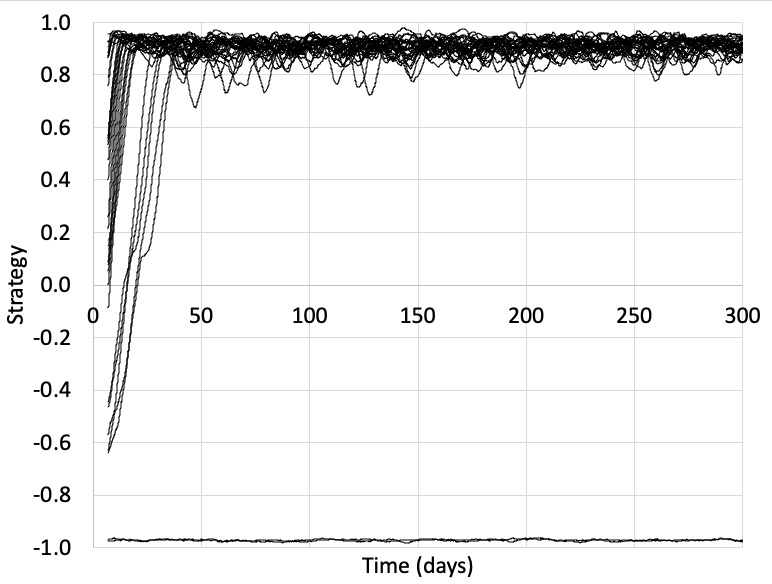}
\end{center}
\caption{Time-series of co-evolving seller strategies from one of the eight experiments for which the buyer strategies were illustrated in Figure~\ref{fig:prsh_coev_M1P1init_s-hat_300}: qualitatively, all eight experiments have time series essentially the same as this one, so only the one is illustrated here. The vast majority of sellers rapidly shift their strategy-values to around $+0.9$, but in any one experiment a small number of sellers instead settle on strategy values close to $-1.0$. In all cases, these two modes are stable for the remainder of the duration of the experiment.  
}
\label{fig:sell_strat_300}
\end{figure}

Thus far, to save space, only the co-evolutionary trajectories of the strategies in the population of buyers have been shown. Naturally, each of the eight buyer-strategy time-series graphs shown in Figure~\ref{fig:prsh_coev_M1P1init_s-hat_300} has a corresponding seller-strategy
time-series graph, but in this specific set of experiments there was much less variation in the outcomes for the seller population: rather than showing all eight,  Figure~\ref{fig:sell_strat_300} shows one representative example; qualitatively, the other seven are all essentially identical to this.

\begin{figure}[hp]
\begin{center}
\includegraphics[width=0.45\linewidth]{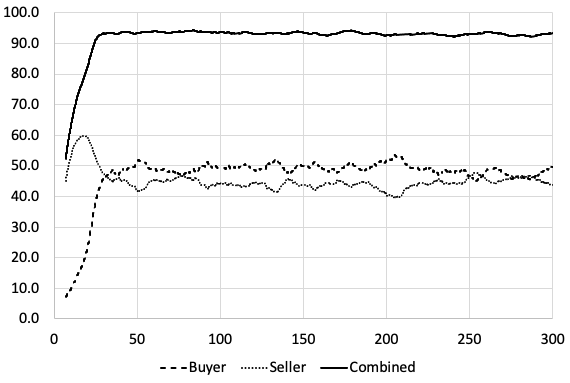}
\includegraphics[width=0.45\linewidth]{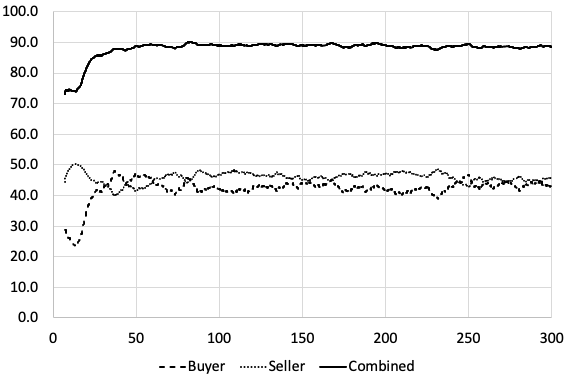}
\includegraphics[width=0.45\linewidth]{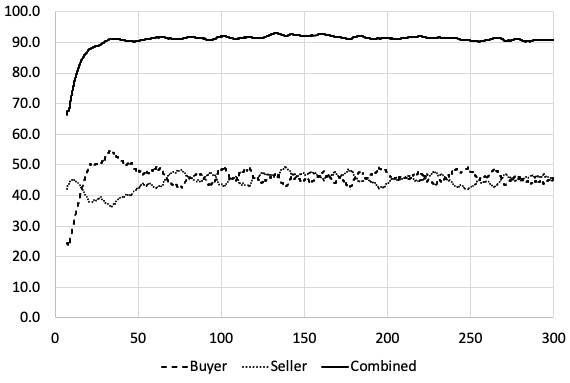}
\includegraphics[width=0.45\linewidth]{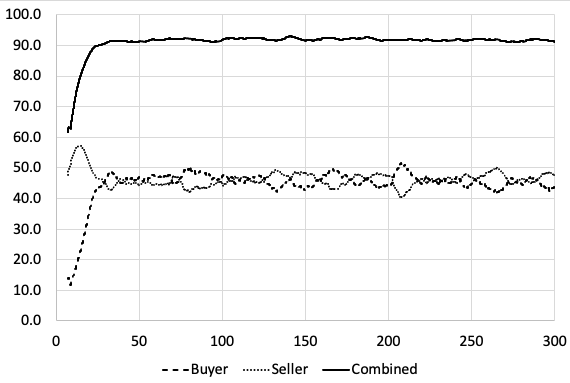}
\includegraphics[width=0.45\linewidth]{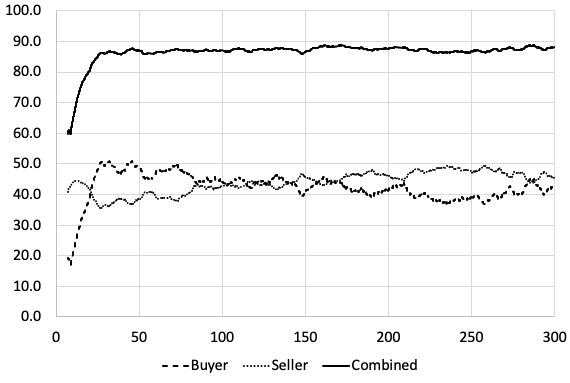}
\includegraphics[width=0.45\linewidth]{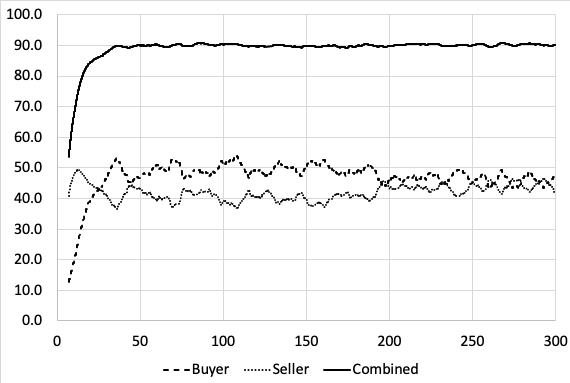}
\includegraphics[width=0.45\linewidth]{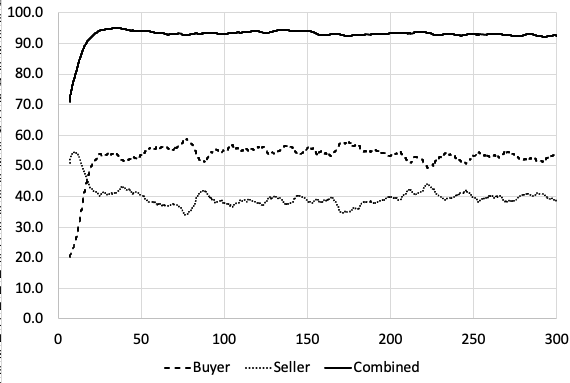}
\includegraphics[width=0.45\linewidth]{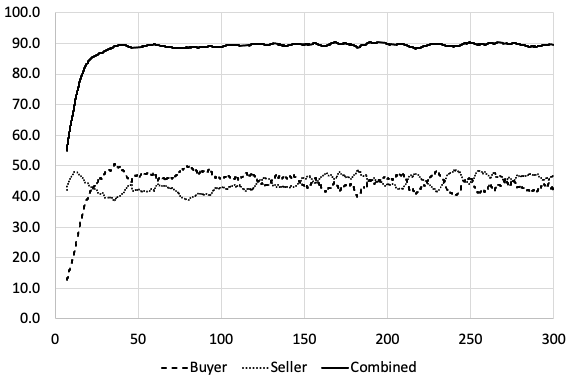}
\end{center}
\caption{Total extraction of surplus/profit for the eight 300-day experiments illustrated in Figure~\ref{fig:prsh_coev_M1P1init_s-hat_300}: horizontal axis is time in days; vertical axis is total profit extracted by a group of traders. Data-lines show collective total profit extracted by the 30 buyers, collective total profit extracted by the 30 sellers, and total profit extracted by the entire set of 60 traders. In each case, after the initial adaptive transient over the first 50 days or less, the buyers' and seller's profit levels stabilise to an approximately constant-sum relationship, where if buyers' profits go up then sellers' profits go down, and {\em vice versa}. The sum that the two populations' profit-levels adds up to is notably unvarying within any one experiment, but varies across experiments: for example, the experiments at upper-left and lower-left both have the sum consistently around 93-95, whereas the upper-right and the left-hand experiment in the third row from the top both never see their sum go above 90. See text for further discussion. 
}
\label{fig:prsh_coev_M1P1init_Prof_300}
\end{figure}

\begin{figure}
\begin{center}
\includegraphics[width=0.8\linewidth]{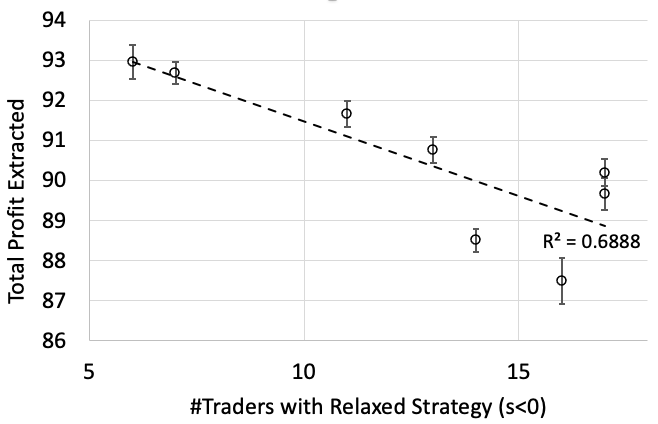}
\end{center}
\caption{Inverse relationship between the percentage of traders in the market playing relaxed strategies (i.e., $s_i<0$) and total profit extracted by all the traders in the market: horizontal axis is percentage of traders with $s_i<0$; vertical axis is profit extracted over the final 50 days' trading (i.e., days 250-300) in the experiment. Markers show the arithmetic mean over that period, with error bars at $\pm$ one standard deviation, for the eight experiments illustrated in Figure~\ref{fig:prsh_coev_M1P1init_s-hat_300}. The dashed line shows linear regression; $R^2\approx 70\%$.
}
\label{fig:run_300_relaxed_corr}
\end{figure}

The co-evolutionary dynamics of strategy values in these model markets is not the only factor of interest: another equally significant concern is the efficiency of the markets populated by traders with co-evolving strategies: something that is illustrated in Figure~\ref{fig:prsh_coev_M1P1init_Prof_300} which shows, for each of  
the eight 300-day experiments illustrated in Figure~\ref{fig:prsh_coev_M1P1init_s-hat_300}, the total surplus/profit extracted by the traders. Data-lines show collective total profit extracted by the 30 buyers (denoted here as $\pi_B$), collective total profit extracted by the 30 sellers (denoted here as $\pi_S$), and total profit extracted by the entire set of 60 traders (denoted here as $\pi_T = \pi_B + \pi_S$). In each case, after the initial adaptive transient over the first 50 days or less, the buyers' and seller's profit levels stabilise to an approximately constant-sum relationship, where if $\pi_B$ goes up then $\pi_S$ goes down, and {\em vice versa}. The sum $\pi_T$ that the two populations' profit-levels add up to is notably unvarying within any one experiment, but the value that $\pi_T$ settles on varies across experiments: for example, the experiments at upper-left and lower-left both have $\pi_T \approx 93-95$, whereas the upper-right and the left-hand experiment in the third row from the top both never see $\pi_T$ go above 90. The underlying reason for this variation in total profit extracted is illuminated in Figure~\ref{fig:run_300_relaxed_corr}, which shows the inverse relationship between the number of traders with `relaxed' strategy values ($s_i<0$) in the terminal strategy set and the total profit extracted: the more relaxed traders there are present in the market, the less profit extracted; despite their constant striving to improve profitability, traders with strategy values in the relaxed mode seem to be stuck on a local maximum in the fitness landscape.  

Although the time-series of co-evolving strategy values and histograms of strategy frequency distributions have served the purposes of this discussion thus far, there is a need for more sophisticated visualization and analysis techniques. Our very first studies studies of co-evolutionary dynamics with a preliminary $k=2$ PRSH-like system, reported in \cite{alexandrov_cliff_figuero_2022} explored the prospects of producing phase portraits, graphical characterisations of the global dynamics of the system, for market sessions in which there are only two evolving traders, each adjusting their $s$-values with the intent of improving their profitability, while all other traders play fixed strategies: in such a two-PRSH market the phase-space of interest is two-dimensional, just the two evolving strategies, and hence very easy to plot as a 2D graphic. But for the all-PRSH $N_T=60$ market sessions studied here, we  need a useful way of plotting the trajectory of the dynamical system through its 60-dimensional real-valued phase-space: that is, the strategy vector $\vec{S}(t) \in [-1.0,+1.0]^{N_T} \in {\mathbb R}^{N_T}$. 

Thankfully, in recent decades researchers in physics have developed a set of visualisation and analysis tools and techniques for such high-dimensional real-valued dynamical systems: the dynamics of such systems can be characterised visually, as a square array of pixels, via the creation of a {\em recurrence plot} (RP), which will often display macro-scale features that are obvious to the human eye; and then straightforward image-processing techniques can be used to generate quantitative statistics that summarise the nature of the RP and the features within it, an approach known as {\em Recurrence Quantification Analysis} (RQA). For readers unfamiliar with RPs and RQA, Appendix~\ref{appendix:RP} presents a brief introduction. 

\begin{figure}
\begin{center}
\includegraphics[trim=80 30 80 40,clip,width=0.8\linewidth]{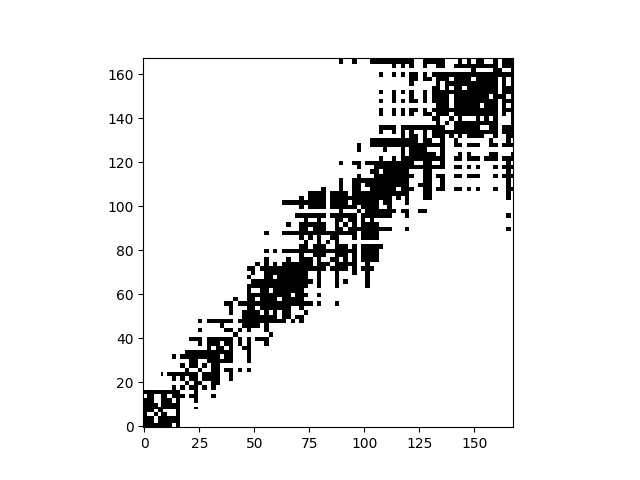}
\end{center}
\caption{
Example recurrence plot (RP) for a PRSH co-evolutionary market session: in this experiment (as with the experiments illustrated in Figures~\ref{fig:prsh_coev_M1P1init_s-hat_300} and~\ref{fig:prsh_coev_M1P1init_Prof_300}) there are 30 PRSH buyers and 30 PRSH sellers (i.e., $N_T=60$) each co-evolving their individual strategy $s$-values, so the collective state of the system of co-evolving strategy values at time $t$ is a strategy-vector $\vec{S}(t)\in[-1.0, +1.0]^{60} \in {\mathbb R}^{60}$. The traders interact continuously, simulated at 60Hz, trading around the clock 24 hours per day, but the $\vec{S}$ strategy-vector is recorded only once every hour. This RP shows the first 7 days of the market session (i.e., $7\times24=168$ hours): numeric labels on the axes are hour-number. The state $\vec{S}(t)$ is considered a recurrence of the state $\vec{S}(t-\Delta_t)$ when $|\vec{S}(t)-\vec{S}(t-\Delta_t)|<\epsilon$, using $\epsilon = \sqrt{60\times0.05^2} = 0.387$ (here the diameter of the phase space, is $\sqrt{60\times 2^2}=15.492$, so the value of $\epsilon$ used here is $\approx$2.5\% of that diameter). See text for further discussion. 
}
\label{fig:prsh_RP168x168}
\end{figure}

Figure~\ref{fig:prsh_RP168x168} shows an RP for a single $N_T=60$ all-PRSH market session lasting for 7 days of continuous round-the-clock trading, with the strategy-vector $\vec{S}(t)$ recorded hourly, resulting in a $168\times168$-pixel plot (i.e., $7\times24=168$) where the time-difference between rows and columns is one hour. In all the RPs plotted here,  $\vec{S}(t)$ is considered a recurrence of the state $\vec{S}(t-\Delta_t)$ when $|\vec{S}(t)-\vec{S}(t-\Delta_t)|<\epsilon$, using $\epsilon = \sqrt{60\times0.05^2} = 0.387$: the maximum distance possible in this system (i.e., the {\em diameter of the phase-space} in the terminology of the physics literature) is $\sqrt{60\times 2^2}=15.492$ 
(e.g., if $[\vec{S}(t)]_i = +1.0; \forall i $ and $[\vec{S}(t-\Delta_t)]_i = -1.0; \forall i $), so the value of $\epsilon$ used here is $\approx$2.5\% of the maximum distance.

As is clear from visual inspection of the RP in Fig.~\ref{fig:prsh_RP168x168}, there are almost always recurrences to the left and below the diagonal line of identity (LOI) and these recurrences are typically short-lasting, being roughly 10 pixels or less (i.e., 10 hours or less) in the first 50 hours of the session, and then lengthening as the session continues, such that by the end of the session the recurrences are recorded as far-distant as roughly 48 hours previously. A commonly-used RQA summary statistic for this kind of observation is the {\em trapping time} (denoted by $TT$: see Appendix~\ref{appendix:RP} for the definition): for the RP in Fig.~\ref{fig:prsh_RP168x168}, the overall $TT\approx 7.25$ hours: i.e., the system typically spends 7.25 hours within $\epsilon$ distance of any particular $\vec{S}(t)$, before co-evolution drives it away from that area of phase-space; and, given the large areas of unshaded area in the RP, we can see that once it co-evolves away from a particular state after a few hours, it never returns to that state (i.e., no further recurrences are recorded), indicating {\em acyclic} evolution -- i.e., continuous ``progress'' of the co-evolutionary dynamic.

\begin{figure}[hp]
\begin{center}
\includegraphics[width=0.49\linewidth]{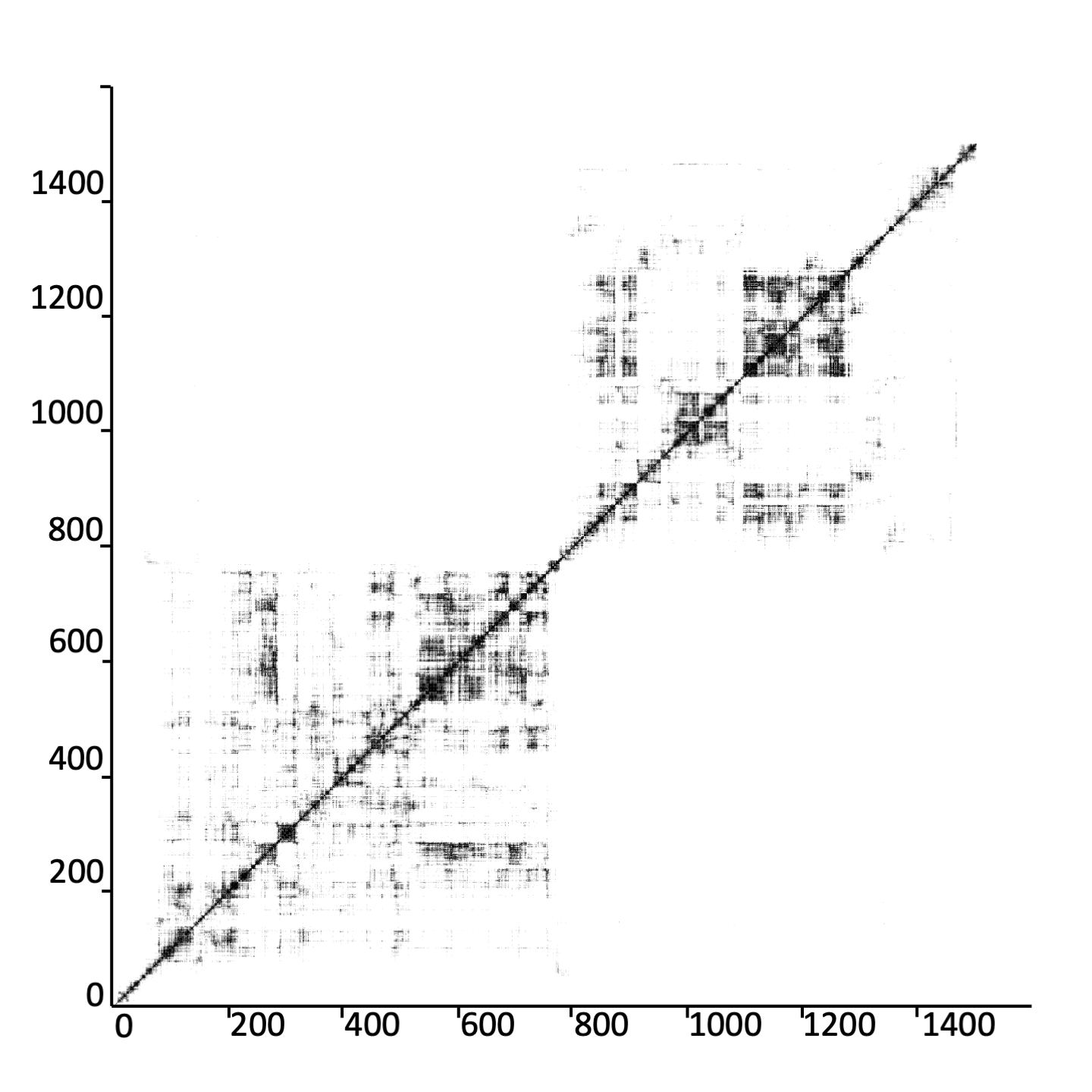}
\includegraphics[width=0.49\linewidth]{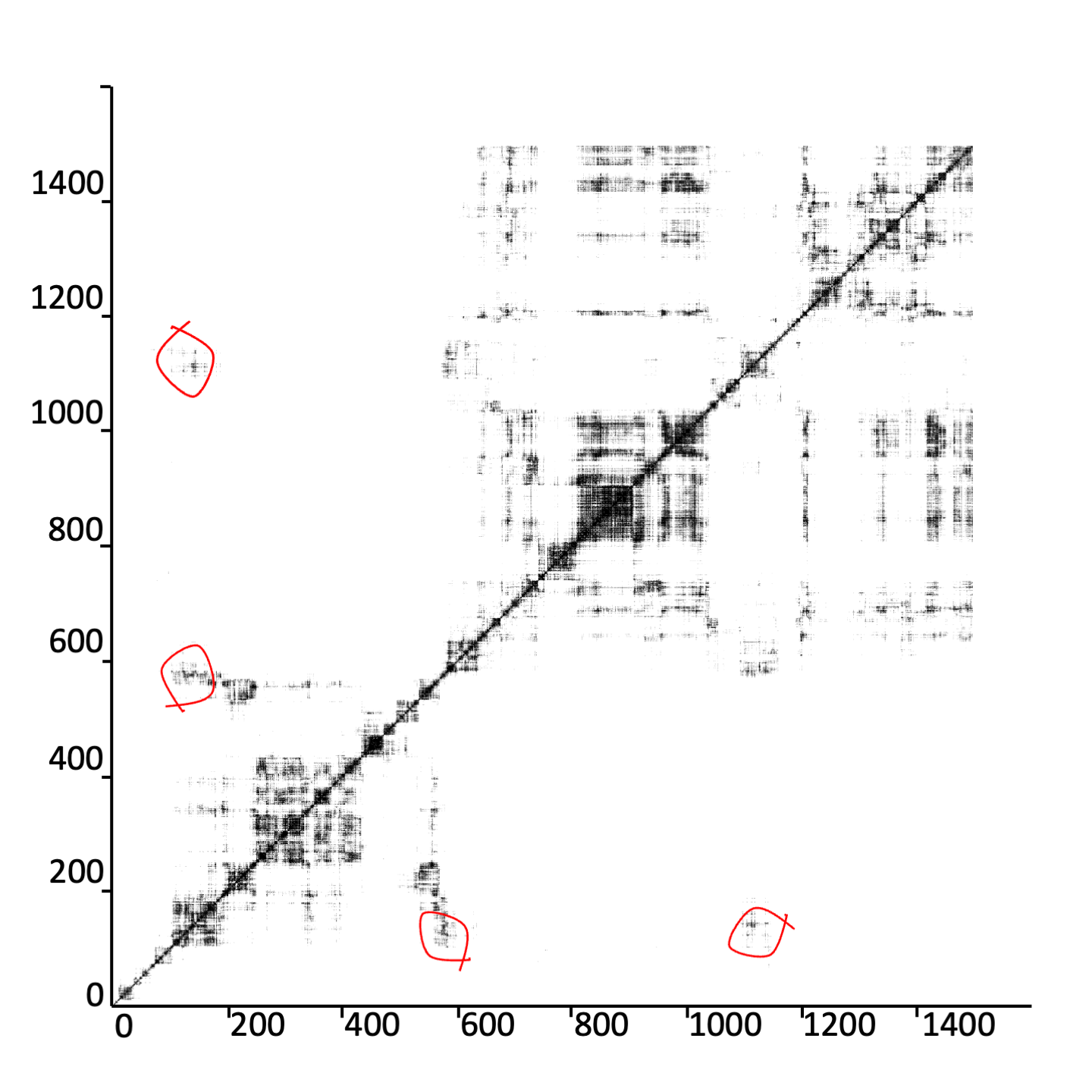}
\includegraphics[width=0.49\linewidth]{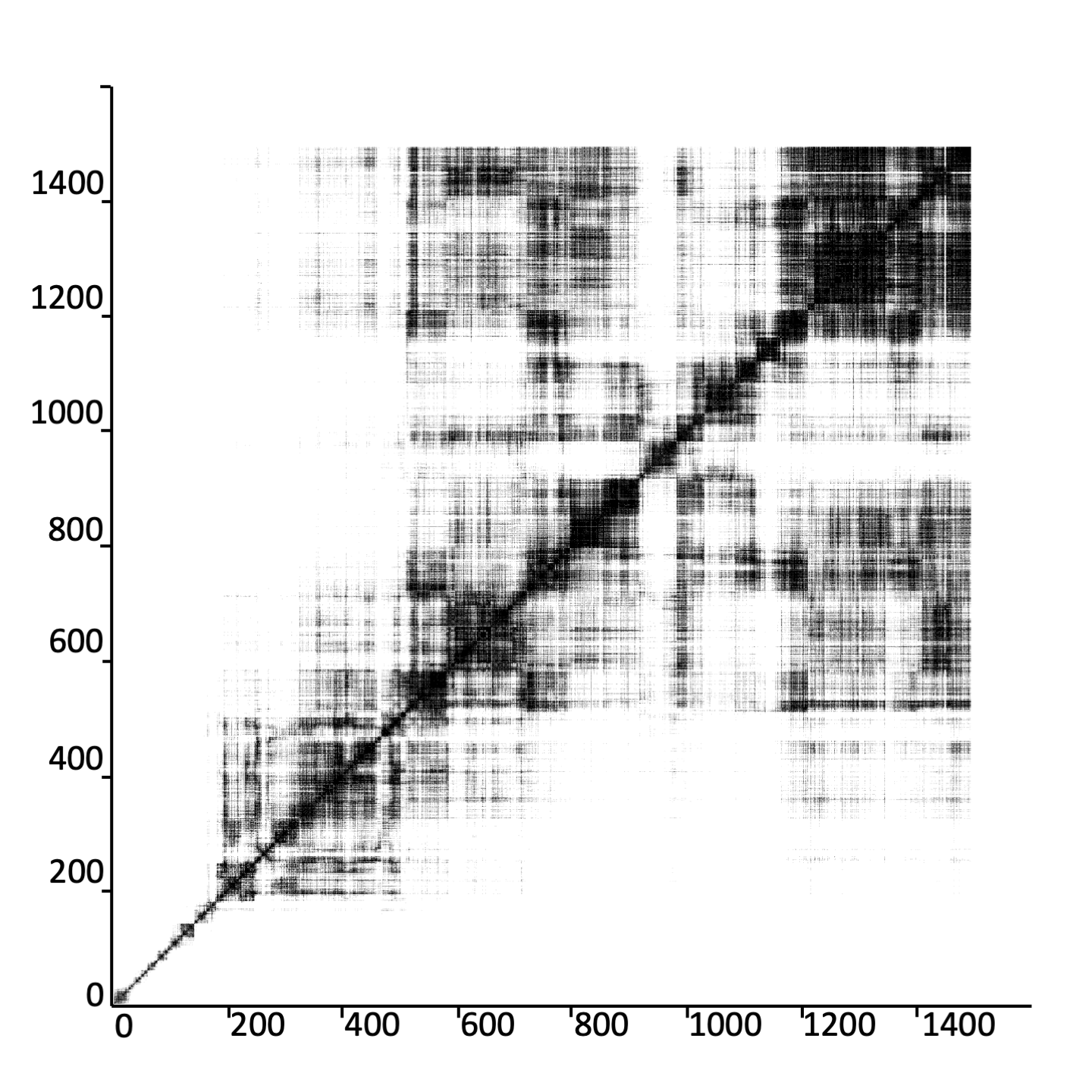}
\includegraphics[width=0.49\linewidth]{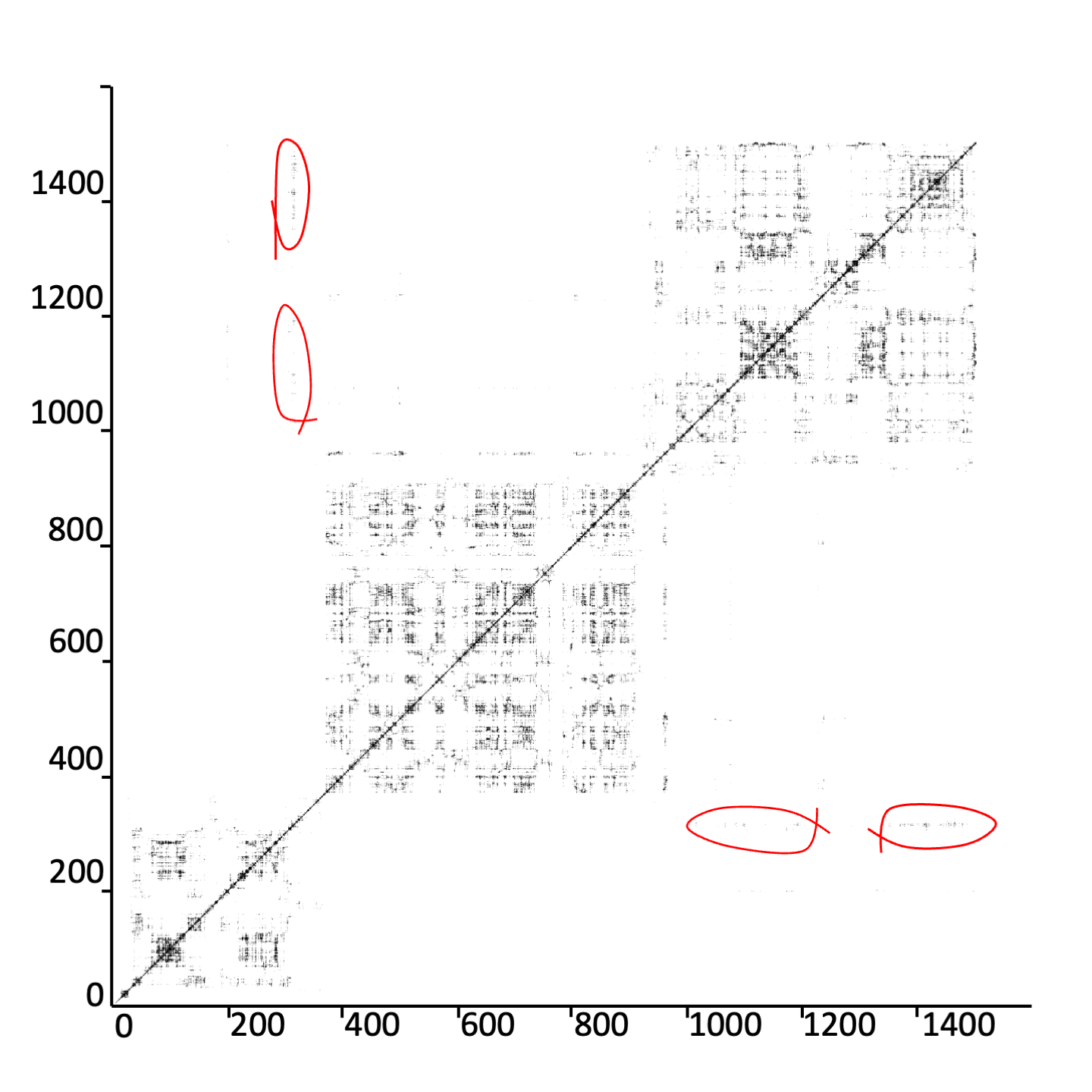}
\includegraphics[width=0.49\linewidth]{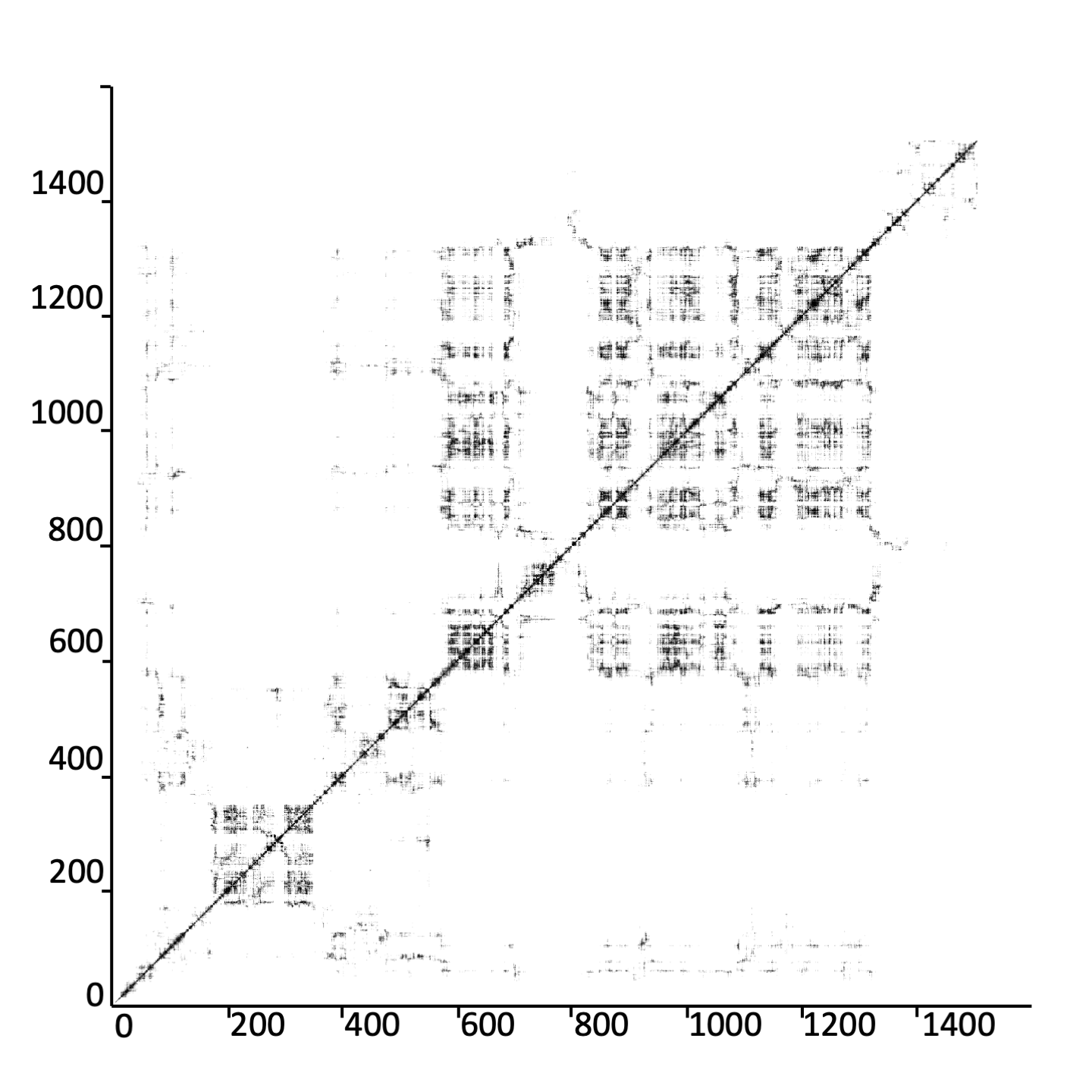}
\includegraphics[width=0.49\linewidth]{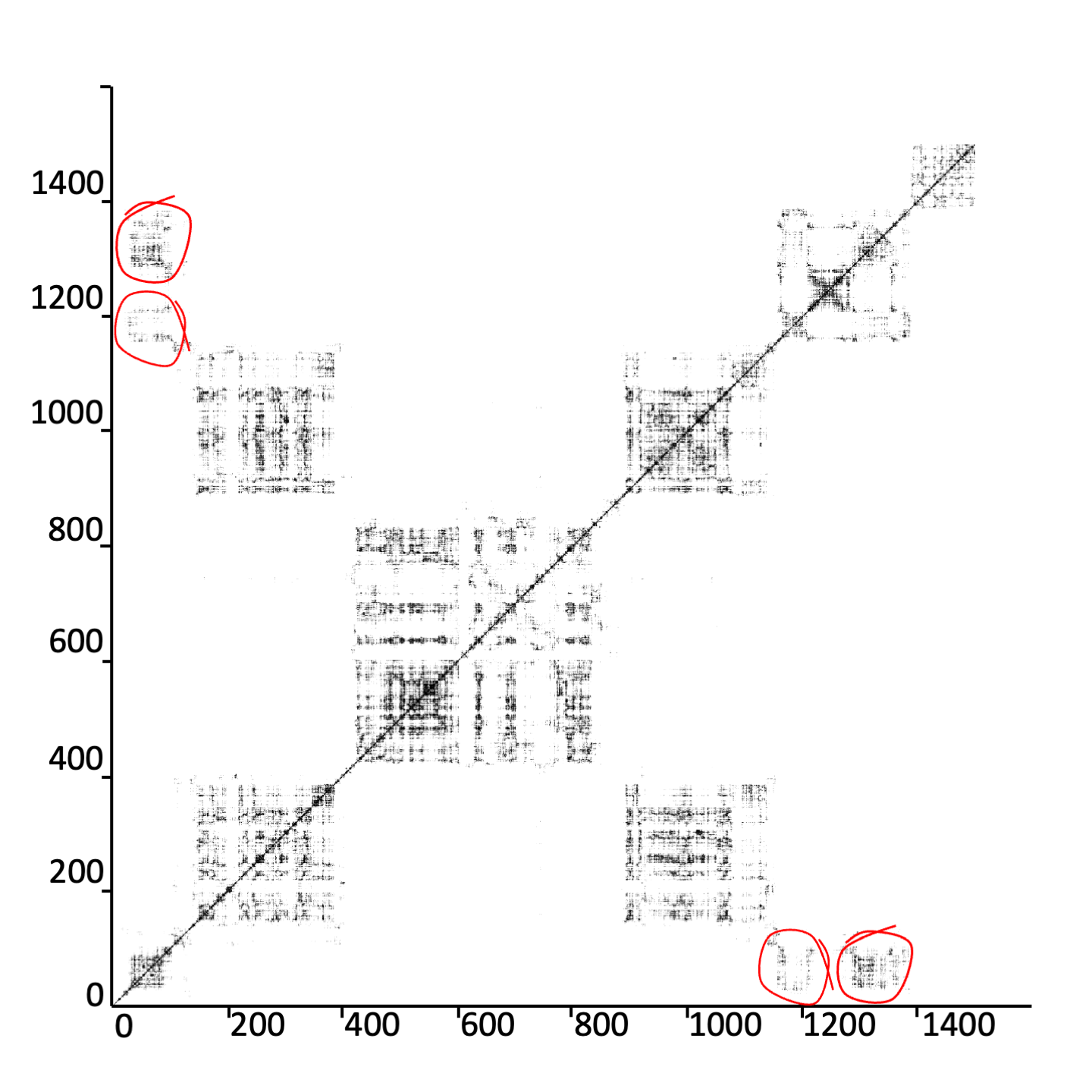}
\end{center}
\caption{Recurrence Plots (RPs) for six {\sc iid} market sessions, each running for ~1,500 simulated days of continuous (24hr/day) trading, each simulated at 60Hz, and each involving multiple transactions per second, i.e. involving on the order of one billion transactions per 1,500-day session. Simulating each market session took approximately 280 hours of wall-clock continuous CPU time on a 16GB Apple Mac Mini (M1 Silicon, 2020), with data frames recorded once per simulated hour, yielding complete RPs that are  36,000$\times$36,000 pixels. For each RP, the numeric labels on both axes shows the number of days elapsed. The RP at upper-left shows the population of traders drifting in one region of strategy space over days $\approx100$ to $\approx200$, then another region over days $\approx200$ to $\approx700$, before evolving into a new region that holds from days $\approx900$ to $\approx1300$, and then continuing to evolve along a transient into previously unvisited areas of strategy space: this can reasonably be described as {\em acyclic} evolution. However in all five of the other sessions, there are clear recurrences, i.e.\ evidence of {\em cyclic} evolution: in the plot at mid-left, the region of strategy-space visited around days $\approx300$ to $\approx500$ is revisted in days $\approx1200$ to $\approx1500$; in the plot at lower-left, the region of strategy-space first visited over days $\approx10$ to $\approx 100$ is revisited sporadically around roughly days 400--600, 700--900, and 1000--1300 as evidenced by the corresponding thin ``trail of dust'' in the RP; for the three plots in the right-hand column, regions of strategy-space first visited in the opening 100-300 days are returned to after many hundreds of days spent in other regions: the recurrences have been highlighted with freehand-drawn ellipses. The lower-right plot is notable in that it shows a recurrence after a transit of more than 1,000 days of co-evolution. 
}
\label{fig:prsh_long_RP_plots}
\end{figure}

Figure~\ref{fig:prsh_long_RP_plots} shows a set of six RPs, from six {\sc iid} market sessions with all parameters set to the same values as used in the experiments illustrated in Figures~\ref{fig:prsh_coev_M1P1init_s-hat_300} and~\ref{fig:prsh_coev_M1P1init_Prof_300}, except these six experiments have each been left to run for 1,500 days. As before, $\vec{S}(t)$ data is recorded hourly, and the traders interact second-by-second simulated at 60Hz, trading around the clock, 24hrs/day; and hence these RPs in their full incarnation are $36000\times36000$ pixels (i.e., $1500\times24=36000$), which of necessity are then downsampled for printed reproduction here.  As is discussed in the caption to Figure~\ref{fig:prsh_long_RP_plots}, five of the six sessions show clear evidence of the co-evolutionary process being {\em cyclic}, in the sense that the system is continuously co-evolving, taking a very large sequence of adaptive steps in the 60-dimensional strategy-space, but eventually it returns to points in strategy space that it previously occupied at an earlier time in the session. And, surprisingly, the path-length of these cyclic transits can be extremely long: more than 1,000 days in one instance. And remember that each trading day in the session is simulated at 24hrs/day, at 60 frames per second resolution (i.e., the simulation timestep is 0.0167s), so the 1,000-day cycle occurred after 5.18Bn timesteps, during which more than a billion transactions will probably have taken place. Simulations run for shorter durations would simply not have revealed these long-term cycles. 

\section{Discussion and Conclusion}

The results presented here are the first from market simulations populated wholly by co-evolving parameterised-response zero-intelligence (PRZI) traders using stochastic hill-climbing as their strategy optimization process (i.e., PRSH), and there are three notable points to highlight:

\begin{itemize}
\item
Despite interacting with each other at sub-second time-resolution, such minimally simple adaptive trader models can exhibit surprisingly rich dynamics, over extremely long timescales, with sequences of punctuated equilibria and with the system's co-evolutionary dynamic cycling back to previously-visited points in its phase-space over periods measured in hundreds of days of simulated trading, in which millions of transactions occur.
\item
The stable attractors in strategy-space are reasonably often neither at the extreme points of the range (i.e., $s_i=\pm1.0$) nor at the mid-point ($s_i=0.0$) but instead are at `hybrid' points along the strategy-space, resulting in trading behaviors (quote-price distributions) with no precedents in the prior ZI-trader literature.
\item
Even though each trading entity is forever engaged in attempting to improve its profitability or efficiency, forever making local adjustments to its own trading strategy, system-level inefficiencies can lock in and persist, apparently indefinitely, because some number of the entities stay trapped on local maxima in the fitness landscape. 

\end{itemize}

There are a wide range of factors that could be explored in further work. For example: the particular form of adaptation used here, the simple stochastic hill-climber of PRSH, will affect the co-evolutionary dynamics; i.e., it might be more likely to result in traders being stuck on local maxima in the fitness landscape, in comparison to other more sophisticated adaptation/optimisation techniques.\footnote{See \cite{cliff_2022_prde} for recent results which show that switching to a different optimization technique does indeed result in fewer traders being trapped on local maxima.}  Also the nature of the supply and demand curves in the market can be expected to affect the dynamics: in the experiments reported here, there was an obvious asymmetry in response, with the vast majority of the population of sellers rapidly co-evolving to be super-urgent (as shown in Figure~\ref{fig:sell_strat_300}) and the buyers then co-evolving toward multi-modal distributions of mainly relaxed strategies in response; with a different supply/demand schedule, this asymmetry could plausibly be reversed. 
One compelling avenue for further research is to conduct experiments that explore the interplay between adaptive PRZI traders such as PRSH, and human traders, interacting and co-adapting in the same market (see \cite{bao_etal_2022} for a recent review) and/or to study markets populated by heterogenous mixes of adaptive strategies, pitting adaptive PRZI traders against other adaptive traders with higher-dimensional strategy-spaces (e.g.\cite{cliff_2009_zip60}); and another is to revisit the possibilities for the market's auction mechanism to be co-evolving along with the set of strategies played by the traders active in that market (see, e.g.: \cite{walia_byde_cliff_2003,phelps_mcburney_parsons_2010}). 
Future papers will explore these and other issues.


\section*{Conflict of interest}

The author declares that he has no conflicts of interest.

\section*{Acknowledgements}

I am very grateful to the anonymous reviewers whose suggestions for changes to an earlier version of this paper improved it significantly, and to Conor Mullaney who pointed out an error in the revised version, now corrected. 

\clearpage
\appendix

\section{Brief Introduction to Recurrence Plots}
\label{appendix:RP}

\begin{figure}[ht]
\begin{center}
\includegraphics[trim=5cm 5cm 5cm 5cm,clip=true,scale=0.2]{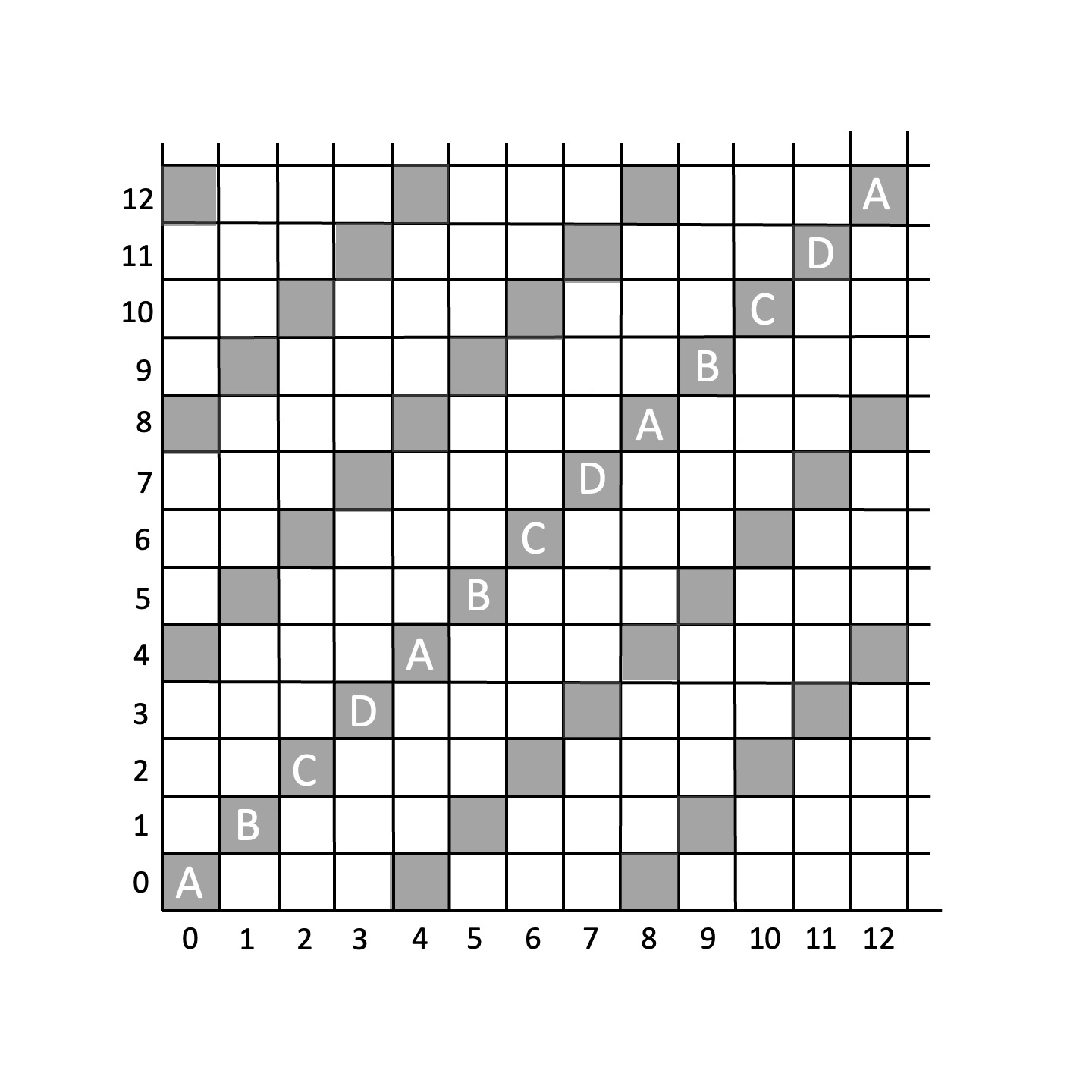}
\end{center}
\caption{Illustrative Recurrence Plot (RP) for a system whose trajectory through phase-space is a four-state limit-cycle, endlessly looping through the sequence ${\sf A} \rightarrow {\sf B} \rightarrow {\sf C} \rightarrow {\sf D} \rightarrow {\sf A} \rightarrow \ldots$. Vertical and horizontal axes both show (discretized) time, and by convention the RP origin point is at lower left. The state at each timestep is indicated by a letter in each cell along the {\em Line Of Identity} (LOI: the diagonal line from lower-left corner to upper-right corner). If the system state (i.e., its point position in phase-space) at time $t_h$ recurred at time $t_v\geq t_h$ then the cells at coordinates $(t_h,t_v)$ and $(t_v,t_h)$ are shaded, and otherwise are left blank thereby signalling no recurrence. In this example the periodically recurring limit cycle leads to diagonal lines at fixed intervals in the RP. The quantitative analysis of patterns of pixellation in RPs, an approach known as {\em Recurrence Quantification Analysis} (RQA), typically involves the calculation of statistics involving the distributions of vertical and/or diagonal lines in the RP.}   
\label{fig:RP_limitcycle}
\end{figure}

\begin{figure}
\begin{center}
\includegraphics[trim=40 40 40 40,clip,width=0.7\linewidth]{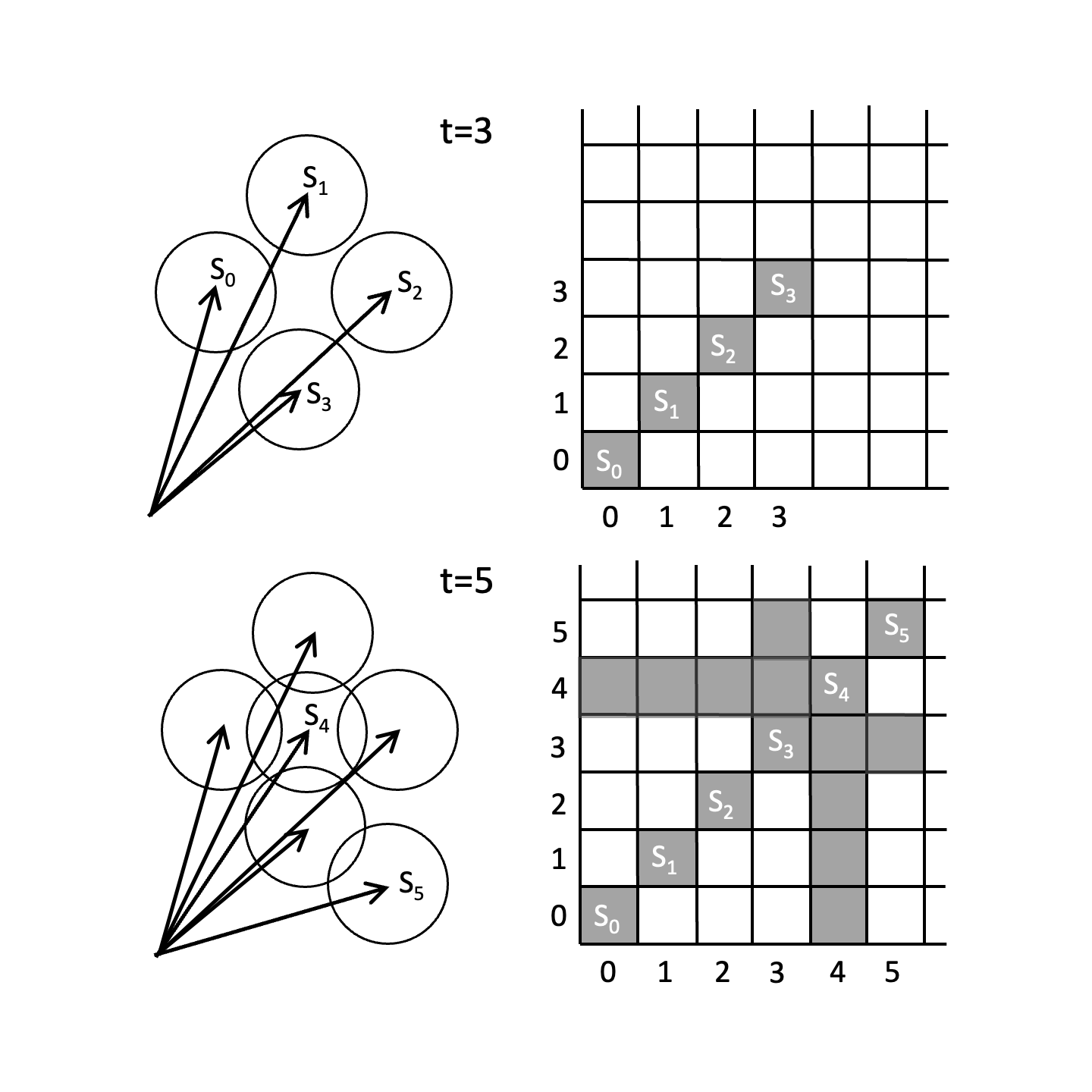}
\end{center}
\caption{
Illustrative synthetic recurrence plot (RP) for a $D-$dimensional dynamical system with state vector $\vec{S}(t)\in{\mathbb R}^D$ that starts at time $t=0$ in state $\vec{S}(0)={\sf S}_0$ and then over the next three timesteps transitions through states ${\sf S}_1$ to ${\sf S}_3$ with no recurrences. The upper pair of figures, labelled $t=3$, illustrates the set of non-recurring state-vectors on the left, and the corresponding RP on the right. Here the end-point of each state-vector is the centre of a $D-$ball (i.e., a solid $D-$dimensional hypersphere) of diameter $\epsilon$, such that if any two balls intersect then the distance between the two vector end-points must be less than $\epsilon$, which is thus counted as a recurrence. As there have been no recurrences by $t=3$, the RP plot only shows shaded cells on the LOI. The lower pair of figures, labelled $t=5$, illustrates the situation after the system has transitioned through state  ${\sf S}_4$ to state ${\sf S}_5$: the ball for ${\sf S}_4$ intersected with the balls for each of states ${\sf S}_0$ to ${\sf S}_3$, so the single state ${\sf S}_4$ is recorded as a recurrence of each of the states ${\sf S}_0$ to ${\sf S}_3$, giving rise to a horizontal line of recurrences on the RP at cells $C_{0,4}$--$C_{3,4}$;  then ${\sf S}_5$ intersects only with ${\sf S}_3$, shown on the RP  as a single shaded cell at $C_{3,5}$. 
}
\label{fig:RP_explainer_horizontal}
\end{figure}

\begin{figure}[htbp]
\begin{center}
\includegraphics[trim=5cm 5cm 5cm 5cm,clip=true,scale=0.2]
{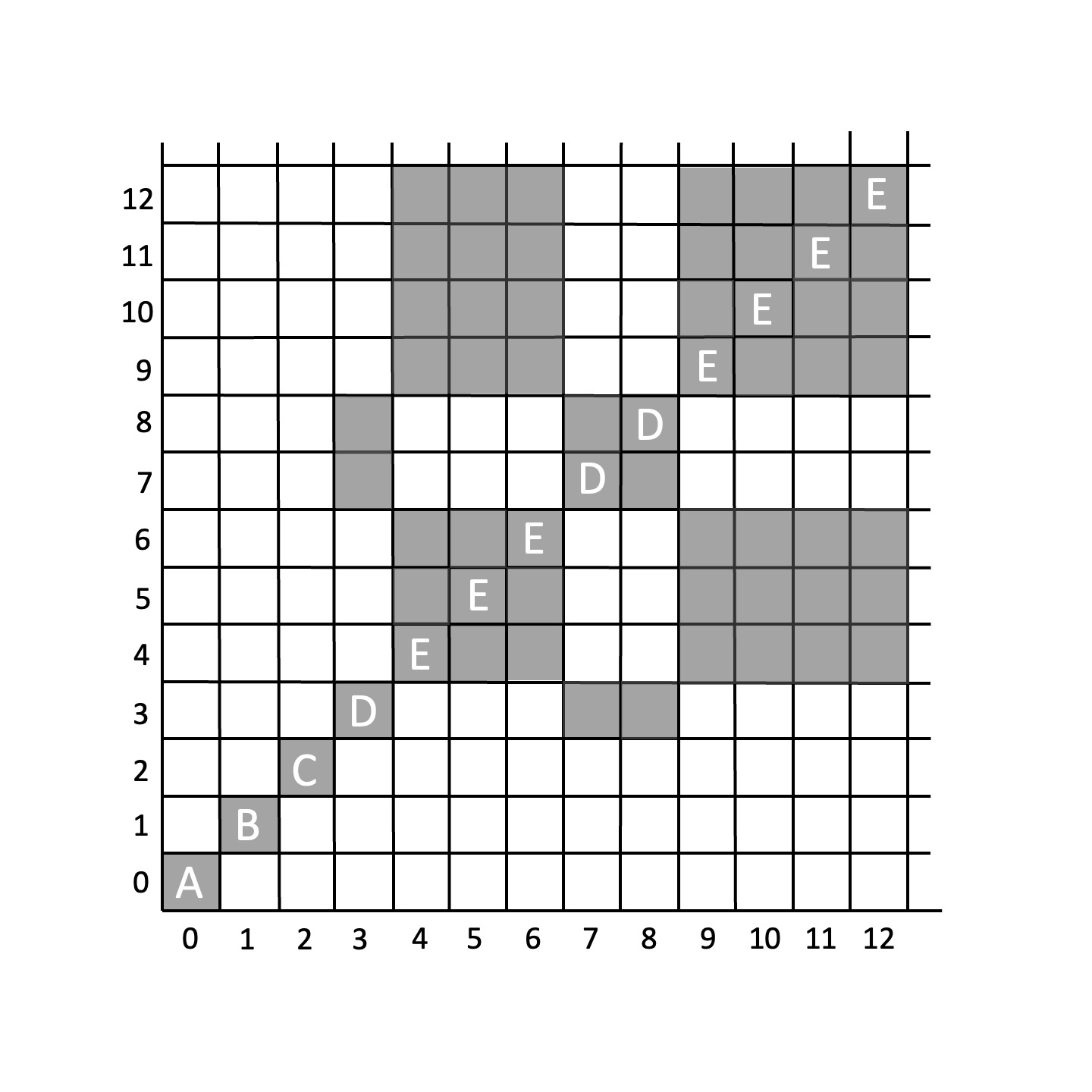}
\end{center}
\caption{Illustrative RP for a system whose trajectory through phase-space involves an initial transient of five states  ${\sf A} \rightarrow {\sf B} \rightarrow {\sf C} \rightarrow {\sf D} \rightarrow {\sf E} $ but which then holds in state {\sf E} for three timesteps, reverts to state {\sf D} for two timesteps, and then settles back into state {\sf E} for the remainder of the plot. Holding at a converged state such as {\sf D} and {\sf E} in this example gives rise to rectangular patches of shading on the RP, giving rise to ``plaid'' or ``tartan'' patterns.}
\label{fig:RP_points}
\end{figure}

For the benefit of any readers unfamiliar with the recurrence plots (RPs) used in Figures~\ref{fig:prsh_RP168x168}  and~\ref{fig:prsh_long_RP_plots}, the diagrams in Figures~\ref{fig:RP_limitcycle}, \ref{fig:RP_explainer_horizontal}, and~\ref{fig:RP_points} illustrate key aspects of this visualization technique for characterising high-dimensional dynamical systems: in their original and simplest incarnation, RPs are square arrays of cells or pixels, that are binary-shaded (e.g.: the pixels are either black or white), with a cell at column $c$ and row $r$ (denoted here by $C_{c,r}$) being shaded if the state of the system at the time associated with row $r$ is a recurrence of a previously-observed system state that occurred at the time associated with column $c$; otherwise unshaded.

In systems where the state at any one time is one of a small number of discrete values, recurrence would usually be defined as strict equality of states. But in many dynamical systems of practical interest, the system state at time $t$ is a $D-$dimensional real-valued vector $\vec{S}(t)$, and for creating an RP any subsequent state $\vec{S}(t+\Delta_t)$ that is within a $D-$dimensional solid hypersphere (i.e., a $D-$ball) centered on $\vec{S}(t)$ with radius $\epsilon$ is considered to be a recurrence of $\vec{S}(t)$. Naturally, the choice of $\epsilon$ is significant: if too large, each new state is registered as a recurrence of all previous states; if too small, it is possible that no recurrences are ever recorded. The RP origin point is normally displayed at lower left, and the diagonal line of cells  $C_{c,r:c=r}$ , referred to as the {\em Line of Identity} (LOI), is always shaded because the distance from any state to itself is zero.

Once an $N$$\times$$N$ RP is created, summary statistics can be calculated by doing simple image-processing such as computing the frequency distribution of lengths of vertical and diagonal lines in the RP, and then calculating summary statistics from those distributions: this approach is known as {\em Recurrence Quantification Analysis} (RQA).  For example, the {\em trapping time} statistic (conventionally denoted by $TT$), 
given $P(v)$ the frequency distribution of vertical lines of length $v$ in the RP,
measures the RP's average length of vertical lines at least as long as $v_{\text{min}}$ (usually $v_{\text{min}}=2)$:
\[
TT=\left( \sum_{v=v_{\text{min}}}^N vP(v) \right) / \left( \sum_{v=v_{\text{min}}}^N P(v) \right)
\]
So for example if an RP has a $TT$ of 6, and the time delta between successive rows/columns on the RP is one hour, then the trapping time is six hours, indicating that on average the system remains within $\epsilon$ of any particular state for six hours. 

For further details of RPs and RQA, see e.g.
\cite{eckmann_etal_1987,marwan_etal_2002,marwan_meinke_2004,march_etal_2005,marwan_etal_2007,marwan_2008, webber_marwan_2015,tolston_etal_2020}.

\bibliographystyle{apalike}
\bibliography{../../dc_bibliography}   

\end{document}